\documentclass[]{aa}

\usepackage[pdftex]{graphicx}
\usepackage{epstopdf}
\usepackage{graphicx}
\usepackage{epsfig}
\usepackage{amsmath}
\usepackage{wasysym}

\usepackage{marvosym}
\usepackage{txfonts}
\pdfoutput=1
\usepackage[pdftex]{color,xcolor}
\usepackage[pdftex]{graphicx}
\usepackage{epstopdf}
\usepackage{mathtools}
\usepackage{natbib}
\usepackage{breqn}
\usepackage{setspace}
\bibpunct{(}{)}{;}{a}{}{,}

\usepackage[]{hyperref}              
\hypersetup{pdfauthor=Christoph Mordasini}
\hypersetup{pagebackref=true, breaklinks=true,colorlinks=true,urlcolor=blue,linkcolor=blue,citecolor=blue,anchorcolor=blue,pagecolor=blue, bookmarks=true,bookmarksopen=true}

\usepackage{etoolbox}

\makeatletter

\patchcmd{\NAT@citex}
  {\@citea\NAT@hyper@{%
     \NAT@nmfmt{\NAT@nm}%
     \hyper@natlinkbreak{\NAT@aysep\NAT@spacechar}{\@citeb\@extra@b@citeb}%
     \NAT@date}}
  {\@citea\NAT@nmfmt{\NAT@nm}%
   \NAT@aysep\NAT@spacechar\NAT@hyper@{\NAT@date}}{}{}

\patchcmd{\NAT@citex}
  {\@citea\NAT@hyper@{%
     \NAT@nmfmt{\NAT@nm}%
     \hyper@natlinkbreak{\NAT@spacechar\NAT@@open\if*#1*\else#1\NAT@spacechar\fi}%
       {\@citeb\@extra@b@citeb}%
     \NAT@date}}
  {\@citea\NAT@nmfmt{\NAT@nm}%
   \NAT@spacechar\NAT@@open\if*#1*\else#1\NAT@spacechar\fi\NAT@hyper@{\NAT@date}}
  {}{}

\makeatother

\def\rbare{R_{\rm bare}}
\def\mbare{M_{\rm bare}}
\def\mearth{M_\oplus}
\def\rearth{R_\oplus}
\def\msun{M_\odot}

\def\mcore{M_{\rm c}}
\def\menve{M_{\rm e}}

\def\f1{f_{\rm I}}
\def\fice{f_{\rm ice}}

\def\beq{\begin{equation}}
\def\eeq{\end{equation}}

\def\t2{\tau_{\rm II}}

\def\sigmas0{\Sigma_{\rm s,0}}

\newcommand{\lj}{L_{\textrm{\tiny \jupiter}}}

\def\s0{S_0}

\newcommand{\lsun}{L_{\odot}}
\def\apjl{ApJ}                % Astrophysical Journal, Letters
\def\apjs{ApJS}               % Astrophysical Journal, Supplement
\def\ao{Appl.Optics}          % Applied Optics
\def\aap{A\&A}                % Astronomy and Astrophysics
\def\({\left(}
\def\){\right)}
\def\<{\left<}
\def\>{\right>}

\begin{document}

\title{Planetary evolution with atmospheric photoevaporation I. \\   Analytical derivation and numerical study of  the evaporation valley and transition from super-Earths to sub-Neptunes}

\subtitle{}
\author{C. Mordasini}
\institute{Physikalisches Institut, University of Bern, Gesellschaftsstrasse 6, CH-3012 Bern, Switzerland} 
\offprints{Christoph Mordasini, \email{christoph.mordasini@space.unibe.ch}}
\date{Received 25.3.2019 / Accepted 5.2.2020}

\abstract 
{Observations have revealed in the Kepler data a depleted region separating smaller super-Earths from larger sub-Neptunes. This can be explained as an evaporation valley  between planets with and without H/He that is caused by atmospheric escape.}{We want to analytically derive the valley's {locus} and understand how it depends on planetary properties and stellar XUV luminosity. We also want to derive constraints for planet formation {models}.}{First, we conduct  numerical simulations of the evolution of close-in low-mass planets with H/He undergoing escape. We make parameter studies with grids {in} core mass and orbital separation, and vary the post-formation H/He mass, the strength of evaporation, and the {atmospheric and} core composition. Second, we develop an analytical model for the valley locus.}
{We find that the bottom of the valley quantified by the radius of the largest stripped core $R_{\rm bare}$ at a given orbital distance depends only  weakly on  post-formation H/He mass. The reason is that a high initial H/He mass means that there is more {gas} to evaporate, but also that the planet density is lower, increasing mass loss. Regarding the stellar XUV-luminosity, $R_{\rm bare}$ is found to scale as $L_{\rm XUV}^{0.135}$. The same weak dependency applies to the efficiency factor $\varepsilon$ of energy-limited evaporation. As found numerically and analytically,  $R_{\rm bare}$ varies a function of orbital period $P$ for a constant $\varepsilon$  as $P^{-2 p_{\rm c}/3}\approx P^{-0.18}$ where  $M_{\rm c} \propto R_{\rm c}^{p_{\rm c}}$ is the mass-radius relation of  solid cores.  $R_{\rm bare}$ is about 1.7 $R_{\oplus}$ at a 10-day orbital period for an Earth-like composition, increasing linearly with ice mass fraction.}
{The numerical results are explained very well with the analytical model where complete evaporation occurs if the temporal integral over the stellar XUV irradiation absorbed by the planet is larger than binding energy of the envelope in the gravitational potential of the core. The weak dependency on the post-formation H/He means that the valley {does not strongly} constrain gas accretion during formation. But {the} weak dependency on primordial H/He mass, stellar $L_{\rm XUV}$ {and $\varepsilon$} could be the reason why {observationally} the valley is {so clearly} visible, and why {theoretically} various models find  similar results. At the same time, given the large observed spread of $L_{\rm XUV}$, the dependency {on it} is still strong enough to explain why the valley is not completely empty.} 

 \keywords{Stars: planetary systems -- Planets and satellites: formation -- Planets and satellites: interiors -- Planets and satellites: physical evolution -- Planets and satellites: atmospheres -- Planets and satellites: composition}

\titlerunning{The evaporative transition from super-Earths to sub-Neptunes}
\authorrunning{C. Mordasini}

\maketitle

\section{Introduction}
An intriguing result of the starting geophysical characterization of extrasolar planets is the large diversity in mean densities of low-mass and intermediate-mass planets. This has been detected observationally in the last few years \citep[e.g.,][]{weissmarcy2014,jontof-hutterford2016,guentherbarragan2017}. At planetary masses of about 1 to 20 Earth masses, observations suggest a variation of more than one order in magnitude in mean density \citep[e.g.,][]{hatzesrauer2015,jinmordasini2018}. This implies that the planetary bulk compositions must vary widely from rocky, Earth-like interiors for dense planets like CoRoT-7b or Kepler-93b \citep{legerrouan2009,dressingcharbonneau2015} to low-mass planets that contain significant amounts of  H/He like the Kepler-11 planets \citep{lissauerjontof2013}. The analysis and interpretation of this transition from solid Earth-like rocky super-Earth (and potentially icy planets) to low-mass sub-Neptune planets with primordial H/He has subsequently attracted a lot of attention on the theoretical statistical side \citep[e.g.,][]{rogers2015,wolfganglopez2015,chenkipping2017}. 

In order to understand these planetary density measurements at present day, it is necessary to understand how the planets have evolved in time in the past. For this purpose, numerical models were developed that combine the long-term post-formation thermodynamical evolution (cooling and contraction) of low-mass planets with atmospheric escape \citep{owenwu2013,lopezfortney2013,jinmordasini2014,chenrogers2016}. 
Atmospheric escape is a process that has been observed for several types of planets like Hot Jupiters \citep[e.g.,][]{vidal-madjaretangs2003} or close-in Neptunian planets \citep[e.g.,][]{ehrenreichbourrier2015,bourrierlecavelier2018}.

These theoretical models were then used in parameter studies to assess the effect of evaporation as a function of planet mass and orbital distance. This showed that the evaporation of primordial H/He envelopes is a process of prime importance for shaping the radii of  typical ``Kepler'' planets, i.e., close-in, low-mass planets with radii smaller than about 4 $\rearth$ at orbital distances of less than a few 0.1 AU. Even more remarkably, despite the different focus and independency of the aforementioned theoretical models, all these studies similarly predicted in a rather rare congruence of theoretical models that on a population-level, atmospheric escape should lead to a characteristic imprint: a depletion of the number of planets in a specific region of the orbital distance - planet radius plane that runs diagonally downwards with increasing distance. This feature was {called} the ``evaporation valley'' by \citet{jinmordasini2014}. The corresponding 1D radius distribution was found to be bimodal, with a local minimum separating smaller super-Earth planets that have lost all H/He from larger sub-Neptune planets that kept some of it. This defines also more clearly these two previously blurred planet types.
  
As pointed out by \citet{owenwu2013}, the Kepler radius distribution known in 2013 contained already a hint of a bimodality compatible with the theoretically predicted one, but its significance was unclear, and the minimum was much less prominent than the deep minimum predicted theoretically for example in \citet[][their Figs. 13 and 14]{jinmordasini2014}. Observationally, a limiting factor at that time were the relatively high uncertainties in the planetary radii because of poorly constrained host star properties in the original Kepler input catalogue. The interest was therefore significant when \citet{fultonpetigura2017} showed that for better-constrained stellar parameters \citep{petigurahoward2017,johnsonpetigura2017}, there is indeed a deep valley and minimum also in the observational data. 

\citet{owenwu2017} and \citet{jinmordasini2018} then showed that the observed locus of the {radius} valley and the corresponding minimum in the 1D radius distribution not only agree with the previously theoretically predicted evaporation valley, but that the locus even shows that the dominant composition of the cores should be Earth-like (iron and silicates) without much ice, a strong constraint for formation models. Thus, this represents one of the not so numerous cases where a prediction from planet formation and evolution theory was later clearly confirmed by observations.

{Besides photoevaporation, several alternative hypotheses have been proposed for the origin of the observed radius gap. First, in the core-powered mass-loss  hypothesis, a planetary core's internal luminosity drives the loss of its atmosphere \citep{ginzburgschlichting2016}. Core-powered mass-loss is also able to reproduce the observed radius valley as a function of orbital period \citep{ginzburgschlichting2018,guptaschlichting2018}. Similar to photoevaporation studies, \citet{guptaschlichting2018} furthermore also find that the observations are consistent with predominantly rocky cores. Second, the valley could also be caused by two distinct formation pathways for  planets above and below the gap, with those above the valley being water-worlds \citep{zengjacobsen2019}. Third, planetesimal impacts can also create a similar radius gap  depending on whether atmospheres grow or deplete in collisions  \citep{wyattkral2019}.}

\citet{fultonpetigura2017} did not yet determine the slope of the gap as a function of orbital distance (or irradiation flux). This is however of central importance, as a transition from solid to planets with primordial H/He because of evaporation leads to a negative slope, whereas a formation in a gas-free environment after disk dispersal should lead to a positive slope \citep{lopezrice2018}. The paper of \citet{vaneylen2018} who used astroseismology to even better constrain the stellar radii finally showed that the valley has a negative slope, consistent with an evaporative transition from super-Earth to sub-Neptunes, where the lower boundary of the valley which corresponds to the largest bare core $R_{\rm bare}$ as function of semimajor axis $a$ is approximately found at
\beq
R_{\rm bare}(a)\approx1.6\pm0.1 \times  \left(\frac{a}{0.1\mathrm{AU}}\right)^{-0.08^{+0.02}_{-0.05}} \rearth
\eeq
For the middle of the gap, \citet{vaneylen2018} found a steeper slope, scaling approximately as $a^{-0.15\pm0.05}$. For comparison, \citet{lopezrice2018} find theoretically a dependency like $a^{-0.22}$.
 
\citet{fultonpetigura2018} confirmed the existence of the gap and quantified how empty it is using even better constrained stellar radii, thanks to new Gaia parallaxes (see also \citealt{bergerhuber2018}). They also found that the characteristic features (maxima of Super-Earths and sub-Neptunes, valley) shift to larger radii for more massive stars. This trend is also seen if one considers a wider stellar mass range \citep{wu2018}. 
{Recently, \citet{martinezcunha2019} find that the valley has a slope scaling as $a^{-0.17\pm0.03}$ based on a spectroscopic analysis of the California-Kepler Survey, in agreement with the findings of  \citet{vaneylen2018}. On the other hand, based on a machine learning approach, \citet{MacDonald2019} report a much steeper slope, going as  $a^{-0.48^{+0.13}_{-0.17}}$.} The theoretical results found in the present paper are compared to the observational results below in Sect. \ref{sect:compobs}.

Coming back to the predictions of theoretical models, one can consider for example the results of  \citet{jinmordasini2014}.  In this work, the population-wide effect of evaporation was studied with a model that does not only simulate the long-term evolution under the effect of evaporation but self-consistently couples this to a planet formation model for the accretion of the solid core and H/He, orbital migration, and disk evolution \citep{alibertmordasini2005,mordasinialibert2012b}. This yields the initial conditions for the evolution (for as recent study also combining formation and evolution, see \citealt{carreraford2018}). Note that in the present paper, ``core'' is always used in the astrophysical, and not geophysical sense. It thus denotes the entire solid part of the planet, including the metallic, silicate, and possibly ice parts, but not the H/He envelope.

\begin{figure}[tb]
    \centering
    \includegraphics[width=0.5\textwidth]{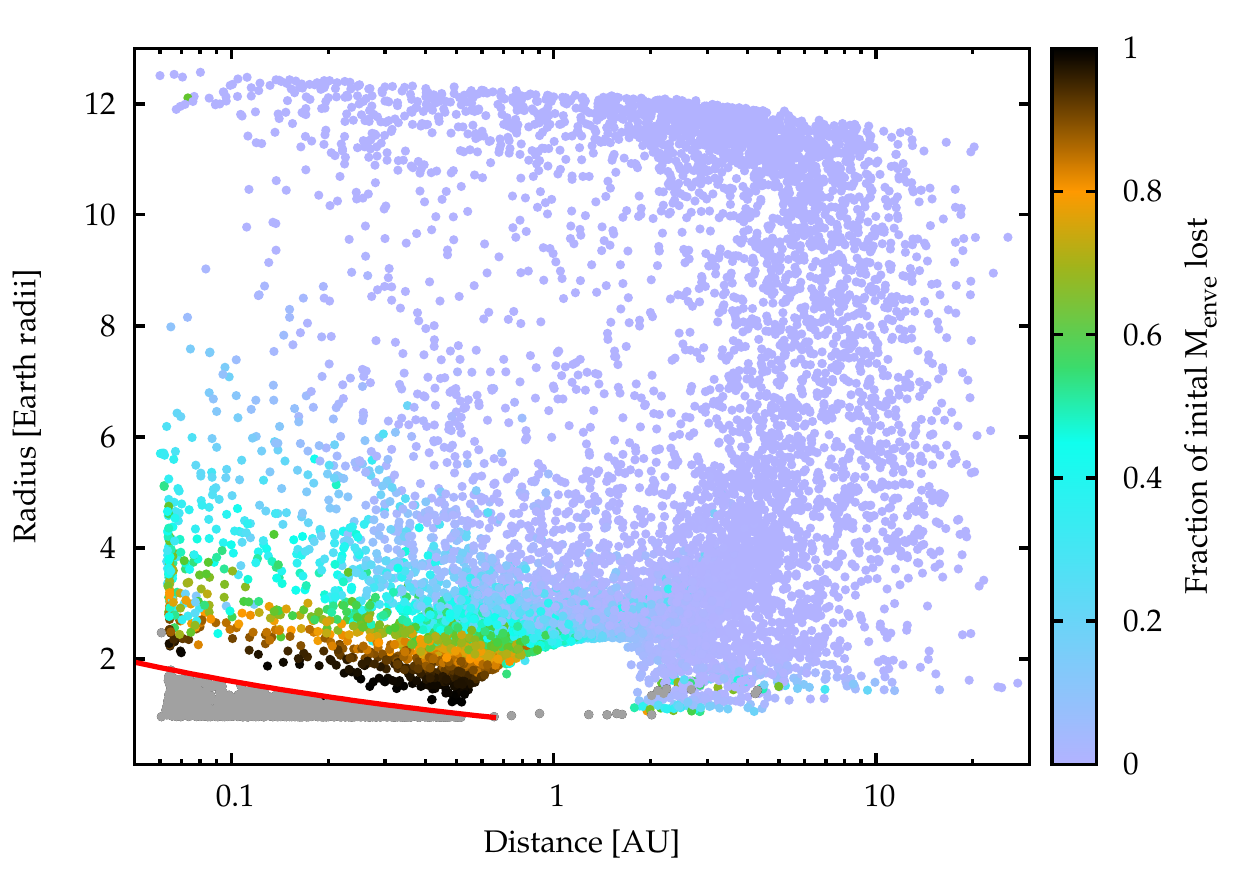}
   \caption{Distance-radius diagram from an {early} population synthesis calculation around 1 $\msun$ stars at 5 Gyr.  Colors give the fraction of the initial H/He envelope mass that was lost. Gray dots are planets that have lost all H/He in the ``triangle of evaporation''. The red line shows the upper limit at $R_{\rm bare}(a)$ given by Eq. \ref{eq:rbarefromoldsynt}. Just above the line, there is the  valley of evaporation. Adapted from  \citet{jinmordasini2014}.}\label{fig:pop}
\end{figure}

Figure \ref{fig:pop} shows an adapted version of a $a-R$ plot of \citet{jinmordasini2014}. Analogous results were found by \citet{lopezfortney2013} and \citet{owenwu2013}. The main result visible in the figure is that close-in low-mass planets lose all H/He whereas more massive planets at larger distances keep it. This leads to a diagonal separation of solid planets from planets with H/He in the orbital distance - planetary radius (or mass) plane, or in other words, a triangle-like region of planets without H/He envelopes. This ``triangle of evaporation'' \citep{jinmordasini2018} with solid planets is clearly separated from the planets with H/He. These two types of planets have very different mean densities as even a tiny amount of H/He strongly increases the radius \citep[e.g.,][]{valenciaguillot2013}. 

\citet{jinmordasini2014} and \citet{jinmordasini2018} found numerically in the population syntheses that for solar-like stars the most massive planet as a function of distance that has lost all H/He and was stripped to a bare solid planet has a radius $R_{\rm bare}$ of about
\beq\label{eq:rbarefromoldsynt}
R_{\rm bare}(a)\approx 1.6\times\left(\frac{a}{0.1\mathrm{AU}}\right)^{-0.27} \ \rearth 
\eeq
and a corresponding mass $M_{\rm bare}$ of about 
\beq
M_{\rm bare}(a)\approx 6 \times\left(\frac{a}{0.1\mathrm{AU}}\right)^{-1} \ \mearth
\eeq
This is shown by the red line in Fig. \ref{fig:pop}. The quoted values are for an Earth-like composition of the core. For a core composition with 0.75\% ice in mass, the separating values at 0.1 AU are shifted from 1.6 to 2.3 $\rearth$ and from 6 to 8 $\mearth$, with a very similar distance dependency \citep{jinmordasini2018}. This transition mass (radius) can be derived analytically as will be demonstrated in the second part  of the present paper (Sect. \ref{sect:analyticalstudy}) from essential energy comparison arguments\footnote{\citet{owenwu2017} have independently derived a comprehensive analytical theory for the evaporative transition. Except for the comparison with the {subsequentely} published observational studies (like \citealt{fultonpetigura2017,fultonpetigura2018} and \citealt{vaneylen2018}), the analytical derivation  and the numerical calculations shown in the paper here were completed before we became aware of their work. Comparisons with the \citet{owenwu2017} study can be found in Sect. \ref{sect:compOW17}.}

The analytical (and numerical) results  show how the transition from solid planets to planets with H/He and thus the evaporation valley depend on orbital distance, stellar XUV-luminosity, efficiency of evaporation, and post-formation envelope mass. These dependencies (and others, like the stellar mass) are typical for evaporation as the mechanism responsible for envelope loss. In general, they should differ from the imprints and dependencies of other mechanisms that can also lead to the loss of the envelope, like impacts  {\citep{liuhori2015,SchlichtingSari2015,InamdarSchlichting2015,bierstekerschlichting2019,wyattkral2019}} or  mass loss powered by the luminosity of a planet's cooling core { \citep{ginzburgschlichting2018,guptaschlichting2018,guptaschlichting2019}}.

This means that it might become possible to disentangle the contribution of these different processes with (future) surveys like TESS \citep{Rickerwinn2014}, CHEOPS \citep{fortierbeck2014}, and PLATO \citep{rauercatala2014}, potentially combined with RV observations. They will allow to map the mean planetary density in the mass or radius-distance plane, which would be important to improve formation, evolution, and evaporation theory. In the ideal case, they will allow to even understand the dependency on additional relevant parameters like the stellar type and activity ($L_{\rm XUV}$)  or the age of the planets. This would render the observations even more constraining.   

When considering impacts as the reason of envelope loss, it appears likely that it will lead to a fuzzy transition due to the stochastic nature of impacts. Also evaporation will in reality not be as deterministic and of homogeneous efficiency for all planets as in the idealized simulations presented here. Even at fixed mass and distance, it will differ from planet to planet as it depends on the ability of the gas to cool and thus its atmospheric composition \citep[e.g.,][]{johnstoneguedel2018}. It will also differ from planetary system to planetary system, as stars can have a wide variety in their luminosity in the XUV-wavelength domain at young ages \citep{tujohnstone2015}. We do however find numerically, and show analytically, that the transition is rather weakly dependent on these parameters. Nevertheless, systems of several planets are of special interest, as there all planets have experienced the same irradiation history from the host star {\citep[cf.][]{owencampos2019}}. This makes in particular Kepler 36 \citep{carteragol2012} with its two closely spaced planets of very different densities an important benchmark case. As will be discussed in Section \ref{sect:kepler36}, this system can be well explained by the calculations presented here without any special model tuning (as previously shown by \citealt{lopezfortney2013,owenmorton2015}). Connecting compositional constraints to the ones from the dynamic system architecture (existence of mean motion resonances, circular versus eccentric orbits, etc.) could be an interesting pathway to allow further insights into the mechanism that have led to the formation and evolution of a specific system \citep{QuillenBodman2013,DawsonLee2015,carreraford2018}.

In this paper, we want to quantify and understand the shape and locus of the valley of evaporation and its dependency on the initial (post-formation) properties of the planets like the envelope-to-core mass ratio, or the strength of evaporation. We work in the limiting assumptions that all planets start with H/He envelope. This is not what is expected because of impacts and/or a formation after the protoplanetary gas disk is gone, and thus we do not think that the actual transition from solid to  planets with H/He will be as clean as reported here. Rather, by comparing the transition due to evaporation found here with observations, should help to disentangle different mechanism. 

The structure of the paper is as follow: In the first part of the paper starting with  Sect. \ref{sect:model}, we conduct numerical simulations of the thermodynamical evolution of low-mass planets undergoing evaporation. We present our model in Sect. \ref{sect:numericalmodel} and the initial conditions in Sect. \ref{sect:initialconditions}. The results from a parameter study covering different planet masses and distances are shown in Sect. \ref{sect:numresults}. We identify the location of the transition and test its dependency on the initial envelope mass, the strength of evaporation, the composition of the core (rocky/icy), the envelope opacity {and the envelope metallicity}. In the second part, Sect. \ref{sect:analyticalstudy}, we develop a simply analytical model based on energy comparison to explain the numerical results and the location of the observed valley of evaporation. We derive the governing equations in Sect. \ref{sect:derivationmbare} and give the final result in Sect. \ref{sect:finalresultanalytical}. Section \ref{sect:compobs} compares the theoretical results with observations. The summary and conclusions are finally given in {Sects. \ref{sect:summary} and \ref{sect:conclusions}}.

\section{Numerical  study}\label{sect:model}
The basic idea of the numerical study is simple: we use our evolutionary model that includes the effect of evaporation  to conduct a parameter study of the evolution of planets lying in the distance and mass range affected by evaporation visible in Fig. \ref{fig:pop}. For this, we follow the evolution of  a high number of simulated planets on grids of core masses and orbital distances, studying the impact of several important parameters on the transition from solid to planets with H/He. For the initial conditions, namely the post-formation envelope-to-core mass ratio and luminosity, we use as a starting point the results from population synthesis calculations of the formation of planets via core accretion \citep{mordasinialibert2012c}, and then vary these initial conditions. Regarding these planet formation simulations, the goal is to see if by comparing the observed and simulated locus of the valley, we can put constraints on them. One could for example a priori think that the locus and slope of the valley depend on the post-formation envelope mass as a function of a planet's core mass and orbital distance (which is not the case, as we will find below).

\subsection{Simulation setup}\label{sect:numericalmodel}
We first summarize the numerical model, then we describe its initial conditions.

\subsubsection{Planet evolution model with atmospheric escape}
Our planet evolution model \texttt{completo21} was already described in several publications \citep{mordasinialibert2012b,jinmordasini2014,lindermordasini2018}, therefore we here only give a short overview. It models the temporal evolution of planets (their cooling and contraction) by integrating numerically their 1D interior structure, taking into account the loss of H/He because of atmospheric escape. 

Our model assumes that planets consist of a solid differentiated part (the core) consisting itself of iron, silicates, and, if the planet accreted outside of the iceline, water ice. These substances are described by temperature-independent modified polytropic equations of state EOS giving the density $\rho$ as a function of pressure $P$ with the expression \citep{seagerkuchner2007}
\beq
\rho(P)=\rho_{0}+c P^{n}.
\eeq
The parameters $\rho_{0}, c$, and $n$ for iron, perovskite, and water ice are taken from \citet{seagerkuchner2007}. The radius of the core is then found by solving the 1D spherically symmetric equations of mass conservation and hydrostatic equilibrium
\begin{align}
\frac{d m}{d r}&=4 \pi r^{2} \rho &\frac{d P}{dr}&=-\frac{G m}{r^{2}} \rho
\end{align}
where $r$ is the distance from the planet's center, $m$ the enclosed mass, and $G$ the gravitational constant. When solving the equations, a possible pressure exerted by the envelope on the outer boundary (surface of the core) is taken into account which is however only important for massive envelopes (see \citealt{mordasinialibert2012c}). The dependency of the core's radius on the temperature is neglected because the influence is only small \citep{grassetschneider2009}. 

The gaseous envelope consisting of H/He is described by the EOS of \citet{saumonchabrier1995}. The interior structure of the envelope is found in the  1D spherically symmetric approximation by solving the equations of mass conservation, hydrostatic equilibrium, energy generation, and energy transport (with $T$ the temperature)
\begin{alignat}{2}
\frac{\partial m}{\partial r}&=4 \pi r^{2} \rho    &\quad  \quad \frac{\partial P}{\partial r}&=-\frac{G m}{r^{2}}\rho    \\
\frac{\partial l}{\partial r}&=0            & \frac{ \partial T}{\partial r}&=\frac{T}{P}\frac{\partial P}{\partial r}\nabla(T,P)          
\end{alignat}
where we make in the energy equation the approximation that the intrinsic luminosity $l$ is radially constant at a given time, and use total energy conservation to follow the temporal evolution, following the method introduced in \citet{mordasinialibert2012b}.

We use the Schwarzschild criterion to decide whether the energy transport occurs via radiative diffusion or convection, such that $\nabla$ is always the smaller of the radiative and the adiabatic gradient.  For the intrinsic luminosity of the planets, the cooling and contraction of  the H/He envelope and of the core are included. For the latter we assume a constant heat capacity of $10^{7}$ and $6\times 10^{7}$ erg g$^{-1}$ K$^{-1}$ for rocky material and ices, respectively. Our evolution model reproduces the usual results \citep{fortneyikoma2011} for the evolution of the gas and ice giant planets in the solar system \citep{lindermordasini2018}: Jupiter's and Uranus' luminosity at present day is recovered, but Saturn is less bright in the model than observed (potentially because of a helium rain which we do not include), whereas Neptune is predicted to be much more luminous than observed (which could be to a less efficient energy transport because of compositional gradients, which is also not considered in the model). The model also reproduces very well the evolutionary simulations of \citet{lopezfortney2014} for a 5 $\mearth$ planets with a 1\% H/He envelope \citep{lindermordasini2018}.

The outer boundary condition of the H/He envelope (the atmosphere) is described by a two stream double gray irradiated atmosphere in the form of an improved version of the \citet{guillot2010} analytical solution, as described in \cite{jinmordasini2014}. For a planet that has an intrinsic temperature $T_{\rm int}$ (resulting from the planet's intrinsic luminosity) and an equilibrium temperature $T_{\rm equi}$ (resulting from stellar irradiation), the resulting temperature as a function of IR optical depth $\tau$ is
\begin{dmath}
T^{4}=\frac{3 T_{\rm int}^{4}}{4}\left(\frac{2}{3}+\tau\right)+\frac{3 T_{\rm equi}^{4}}{4}\left(\frac{2}{3}+\frac{2}{3\gamma} \\ \times \left[1+\left(\frac{\gamma \tau}{2}-1 \right)e^{-\gamma \tau} \right]  +\frac{2 \gamma}{3}\left(1-\frac{\tau^{2}}{2}\right)E_{2}(\gamma\tau)\right)
\end{dmath}
where $\gamma$ denotes the ratio of the mean opacity in the visual $\kappa_{\rm v}$ to the  mean opacity in the thermal infrared  $\kappa_{\rm th}$, while $E_{2}$ is an exponential integral. The parameter $\gamma$ is tabulated in \citet{jinmordasini2014}.  The opacities correspond to a condensate-free gas of solar composition \citep{freedmanlustig-yaeger2014}. 

In our evaporation model (described in \citealt{jinmordasini2014}), atmospheric escape is assumed to be hydrodynamic, and to occur at lower incident EUV-fluxes in the energy-limited regime \citep{watsondonahue1981} where
\beq\label{eq:mdotenergylim}
\frac{d M_{\rm UV,e-lim}}{d t}=-\frac{\varepsilon_{\rm UV} \pi F_{\rm UV} R_{\rm UV}^{3}}{G M}.
\eeq 
In this equation, $M$ is the planet's mass, $\varepsilon_{\rm UV}$ is an efficiency factor, and $R_{\rm UV}$ the planetary radius where EUV radiation is  absorbed which is estimated as in \citet[][]{murray-claychiang2009}. Following this work, the efficiency factor is  set to 0.32, and assumed to be constant. At high EUV-fluxes, the radiation-recombination limited regime is considered, where the mass loss rate is \citep{murray-claychiang2009}
\beq
\frac{d M_{\rm UV,rr-lim}}{d t}=-4 \pi \rho_{\rm s} c_{\rm s} r_{\rm s}^{2}
\eeq
where $\rho_{\rm s}$ and  $c_{\rm s}$ are the density and speed of sound at the sonic point at a radius $r_{\rm s}$. These quantities are also estimated as described in \citet{murray-claychiang2009}. 

Heating both by X-ray and EUV radiation is included, using the criterion of \citet{owenjackson2012} to decide which regime is dominant. In the X-ray driven regime, the escape rate is also modeled with an energy-limited formula. The stellar $L_{\rm XUV}$ as a function of time is taken from \citet{ribasguinan2005}, but we explore in the simulations below scenarios with a 10 times higher and lower evaporation rate, which could be caused by correspondingly higher and lower stellar XUV luminosities, to account for the observed spread \citep{tujohnstone2015}. 

For some initial conditions of low-mass cores with massive envelopes, we (numerically) find at the beginning of the simulation an outer radius that exceeds the planet's Hill sphere radius, indicating that overflow occurs. In this case, we remove at each timestep gas outside of $R_{\rm Hills}$, which leads to a very fast ($\sim10^{4}$ yr) reduction of the envelope mass and outer radius, until it falls below $R_{\rm Hills}$. This might be manifestation of the boil-off regime described by \citet{owenwu2016} (see also \citealt{ginzburgschlichting2016} and \citealt{fossatierkaev2017}). 

The exact temporal behavior during this overflow might not be well described  in our model because of (a) its underlying hydrostatic approximation, and (b) the radially constant luminosity. However, numerical experiments varying the treatment of this phase (for example the fraction of the envelope mass outside of  $R_{\rm Hills}$ removed at each timestep) show that in any case, such planets suffer eventually a complete loss of the envelope, typically on short timescales of less than about 1 Myr, meaning that the final result at, e.g., 5 Gyr, is not affected. 

When comparing to observations, it should be kept in mind that despite the inclusion of several sub-regimes, our evaporation model still a simplified, parameterized (e.g., via the constant $\varepsilon_{\rm UV}$) description of the actual loss process. The evaporation model will be improved in future work to include the results of \citet{owenalvarez2016} on the occurrence of different evaporation regimes. Given the rather weak dependency on the details of the evaporation model found analytically and numerically, and the general agreement of our results with other more sophisticated evaporation models \citep[see comparison in][and Sect. \ref{sect:compOW17}]{jinmordasini2014}, we however do not expect these factors to fundamentally change the results found here. 

\subsubsection{{Simulations with enriched atmospheres}}\label{sect:simswithZ}
{For a subset of simulations, we take into account that the gaseous envelopes of the planets could be significantly enriched in heavy elements instead of being dominated by H/He. In these simulations, the model is modified relative to the description in the previous section with respect to three points:}
\begin{enumerate}
\item {The envelope is assumed to consist of a mixture of H/He described by the EOS of \citet{saumonchabrier1995} and of H$_{2}$O with a mass fraction $Z_{\rm enve}$ described by the ANEOS equation of state \citep{thompson1990}. The results of ANEOS for H$_{2}$O are briefly described in Appendix \ref{appendix:aneos}. $Z_{\rm enve}$ is assumed to be radially and temporally constant in the envelope. While its specific thermodynamic properties will obviously affect the results, the H$_{2}$O should in the current context be seen as representing heavy elements (metals) in general. The H/He and H$_{2}$O  is mixed using the additive volume law \citep[e.g.,][]{baraffechabrier2008a}.}
\item {As in \citet{lopez2017}, in the evaporation model the efficiency factor in the energy-limited regime is varied with $Z_{\rm enve}$ as $Z_{\rm enve}^{-0.77}$ \citep{owenjackson2012}. One should note that this scaling is taken from studies of the X-ray heating of protoplanetary disk atmospheres \citep{ercolanoclarke2010} rather than planetary atmospheres. While it seems likely that evaporation rates indeed decrease with increasing metal content because metal atomic lines are important coolants \citep{salzczesla2016,owenmurray2018}, the exact dependencies have not yet  been explored. Our results  based on this simple scaling are thus to be taken with caution. In the radiation-recombination limited domain, the  mean molecular weights are also adjusted depending on $Z_{\rm enve}$ (see \citealt{lopez2017}), with the  mean molecular weights taken directly from the mixed EOS. }
\item {The atmospheric opacities \citep{freedmanlustig-yaeger2014} are calculated for the [M/H] that corresponds to $Z_{\rm enve}$. The conversion from metal mass fraction into [M/H] is made in an analogous way as described in \citet{valenciaguillot2013}. For the $Z_{\rm enve}$=0.1, 0.3, and 0.5 that we consider in the simulations below, this leads to [M/H]=0.86, 1.43, and 1.78.}
\end{enumerate}

\subsubsection{Post-formation H/He envelope mass}\label{sect:initialconditions}
A central initial condition for the evolutionary calculations is the envelope mass of H/He $M_{\rm e,0}$ as a function of planetary properties at the end of the formation epoch when the gas disk disperses (even if we will see later that $\rbare$ only depends very weakly on it). Here we start from the results of population synthesis calculations based on the core accretion paradigm that were presented in \citet{mordasiniklahr2014}. 

In this paper, the effect of the atmospheric opacity during formation on the accreted H/He mass and associated mass-radius relation was studied. The envelope masses were found by explicitly solving numerically the standard 1D internal structure equations (e.g., \citealt{mordasinialibert2012b}). The accretion of planetesimals, orbital migration, and the evolution of the protoplanetary disk are also included in our global model \citep{alibertmordasini2005}. Here we use the nominal population of  \citet{mordasinialibert2012c}. It is characterized by a fixed stellar mass of 1 $\msun$, solar-composition H/He envelopes, and an opacity in the protoplanetary envelope that is given by \citet{freedmanlustig-yaeger2014} for the grain-free molecular opacities, and the \citet{belllin1994} grain opacities reduced by a factor 0.003. This nominal reduction factor was determined in \citet{mordasiniklahr2014} by calibrating the gas accretion timescales with the ones found  with the detailed model of  \citet{movshovitzbodenheimer2010} for the grain dynamics. Such microphysical models for the dynamics of the grains \citep{podolak2003,movshovitzbodenheimer2010,ormel2014,mordasini2014} consider the settling, coagulation,  and evaporation  of the grains in the outer radiative zone/atmosphere of the protoplanets. They predict opacities that are much smaller than in the ISM, and on the order of a few $10^{-3}$ cm$^{2}$/g at the radiative-convective boundary rcb. This is comparable to, but still a bit higher than expected for a completely grain-free atmosphere. 
 
For this work, we are interested in a mean analytical relation. We have therefore determined the mean post-formation envelope mass $M_{\rm e,0}$ as a function of the planet's core mass $\mcore$ and final semimajor axis $a$ by fitting the numerical results with a least-square method. The core mass and orbital distance are the quantities that most clearly and systematically influence $M_{\rm e,0}$. There is a significant spread around the mean relation, as also other factors like the disk lifetime or the accretion rate of planetesimal{s} prior to disk dispersal influence the final $M_{\rm e,0}$. For the fit, we have included planets with $0.1<a<1$ AU and $1<\mcore<10$ $\mearth$. They have envelopes with typical masses of $\sim1-10$\% of their total mass.  Such planets are a very frequent outcome of the formation simulations. This shows that the formation of close-in cores of 5-10 $\mearth$ with only relatively low-mass envelopes is a natural outcome when the structure equations are directly solved - not all these core{s} become giant planets.  By solving the internal structure equations in the formation model, the decrease and/or limitation of the envelope mass due to the luminosity caused by solid accretion as  well as because of the decreasing nebular pressure are automatically included. As impact stripping is not included, all planets are assumed to start with primordial H/He.

\begin{figure}[tb]
    \centering
    \includegraphics[width=0.5\textwidth]{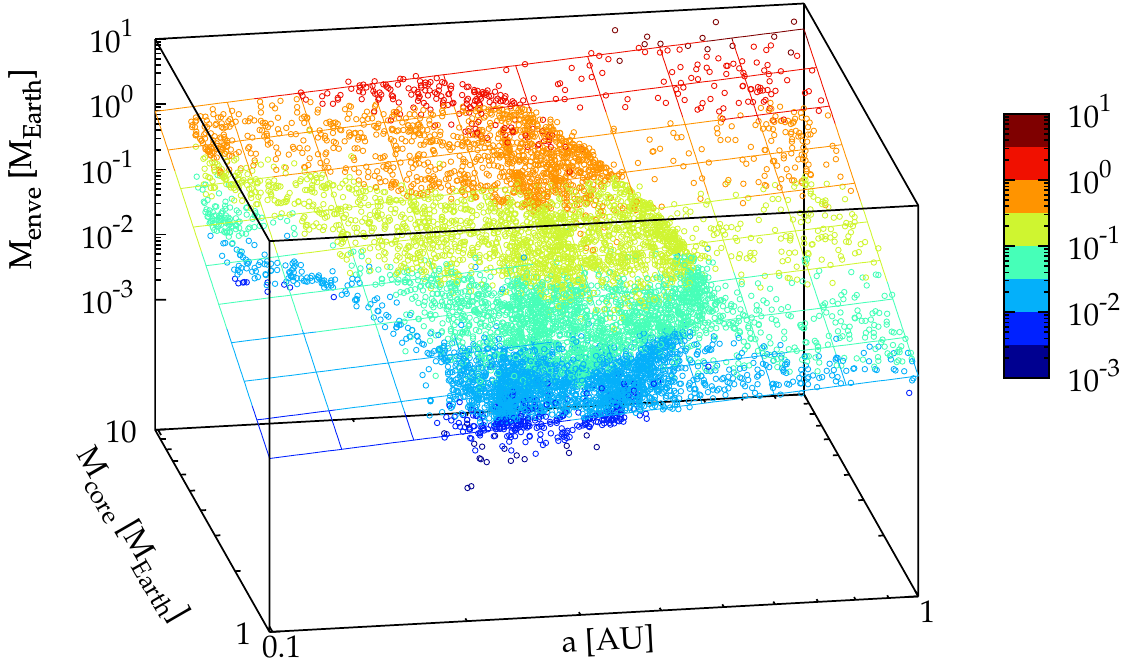}
   \caption{Post-formation envelope mass as a function of core mass and semimajor axis for synthetic close-in low-mass planets {
   (colored points)}. The {colored mesh} is the least-square fit (Eq. \ref{eq:fitme0fromsynt}).  }\label{fig:mxymcoreclose}
\end{figure}

Figure \ref{fig:mxymcoreclose} shows the envelope mass as function of core mass and semimajor axis together with the fit. One sees that the envelope mass increases with core mass and orbital distance, which is expected \citep{ikomahori2012}. The imprint of the inner convergence zone of type I migration \citep{dittkristmordasini2014} is visible in an arc like structure. Quantitatively, the following power law dependency is found for $M_{\rm e,0}$:
\beq\label{eq:fitme0fromsynt}
\frac{M_{\rm e,0}}{\mearth}=0.024  \left(\frac{\mcore}{\mearth}\right)^{2.23}\left(\frac{a}{\rm 1 AU}\right)^{0.72}
\eeq
which corresponds to a roughly speaking quadratic increase with the core mass, and a weaker than linear increase with distance. In terms of gas-to-core mass ratio $M_{\rm e,0}/\mcore$  and normalized at a more relevant semimajor axis of 0.1 AU, this corresponds to
\beq
\frac{M_{\rm e,0}}{\mcore}=0.005 \left(\frac{\mcore}{\mearth}\right)^{1.23}\left(\frac{a}{\rm 0.1 AU}\right)^{0.72}
\eeq
It is interesting to compare this scaling relation with the analytical result of \citet{leechiang2015}. The analytical result is obtained by considering that the planets accrete gas on a timescale given by the envelope's Kelvin-Helmholtz cooling timescale \citep{ikomanakazawa2000}, and that the magnitude of cooling is set at the radiative-convective boundary  rcb \citep{leechiang2015}. For the planets here, the regime of completely dust-free atmospheres in a gas poor nebula at 0.1 is most likely to apply. If we assume that the temperature at the rcb is approximately equal to the nebular temperature as suggested by \citet{leechiang2015}, that the later scales as $a^{-0.5}$ \citep{idalin2004}, and finally that the disk lifetime is 2 Myr (the mean lifetime of our synthetic disks, see also \citealt{haischlada2001}), Eq. 22 of  \citet{leechiang2015} predicts 
\beq
\frac{M_{\rm e,0}}{\mcore}=0.016 \left(\frac{\mcore}{\mearth}\right)^{1.6}\left(\frac{a}{\rm 0.1 AU}\right)^{0.95}.
\eeq
We thus see that the power law exponents are not too different, but that the absolute mass found in the numerical calculations is about a factor 3.5 lower.  These lower envelope masses could be due to (a) a higher opacity (residual grain opacity, neglected in the analytical model), (b) some accretional heating by planetesimals (also neglected in the analytical model), especially as orbital migration brings the protoplanets into regions of the disk still containing planetesimals or (c) the decline of the ambient nebula over time. A detailed comparison of the analytical result and the direct solution of the structure equations is, however, {beyond} of this paper, especially in view of the weak dependency of $\rbare$ on it that we will find further down.

\begin{table*}
\centering
\caption[]{Description and outcome of the 1{4} simulations. The second column gives for the first 7 simulations the initial envelope mass (in units of $\mearth$) as a function of core mass $M_{\rm c}$ (in $\mearth$) and orbital distance $a$ (in AU). The last {7} simulations use the same initial envelope masses as M3 {or M0 (for Z1, Z2, Z3)}, but differ from it in other characteristics shown in the table. $M_{\rm b,0p1}$ is the maximum bare core mass at 0.1 AU, found numerically (``num.'', Eq. \ref{eq:fitmbarefromnum}) and analytically (``an.'', Eq. \ref{eq:mbarelaterocky}), while $e_{\rm m}$ is the slope of the valley in the mass-distance plane found numerically. The analytical model predicts $e_{\rm m}=1$.  $R_{\rm b,0p1}$ and $e_{\rm r}$ are the equivalent quantities in the radius-distance plane (Eqs. \ref{eq:fitrbarefromnum}, \ref{eq:rbarelate}), meaning that $R_{\rm b,0p1}$ is the radius of the most massive planet that can be completely stripped. It corresponds to the bottom of the evaporation valley at 0.1 AU. The analytical model predicts  $e_{\rm r}=p_{\rm c}=0.27$. }\label{tab:sims}
\begin{tabular}{lcccccccc} 
\hline\noalign{\smallskip}
  Name                      & Characteristics             & $M_{\rm b,0p1}$ (num.)  &  $M_{\rm b,0p1}$ (an.) & $e_{\rm m}$   & $R_{\rm b,0p1}$ (num.)  & $R_{\rm b,0p1}$ (an.)  & $e_{\rm r}$   \\ \hline
\noalign{\smallskip}
\hline\noalign{\smallskip} 
 M0 (nominal) & $0.024 M_{c}^{2.23} a^{0.72}$ & 5.82 & 6.5 & 1.05 & 1.67 &  1.6 & 0.28\\ % (nominal S0) 
 M1  & $ 0.03 $ & 8.30 & 6.5 & 1.49 & 1.82 & 1.6 & 0.36  \\ %S9
 M2 & $ 0.03 M_{c}$ & 5.98 & 6.5 &1.07 & 1.68 & 1.6 & 0.30\\ % (S7) 
 M3 (reference) & $ 0.03 M_{c}^{2}$ & 5.52 & 6.5 & 0.99 & 1.65 & 1.6 & 0.27\\ % S6
 M4  & $ 0.03 M_{c}^{3}$ & 5.21 & 6.5& 0.87 & 1.60 &1.6 & 0.25\\ %(S8)
 N1 & $ 0.06 M_{c}^{2}$ & 5.39 & 6.5 & 0.93 & 1.65 &1.6 & 0.28\\ %(S10) 
 N2 & $ 0.3 M_{c}^{2}$ & 5.38 & 6.5 &0.93 &1.66 & 1.6 & 0.28\\ %(S11) 
E1& $\dot{M}_{\rm evap} \times 0.1 $&  2.88 & 2.0 & 0.91 & 1.39 & 1.2 & 0.28\\ % (S15) 
E2 & $\dot{M}_{\rm evap} \times 10 $ & 10.6 & 20.5  & 0.99 & 1.97 & 2.2 & 0.27\\ % (S16) 
O1 & $\kappa_{\rm R} \times 10$ & 7.09 & {-}  & 0.97 & 1.78 & {-}  & 0.28\\ %(S13) 
{Z1} & ${Z_{\rm enve}=0.1}$   &   {5.68}       &   {-}       &  {1.01   }  & {1.58 }& {-}& {0.28 } \\         
{Z2} & ${Z_{\rm enve}=0.3}$   &   {3.97}       &   {-}       &  {1.10   }  & {1.43 }& {-}& {0.30 } \\         
{Z3} & ${Z_{\rm enve}=0.5}$   &   {3.01}       &   {-}       &  {1.14   }  & {1.32 }& {-}& {0.31 } \\         
 I1 & Icy cores ($f_{\rm ice}$=1) & 8.57 & 11.9 & 0.89 & 2.68 & 2.4 & 0.26\\ %(S12) 
\hline
\end{tabular}
\end{table*}

\subsubsection{Post-formation luminosity}
Besides the envelope mass fraction, one also needs to specify the luminosity at the end of the formation phase as an initial condition for the evolutionary simulations. This post-formation luminosity $L_{0}$ was also taken from the aforementioned population synthesis calculations of planetary formation, considering the luminosity as a function of core mass for low-mass planets with masses between 1 to 10 $\mearth$ inside of 1 AU at an age of 10 Myr. One finds a fitting relation of
\beq\label{eq:l0par}
L_{\rm 0}/L_{\rm \jupiter}\approx 0.008 \times \left(\frac{M_{\rm c}}{\mearth}\right)^{2.5}
\eeq
where $L_{\rm \jupiter}$ is the present day intrinsic luminosity of Jupiter (about $8.7\times 10^{-10} \lsun$, \citealt{guillotgautier2014}). Most synthetic planets are within a factor two higher or lower than this mean relation. Given the rather weak dependency of the thickness of the convective zone of the H/He envelope on the age (i.e., luminosity) found by \citet{lopezfortney2013b}, especially when compared to the impact of the envelope mass, we did not investigate the consequences of varying $L_{\rm 0}$. The role of the luminosity for the evaporation valley, in particular when also considering a possible additional luminosity source like ohmic dissipation (that could be strong in low-mass planets, \citealt{puvalencia2017}) should however be addressed in future work.

In Appendix \ref{appendix:lumifit} we give as a side result for higher ages of the planets a fit for the luminosity as a function of time, core mass, and envelope mass. This fit may be used in time-independent internal structure calculations like \citet{rogersbodenheimer2011,dornventurini2017,lozovskyhelled2018} which need the luminosity as an input quantity.

\subsection{Simulations}
We have calculated 1{4} grids of planetary evolution simulations in the $M_{\rm c}-a$ plane, varying the (1)  the post-formation envelope mass in several ways because of the motivation to understand whether gas accretion during formation can be constrained by the locus of the valley, (2) the strength of evaporation which could represent different efficiency factors and/or different stellar XUV-luminosities, (3) the Rosseland mean opacity in the atmosphere, {(4) the metallicity (heavy element enrichment) of the gaseous envelope and (5)} the ice mass fraction in the core.

Table \ref{tab:sims} gives an overview of the simulations. In the second column, the post-formation H/He envelope mass $M_{\rm e,0}$ (in $\mearth$) as a function of core mass $\mcore$ (also in $\mearth$) and potentially the semimajor axis $a$ (in AU) is given. The last {seven} simulations use the same $M_{\rm e,0}$ as M3 {(E1, E2, O1, I1) or M0 (Z1, Z2, Z3)}, but other parameters are varied, as indicated in the table. The 1{4} simulations are described as follows:
\begin{itemize}
\item Simulation M0 is the nominal simulation, where the envelope mass varies as described by Eq. \ref{eq:fitme0fromsynt}. It thus increases roughly quadratically with the core mass and linearly with orbital distance. 
\item The next four simulations (M1-M4) vary the exponent $p_{\rm e}$ in the power law dependency of the envelope mass on core mass, $M_{\rm e}\propto M_{\rm c}^{p_{\rm e}}$  from 0 to 3 (Eq. \ref{eq:memcscaling}). 
\item The simulations N1 and N2 also investigate the impact of the primordial envelope mass. In M1-M4, the normalization constant $M_{\rm e,1}$ (Eq. \ref{eq:memcscaling}) which is the envelope mass of a 1 $\mearth$ core is 0.03 $\mearth$. In N1 and N2 it is instead 0.06 and 0.3 $\mearth$ respectively, i.e., we are studying the effects of envelope masses with are two and ten times as high than in M3. 
\item In simulations E1 and E2 the evaporation rate in all regimes is uniformly multiplied by a factors 0.1 and 10 relative to the evaporation rate normally given by the model, respectively. 
 \item The simulation O1 quantifies the impact of a Rossland opacity in the H/He envelope that is increased uniformly by a factor of 10. Other quantities that likely also depend on the gas composition (EOS, evaporation rate) are unchanged.
\item {In the simulation Z1, Z2, and Z3 the composition of the gaseous envelope is changed from solar composition as in all other simulations to mixtures of H/He with H$_{2}$O with a mass fraction of $Z_{\rm enve}$=10, 30, and 50\%. The EOS, opacity, and evaporation rate are all modified self-consistently, as described in Sect. \ref{sect:simswithZ}. }
\item The last simulation I1 shows the impact of increasing the ice mass fraction in the core to 1. Such completely icy cores (without any silicates and iron) are certainly not expected in reality, but it is instructive for comparison with the analytical model and with \citet{owenwu2017,jinmordasini2018} who both investigated the valley's position as function of ice mass fraction. 
\end{itemize}

\begin{figure*}%![h]
    \centering
    \includegraphics[width=\textwidth]{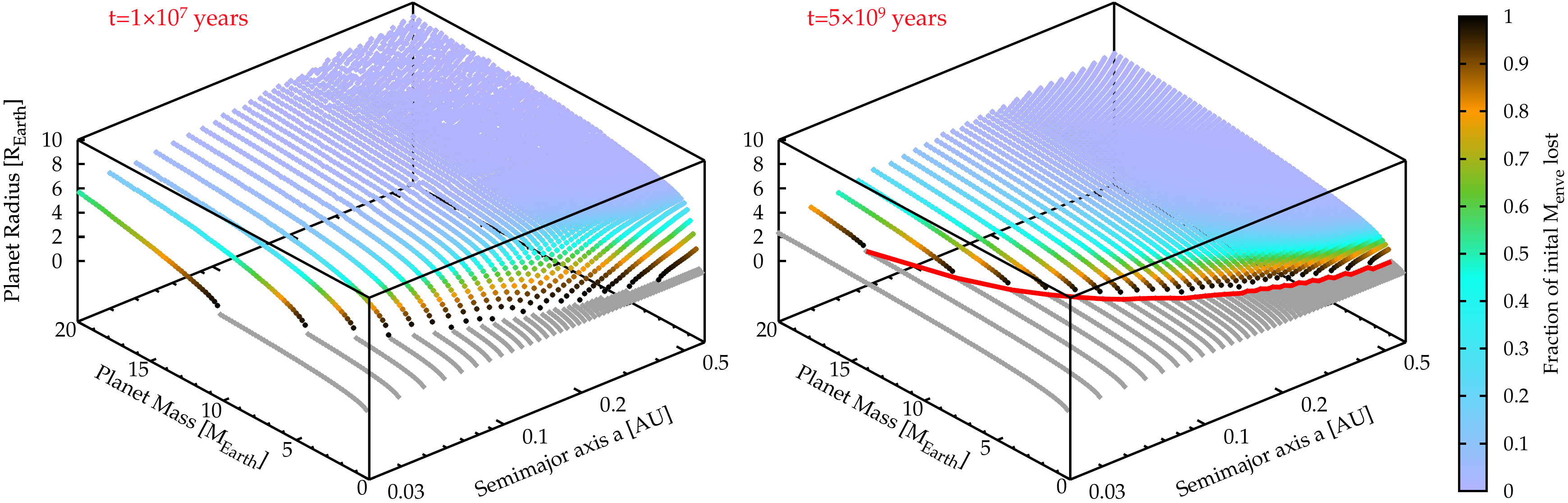}
   \caption{Evolution of the planets of the nominal model M0 in the  distance-mass-radius-time space. The radius (z-axis)  at 10 Myr and 5 Gyr is shown as a function of semimajor axis and mass. The color code  gives the fraction of the initial H/He envelope that was evaporated. Planets that have lost all H/He are plotted in gray. The overall contraction and the growing bare core triangle is visible, extending to larger masses and distances as time goes on. The red curve indicates $\mbare$ and $\rbare$ at 5 Gyr as in Fig. \ref{fig:MRbareS0}.  }\label{S0aMRopt}
\end{figure*}

Regarding the grid, a range in core masses between 1 and 20 $\mearth$ and orbital distances between 0.03 and 0.6 AU was covered in most simulations. Because of computational time reasons, some simulations where conducted on a grid of reduced size.

\subsection{Results}\label{sect:numresults}
The main result of the numerical study is the location of the valley (transition from super-Earth to sub-Neptunes) as a function of orbital distance found in the 1{4} simulations. Given the simulation results, we quantify its location by numerically deriving a least-square power law fit to the highest mass $M_{\rm bare}$ and largest radius $R_{\rm bare}$ as a function of orbital distance $a$ that has lost the entire H/He envelope at an age of 5 Gyr, normalized by the value at 0.1 AU.  This means that the location of the valley is quantified with $M_{\rm b,0p1}$ and $e_{\rm m}$ in
\beq\label{eq:fitmbarefromnum}
M_{\rm bare}(a)=M_{\rm b,0p1} \left(\frac{a}{\rm 0.1 AU}\right)^{-e_{\rm m}}
\eeq 
in the mass-distance plane, and equivalently with $R_{\rm b,0p1}$ and $e_{\rm r}$ in
\beq\label{eq:fitrbarefromnum}
R_{\rm bare}(a)=R_{\rm b,0p1} \left(\frac{a}{\rm 0.1 AU}\right)^{-e_{\rm r}}
\eeq 
in the radius-distance plane. Note that the two fits to obtain the four quantities were made independently. These four quantities can be directly compared with the analytical predictions in the second part of the paper.

The choice of the specific age of 5 Gyr does not affect the results as long as we are considering Gyr-old planets, as most of the atmospheric loss occurs during the first $\sim100$ Myr anyway, after which the triangle of evaporation has already attained almost its final size \citep{jinmordasini2014}. An observational determination of the temporal growth of the triangle at early times - if possible - would however represent an extremely interesting constraint for the various proposed processes of envelope loss. In this context it is interesting to note that the PLATO mission should be able to determine accurate stellar ages thanks to astroseismology. 

We now discuss the outcomes of the 1{4} simulations.

 \subsubsection{Nominal simulation: M0}
 \begin{figure*}
\begin{center}
\begin{minipage}{0.5\textwidth}
	 \centering
        \includegraphics[width=1\textwidth]{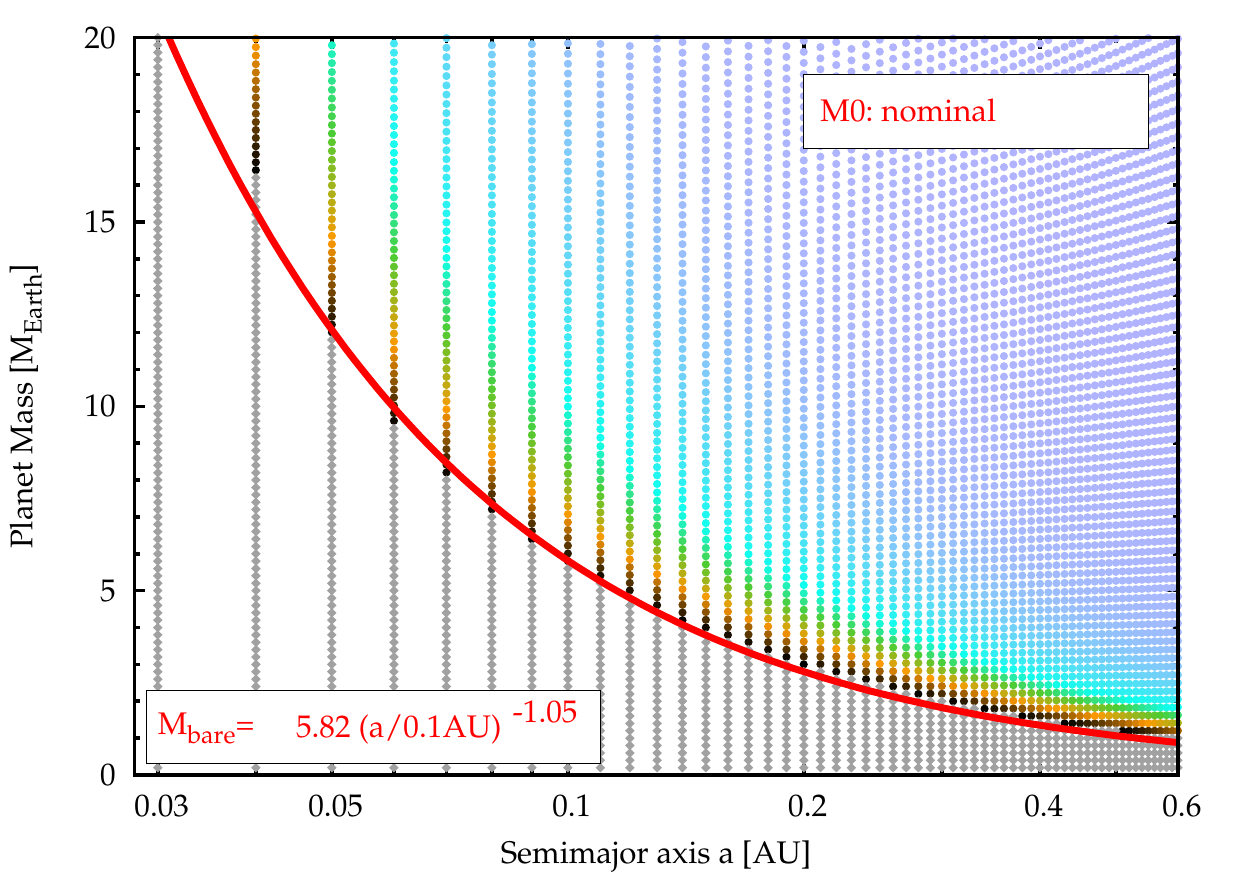}
         \end{minipage}\hfill
     \begin{minipage}{0.5\textwidth}
      \centering
       \includegraphics[width=1\textwidth]{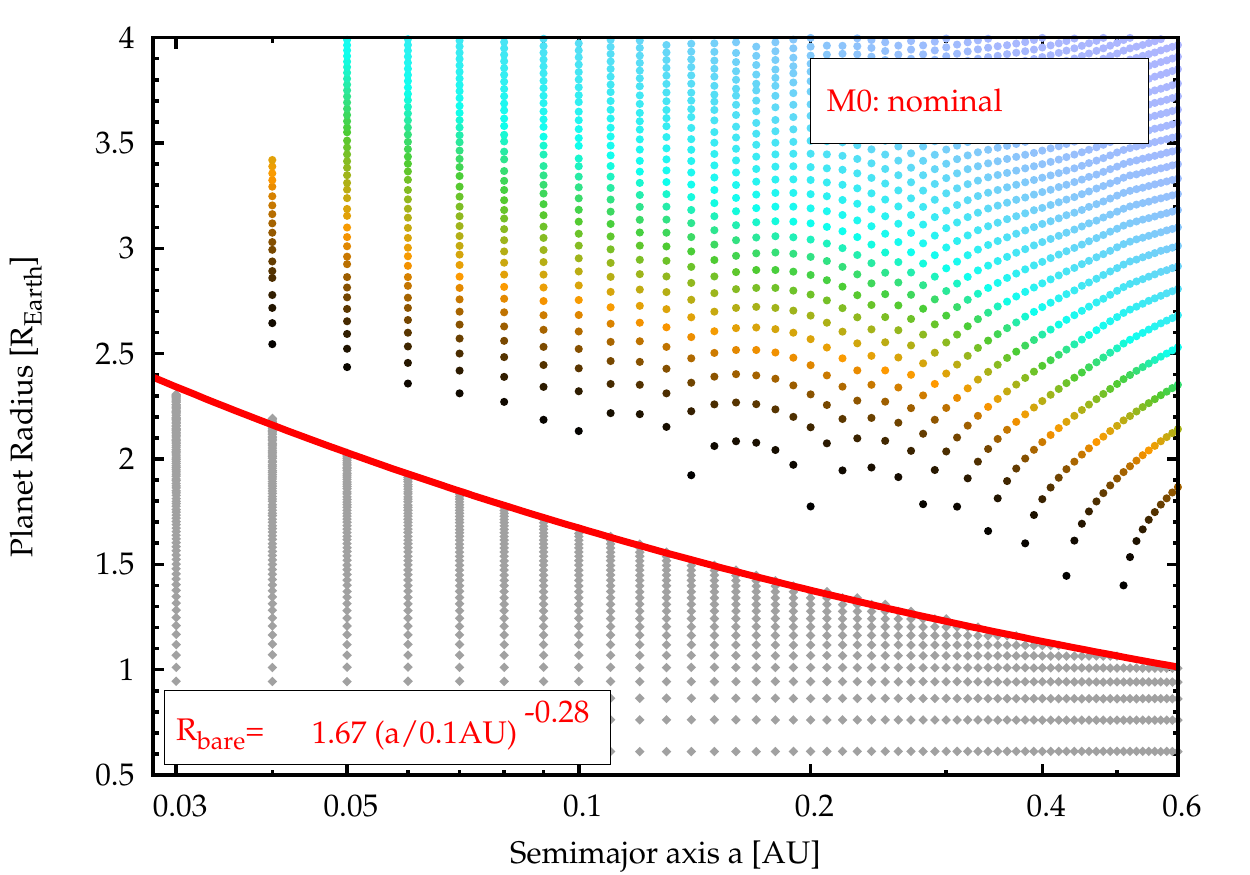}
     \end{minipage}
\caption{Transition from solid planets to planets with H/He in the plane of mass (left panel) and corresponding radius (right panel) versus semimajor axis for the nominal simulation M0 at 5 Gyrs. Colored points are planets that still have primordial H/He whereas gray diamonds are bare rocky planets. The color code shows the fraction of initial H/He that was evaporated (same color scale as in Fig. \ref{S0aMRopt}). The red line is a power law fit to the most massive/largest  planet that has lost the envelope, representing the transition at $\mbare$ respectively $\rbare$, with the fit parameters indicated in the inset ($\mbare$ in $\mearth$, $\rbare$ in $\rearth$).  Planets below  the red line are solid planets in the bare core triangle. In the mass plane the transition is continuous, whereas in the radius plane, there is a gap separating solid planets from planets with gas (the evaporation valley).   }\label{fig:MRbareS0}
\end{center}
\end{figure*}
To illustrate the general outcome of the simulations,  we show in Fig. \ref{S0aMRopt} the radius of the planets in the grid as a function of their orbital distance and (total) mass at 10 Myr and at 5 Gyr. The overall contraction of the radii, the decrease of the H/He mass, and the growing bare core triangle is visible, extending to larger masses, radii, and distances as time goes on. At 5 Gyr, the red curve indicates $R_{\rm bare}$ as found from the least square fit (same as in Fig. \ref{fig:MRbareS0}).

Figure \ref{fig:MRbareS0} shows the location of the valley in the mass-distance and radius-distance {plane} at 5 Gyr.  As expected, there is no gap or valley in the mass-distance distribution, because first, the mass fraction of the H/He envelope is small compared to the core mass in any case (at least for the lower mass cores), and second, the envelope mass is reduced to zero in a continuous fashion, as indicated by the colors. In the radius, there is  in contrast a gap/valley, as expected. As explained for example in \citet{jinmordasini2018}, the valley comes into existence because first, even the addition of a very low-mass H/He envelope significantly increases the radius, and second, the loss of this last remaining envelope occurs on a timescale of only $\sim10^{5}$ years \citep{jinmordasini2014}. This means that if we take a snapshot of the population at 5 Gyr, it is unlikely  (but not impossible) to observe a planet just in this final phase, leading to the depleted region.

The location of the valley, quantified by $M_{\rm b,0p1}$, $e_{\rm m}$, $R_{\rm b,0p1}$, $e_{\rm r}$ (Eqs. \ref{eq:fitmbarefromnum}, \ref{eq:fitrbarefromnum}), is similar to the ones found in \citet{jinmordasini2014} and \citet{jinmordasini2018}. The four parameters are given in the figure, and in Table \ref{tab:sims}. This similarity is not surprising, because the new simulations shown here use the same evaporation model and similar initial conditions.

In Table \ref{tab:sims} we also compare the numerical results for the locus of the transition with the analytical model of Sect. \ref{sect:analyticalstudy}. For a constant efficiency factor $\varepsilon$ in the energy-limited escape rate - as assumed in the numerical model -,  the analytical model predicts for all numerical simulations a power law exponent for the transition mass as a function of orbital distance of $e_{\rm m}$=1 (Eq. \ref{eq:mbarelaterocky}) and for the radius $e_{\rm r}$=0.27  (Eq. \ref{eq:rbarelate}). Numerically, for simulation M0, $e_{\rm m}$=1.05, and $e_{\rm r}$=0.28 is found, i.e., a very similar result.

\subsubsection{{Sub-Neptune} desert vs. (evaporation) valley}
In Fig. \ref{fig:MRbareS0} (and also other simulations shown below), besides the valley, we also note a complete absence of very strongly irradiated planets with H/He inside of  0.04 AU. Inside of this distance, even the most massive core considered in the grid (20 $\mearth$)  loses the entire envelope. Only much more massive giant planets could keep their H/He at these very small distances. This ``{photoevaporation or sub-Neptunian} desert'' which must be distinguished from the {radius} valley was explored observationally for example in \citet{lundkvistkjeldsen2016,mazehholczer2016,bourrierlecavelier2018}. It is another characteristic consequence of atmospheric escape \citep[e.g.][]{lecaveliervidal2004,kurokawanakamoto2014,mcdonaldkreidberg2019}. In this desert, more massive planets are affected for which the H/He initially represents a significant part of the total mass, in contrast to the evaporation valley. Hence, the loss of the envelope leads for these more massive planets to a substantial reduction of the total mass (and not only of the radius as for the valley). This is why the desert shows up also in the mass-distance plot, whereas the valley is only visible in the radii.  While due to the same physical effect, this imprint is thus different in nature from the evaporation valley.

\subsubsection{Scaling of the envelope mass with core mass: M1 - M4}\label{sect:M1M4}

\begin{figure*}
\begin{center}
\begin{minipage}{0.42\textwidth}
	 \centering
        \includegraphics[width=1\textwidth]{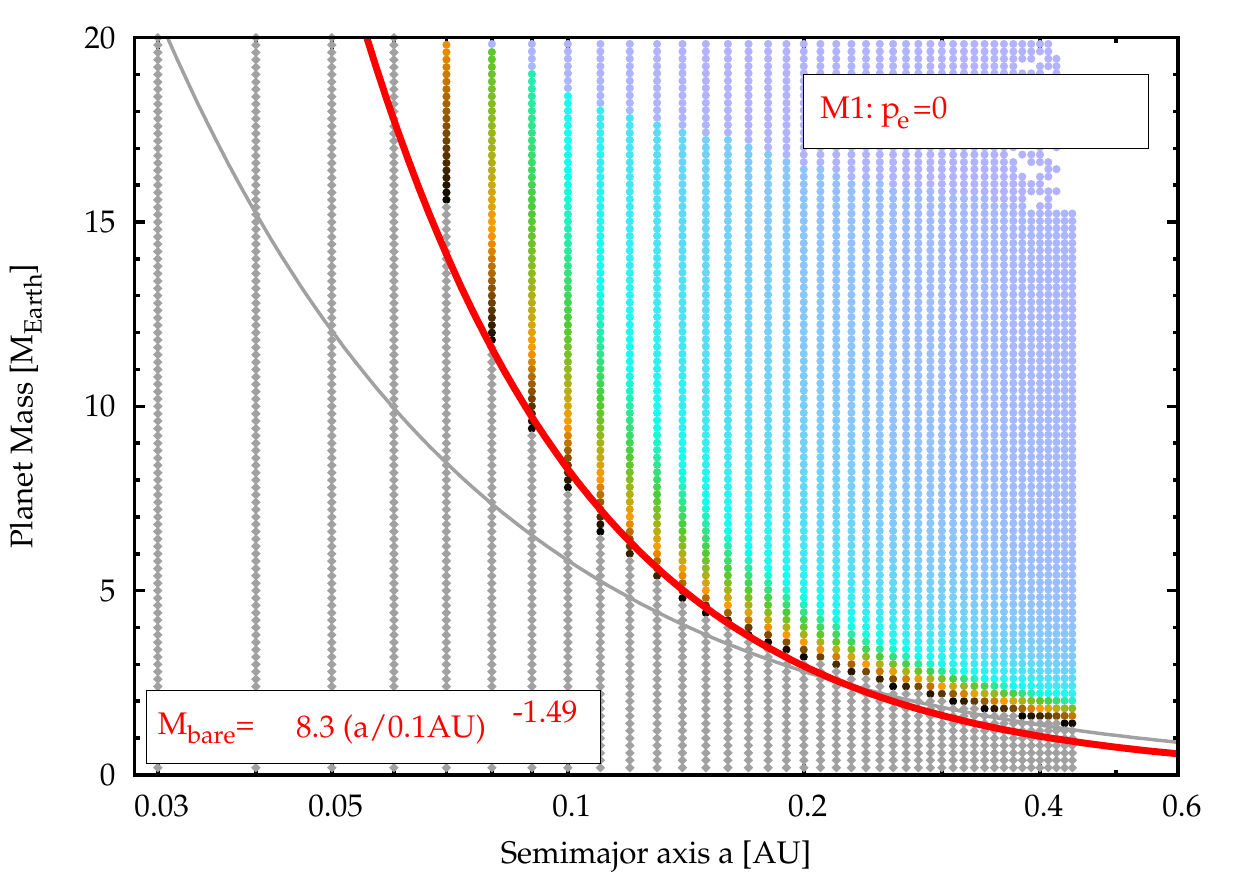}
        \includegraphics[width=1\textwidth]{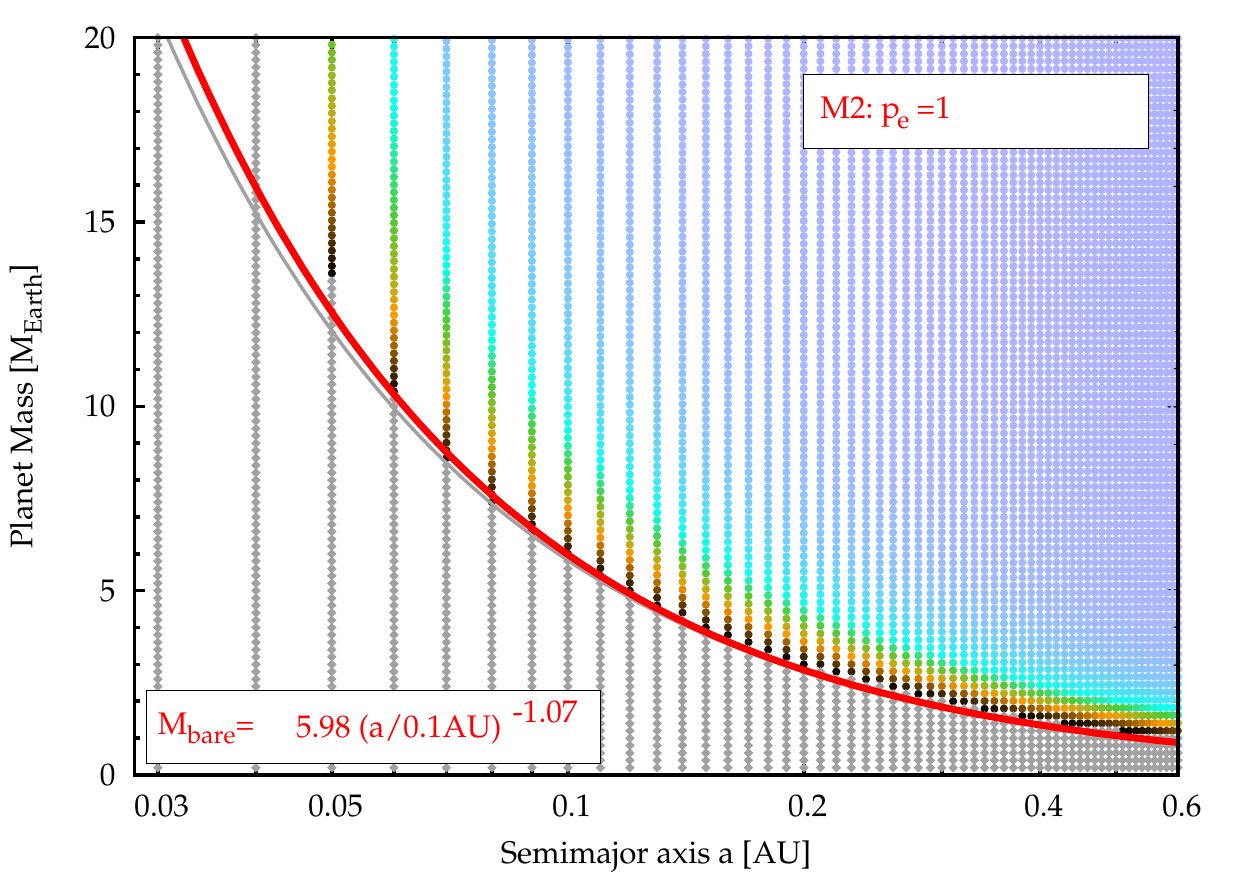}
        \includegraphics[width=1\textwidth]{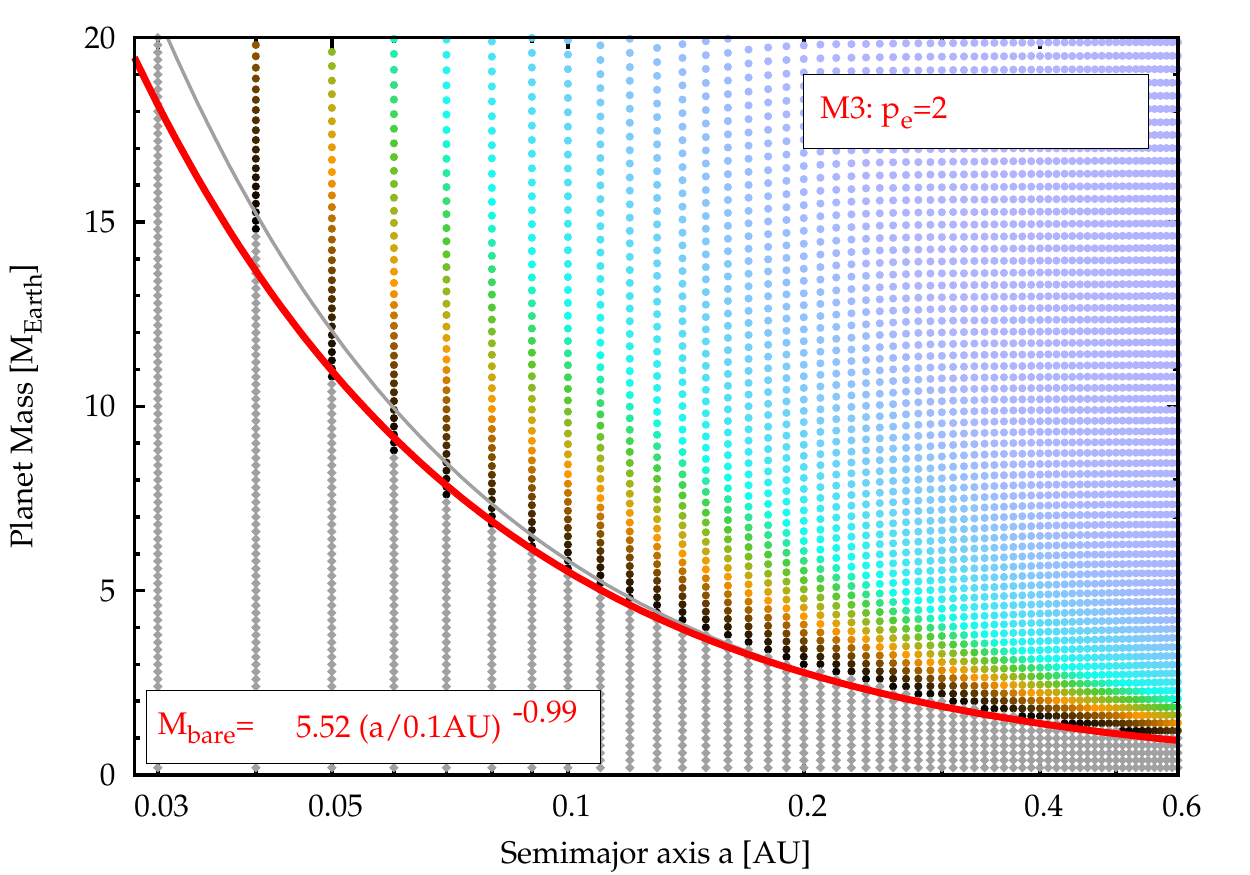}
        \includegraphics[width=1\textwidth]{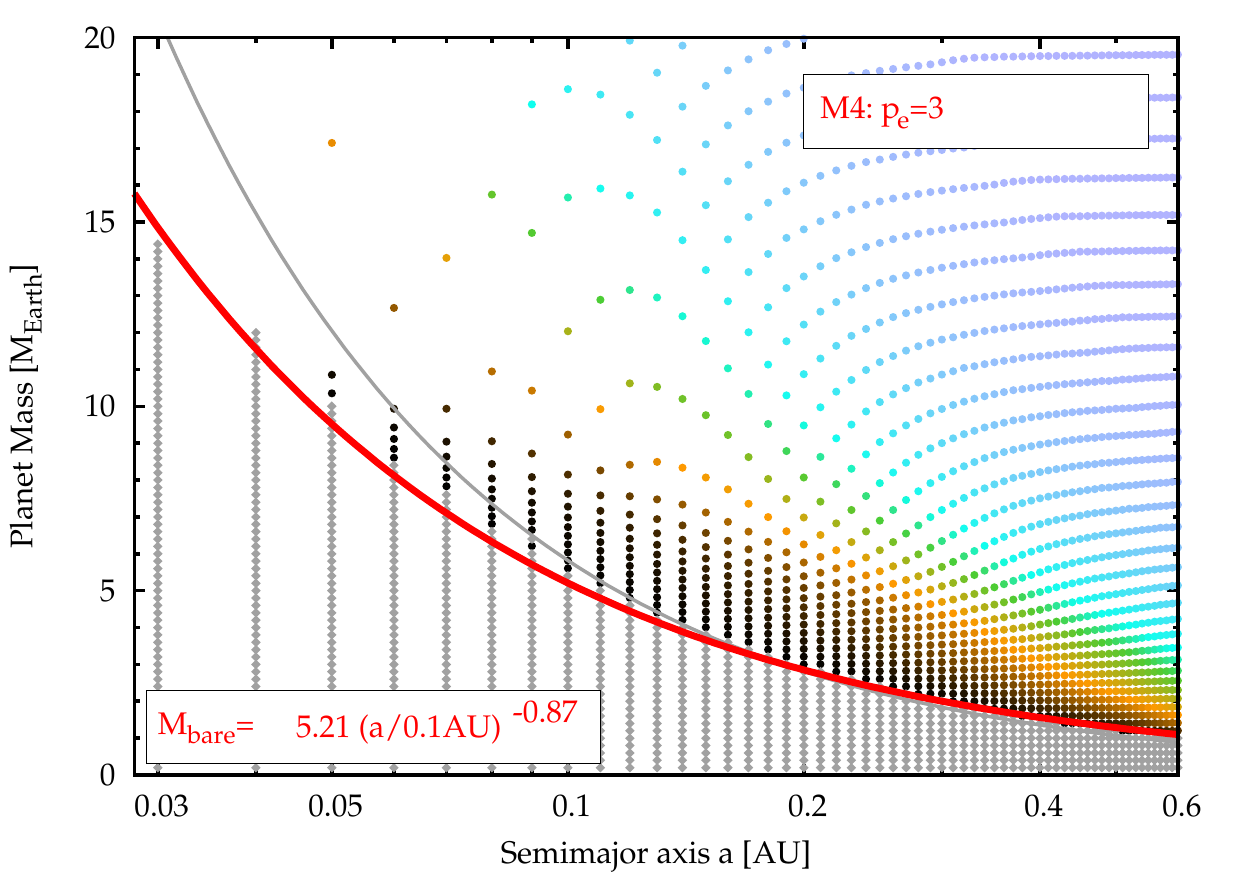}

     \end{minipage}%\hfill
     \begin{minipage}{0.42\textwidth}
      \centering
       \includegraphics[width=1\textwidth]{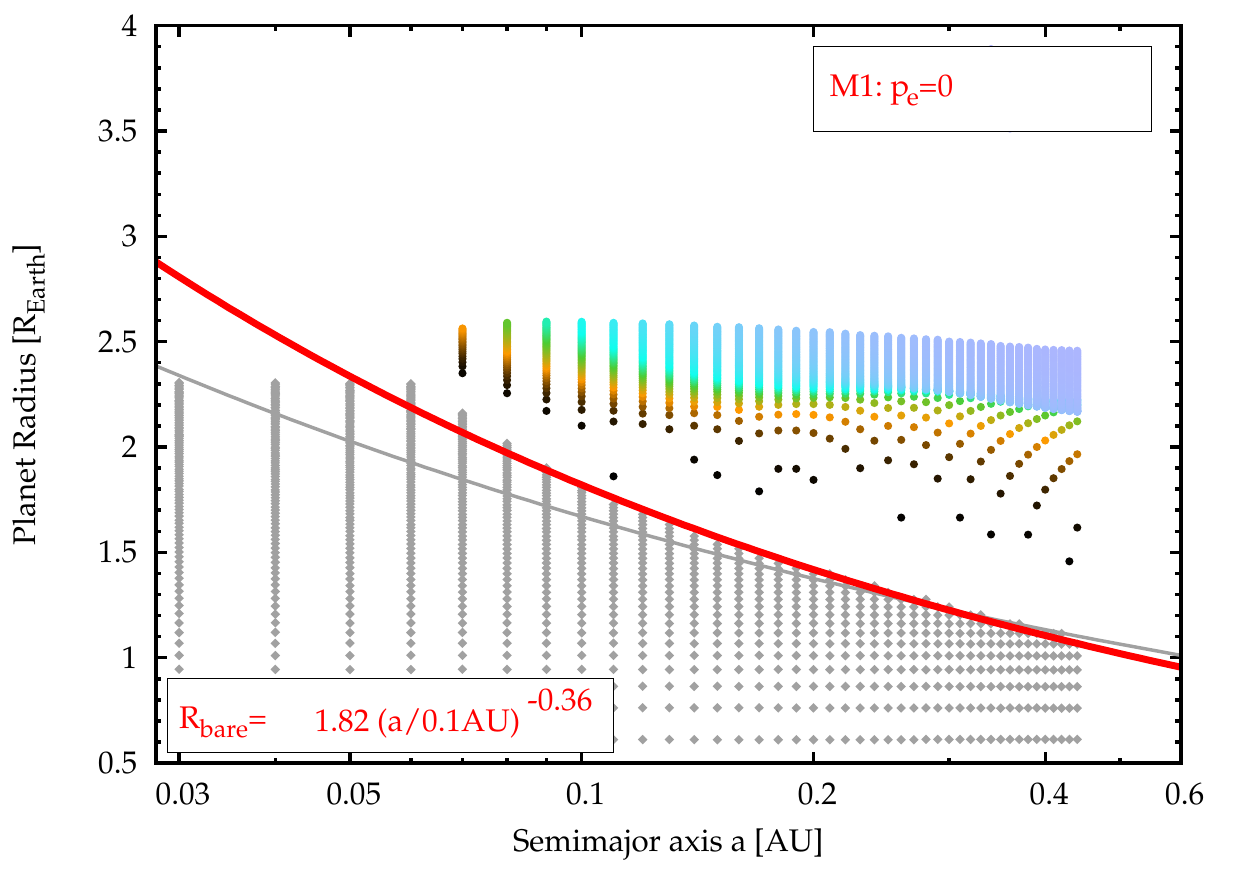}
       \includegraphics[width=1\textwidth]{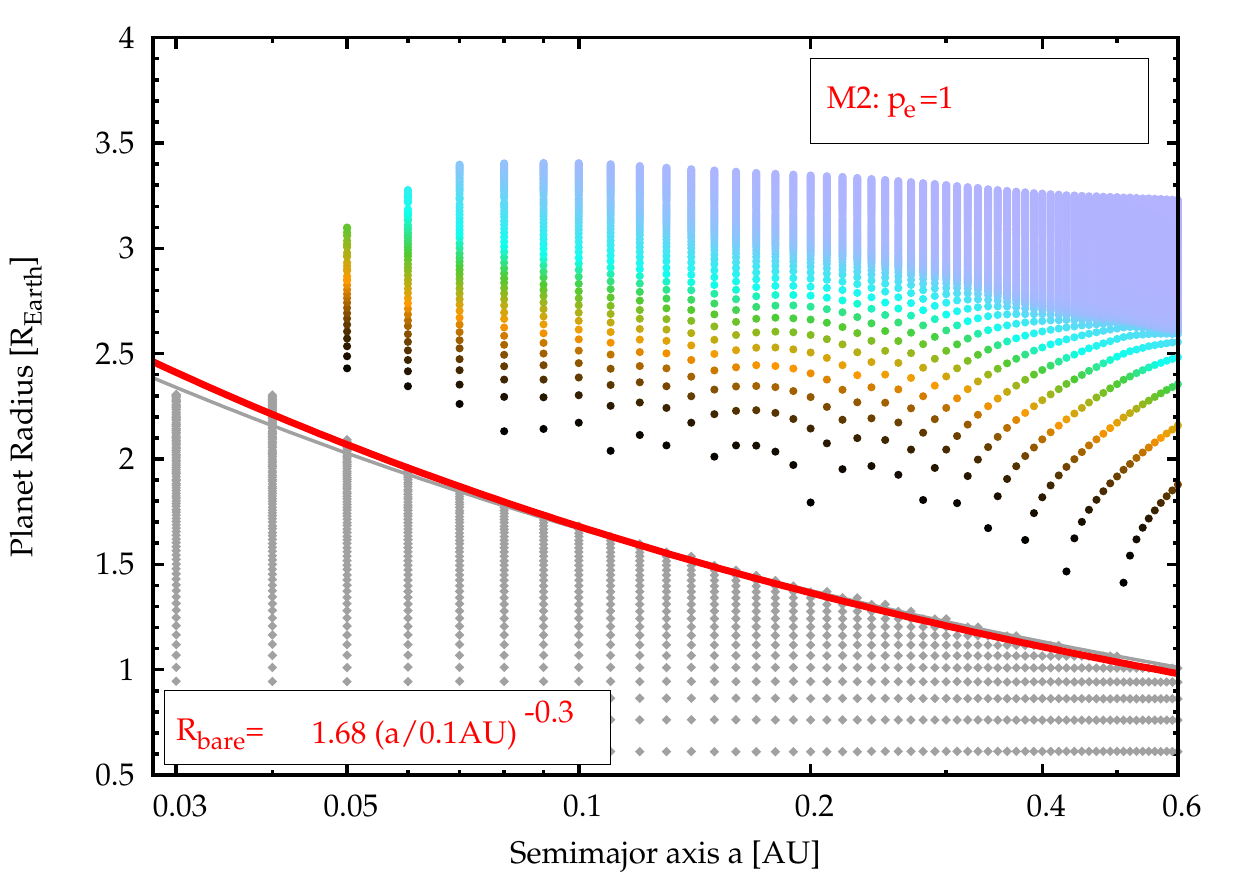}
       \includegraphics[width=1\textwidth]{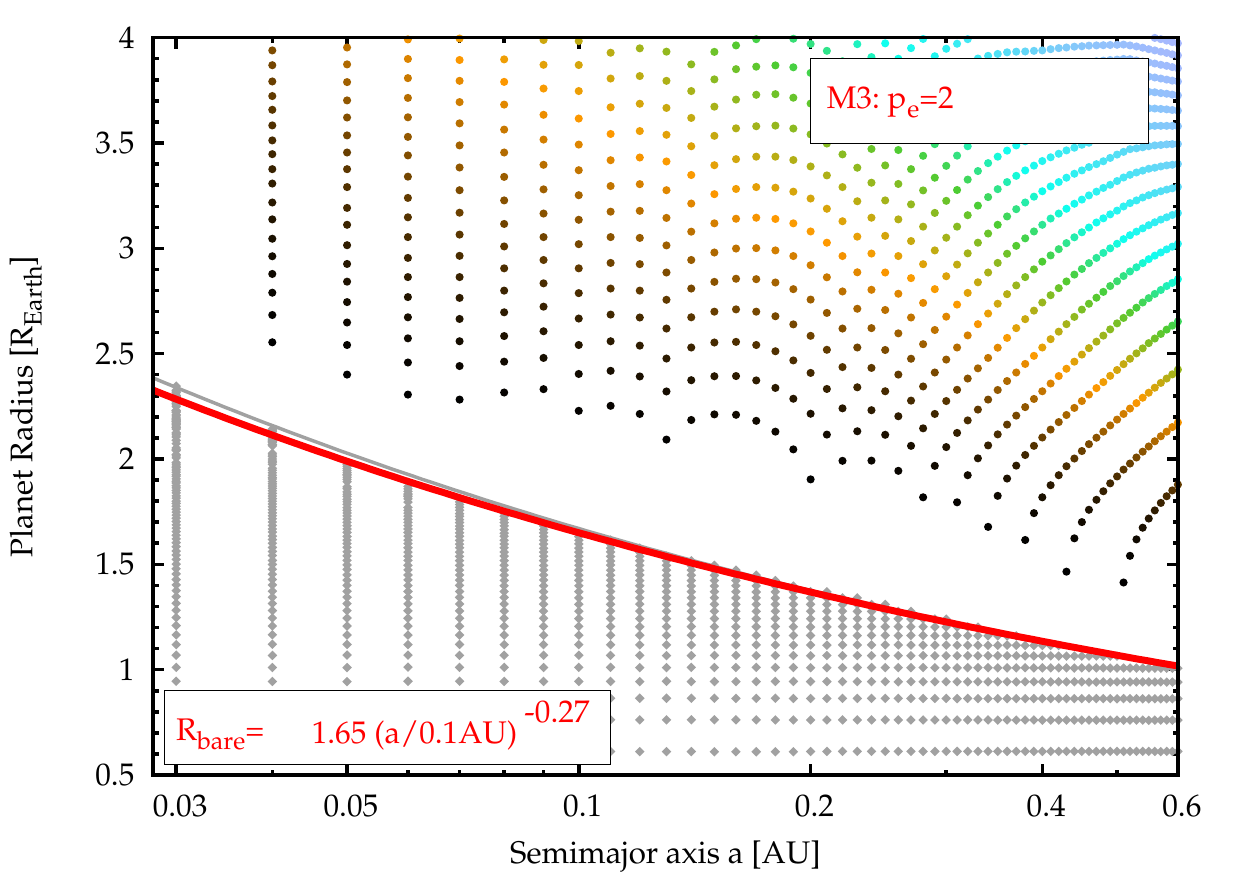}
       \includegraphics[width=1\textwidth]{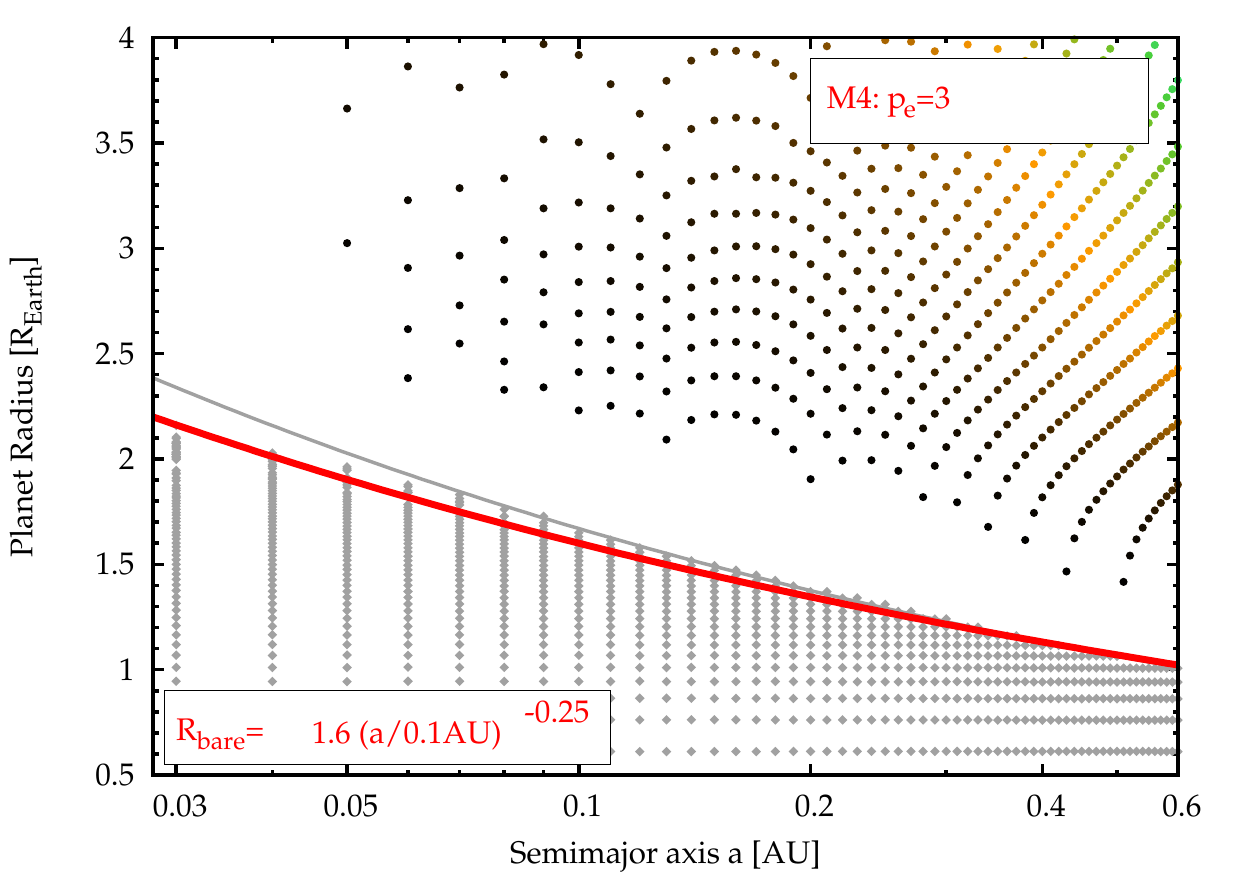}

     \end{minipage}
\caption{Impact of the core-envelope mass scaling on the transition. The figure is analogous to Fig. \ref{fig:MRbareS0}, but for the simulations M1, M2, M3 (reference simulation), and M4. These 4 simulations are identical except for the power law exponent in the scaling of the initial envelope mass $M_{\rm e,0}$ with core mass $\mcore$, $M_{\rm e,0}/\mearth=0.03 (\mcore/\mearth)^{p_{\rm e}}$ with $p_{\rm e}$=0, 1, 2 (reference simulation), and 3. In each case, the transition mass and radius were fitted with a power law for $\mbare$ and $\rbare$, as shown with the thick red line. The thinner gray line shows the transition mass and radius in the nominal simulation M0 to allow direct comparison.}\label{fig:MRbareM1M2M3M4} 
\end{center}
\end{figure*}

To understand whether the valley's position can constrain the post-formation core-envelope mass relation, it is interesting to vary the initial H/He mass. Figure \ref{fig:MRbareM1M2M3M4} shows the location of the valley for four power law scalings of the initial envelope mass $M_{\rm e,0}$ with core mass, $M_{\rm e,0}/\mearth=0.03 (\mcore/\mearth)^{p_{\rm e}}$ with $p_{\rm e}$=0, 1, 2 (reference simulation), and 3.

The plot shows that while there is a  correlation of a decreasing $\mbare$ and $\rbare$ for an increasing $p_{\rm e}$ (more massive envelopes for planets with $\mcore>1\mearth$), we always have $\rbare\approx 1.6 ... 1.8 \times (a/0.1\rm{AU})^{0.25...0.35}$ $\rearth$ despite the very large differences in initial (i.e., post-formation) H/He envelope masses for the more massive cores (e.g., a factor 125 for the 5 $\mearth$ core between M1 and M4). This  shows that the post-formation envelope mass has no significant influence on the final locus of the valley, at least if the initial envelope masses are not very different from what we nominally assume based on formation models (Sect. \ref{sect:initialconditions}). This has the important implication that unfortunately (from a formation point of view), the valley location does not constrain strongly envelope accretion models. We also see that the valley is at a very similar location as in the nominal simulation M0 (shown by gray lines in the plots) where the initial envelope mass depends on the semimajor axis also, in contrast to the four simulation shown here. Clearly, this distance dependency  has no significant effect, neither.

The plot also shows that the relative difference is larger for $\mbare$ than for $\rbare$. This is expected, as the core mass in the form of the local $\mbare$ is the more fundamental quantity  controlling whether a planet can hold on to its H/He envelope than the core radius, as will become clear in the analytical model (Sect. \ref{sect:finalresultanalytical}). Once $\mbare$ is given, the $\rbare$ then follows simply from the weak dependency inherent to the relation of a solid planet's (core) radius and its (core) mass, approximately as $R\propto M^{0.27}$ for an Earth-like composition (see Eq. \ref{eq:rcoremcoreearthlike}).

Compared to the weak dependency found here, the analytical model even predicts that there is no dependency of the valley's location on $p_{\rm e}$ at all (Sect. \ref{sect:finalresultanalytical}). As will become obvious there, the physical reason is that at a given core mass, a higher (lower) post-formation H/He mass on one hand means that there is more (less) material to evaporate, but on the other hand also that the planet has a larger (smaller) radius. Since the (total) mass is essentially given by the core mass, this means also that the planet has a smaller (larger) mean density $\bar{\rho}$. But as shown by Eq. \ref{eq:mdotenergylim}, $\dot{M}_{\rm evap}\propto \bar{\rho}^{-1}$, meaning that the planet with more (less) H/He also loses more (less). As we see here numerically, and show analytically in Sect. \ref{sect:derivationmbare}, the mass-radius relation  $R(\mcore,\menve)$ of low-mass planets with H/He is such that the two effects nearly cancel. 

For the distance dependency, the analytical model also predicts that $\mbare\propto a^{-1}$ ($e_{\rm m}$=1), and $\rbare \propto a^{-0.27}$ ($e_{\rm r}$=0.27), independently of the scaling of the envelope mass with core mass $p_{\rm e}$, which is to good approximation also the case in most of the numerical simulations shown here. For M2 to M4, $e_{\rm m}$=0.87 to 1.07 and $e_{\rm r}$=0.25 to 0.3.  The largest discrepancy between analytical prediction and simulation occurs for M1 where $e_{\rm m}$=1.49 and  $e_{\rm r}=0.36$. In M1, the envelope mass is independent of core mass, meaning that also massive cores $\gtrsim10\mearth$ only have a very low-mass envelopes, in contrast to the predictions from formation theory. This case is addressed in Sect. \ref{sect:companalyticalnumerical}.

\subsubsection{{Constraints on post-formation envelope masses}}

\begin{figure*}
\begin{center}
  \includegraphics[width=0.82\textwidth]{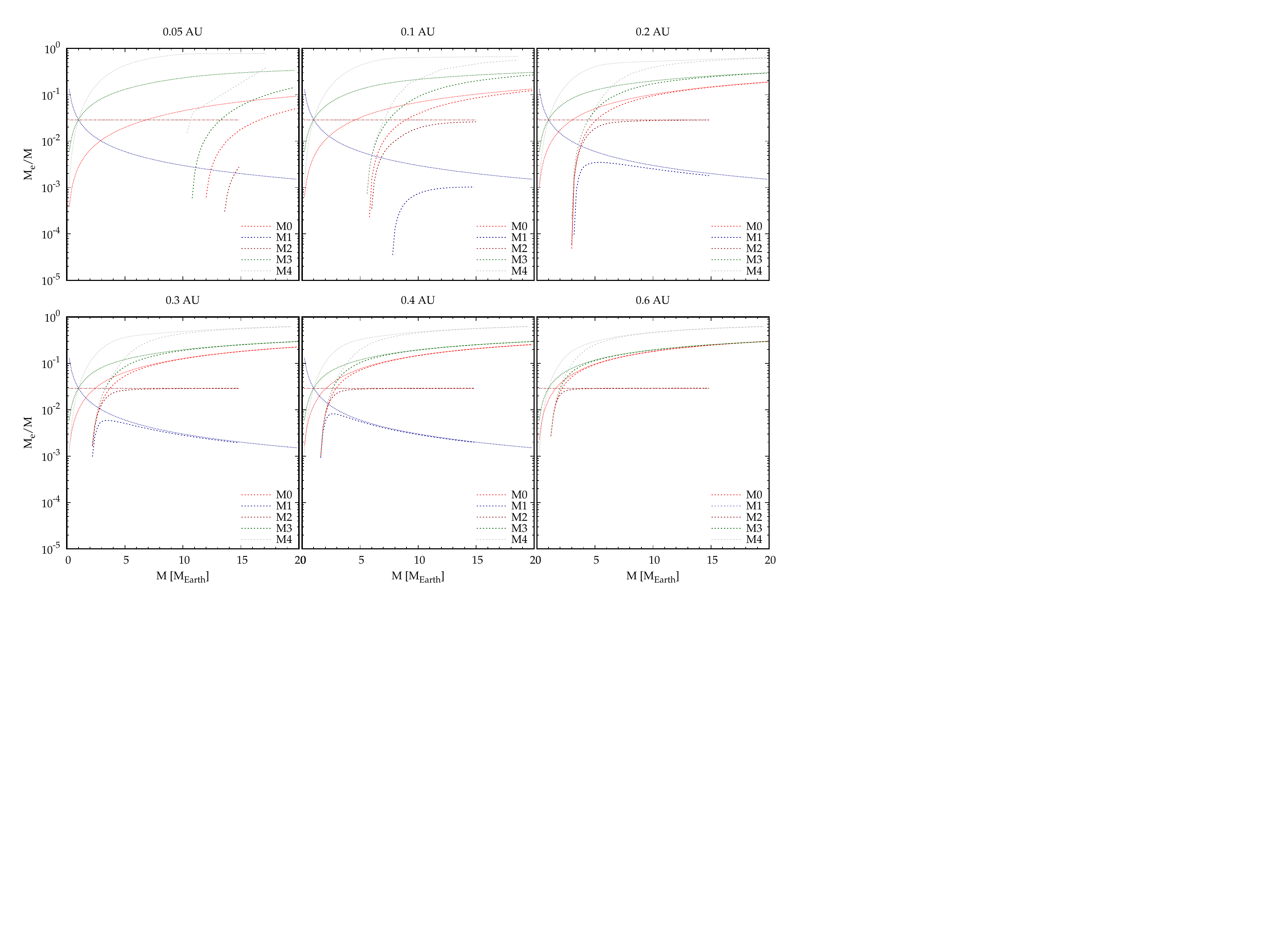}
%\vspace{20ex}
    \includegraphics[width=0.82\textwidth]{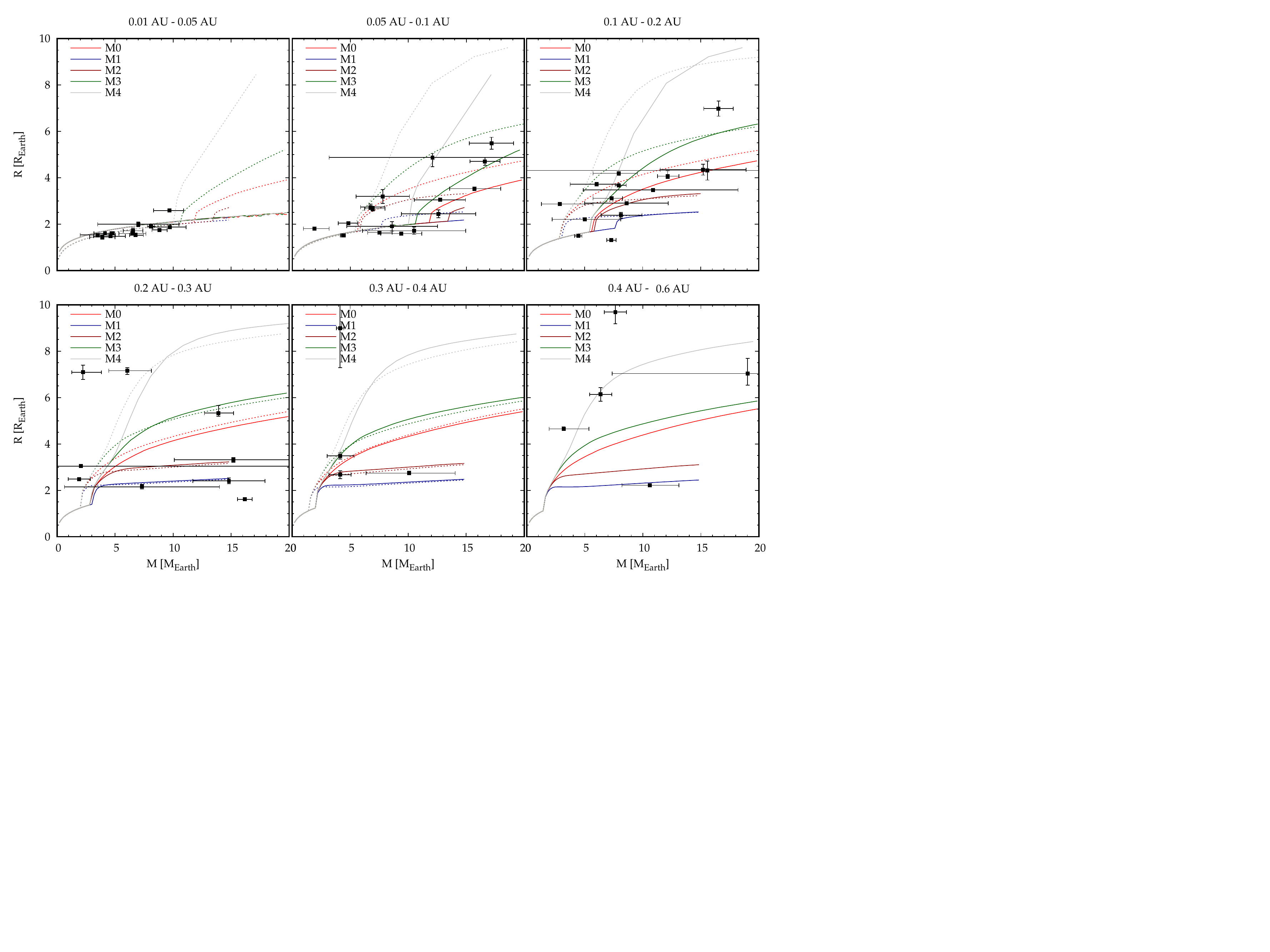}
\caption{{\textit{Upper six panels:} Envelope mass fraction $M_{e}/M$ as a function of planet mass $M$ for 6 different orbital distances. The dashed lines show the envelope mass fraction at 5 Gyr, while the dotted lines show the initial (post-formation) envelope mass fraction $M_{e,0}/M$. The simulations M0-M4 from Figs. \ref{fig:MRbareS0} and \ref{fig:MRbareM1M2M3M4} are shown to display the consequences of different initial conditions. \textit{Lower six panels:} Corresponding mass-radius relations compared to observations. In each panel, the solid and dashed lines correspond to the lower and upper limit of the distance interval, respectively. Black dots with error bars show the observed extrasolar planets orbiting stars with masses between 0.7 and 1.3 $\msun$ in these distance intervals. }}\label{fig:MRdistances} 
\end{center}
\end{figure*}

{While the valley location  is weakly dependent on the post-formation envelope mass, far above the valley (i.e., at larger orbital distances and/or higher planetary masses), the influence of photoevaporation must be weaker and become eventually negligible, so that planets there still have the unaltered initial (post-formation) envelope mass. In Figs. \ref{fig:MRbareS0} and \ref{fig:MRbareM1M2M3M4}, planets that have lost less than about 10 \% of their initial envelope are shown by blue points.}

{To further investigate the relation between initial envelope mass and the one after 5 Gyr of evolution, we show in the upper six panels of Fig. \ref{fig:MRdistances} the envelope mass fraction $M_{e}/M$ as a function of planet mass $M$ for different orbital distances. Both the envelope mass fractions at 5 Gyr as well as the initial fractions are shown. Inspecting the curves shows that first, the smaller the orbital distance, the higher the planets' mass relative to $M_{\rm bare}$ that  still have lost a significant part of their original envelope. Second, the transition from full loss to nearly unaltered envelope masses is more brusque for lower post-formation envelope masses (cases M1, M2) than for very high ones (M3, M4). For example, at 0.2 AU, in the case of very large initial $M_{\rm e}$ for planets more massive than 1 $\mearth$ in simulation M4, a planet needs a mass more than 10 $\mearth$ higher than the $M_{\rm bare}$ at this distance (about 3 $\mearth$) to approximately still have the initial envelope mass. In M2 with much lower initial envelope masses, a mass difference only about half as high is sufficient. The same picture is also seen is in terms of the radii instead of masses (right panels of Figs. \ref{fig:MRbareS0} and \ref{fig:MRbareM1M2M3M4}).  }  

{The lower six panels of Fig. \ref{fig:MRdistances} show the corresponding mass-radius relationships, comparing them to observations. In each panel, the dashed lines correspond to the dashed $M_{e}/M$ lines in the upper group of panels. Black dots with error bars show the observed extrasolar planets orbiting stars with masses between 0.7 and 1.3 $\msun$ with semimajor axes in these six distance intervals. The observational data was downloaded from the NASA Exoplanet Archive on 6. Dec. 2019 and includes all planets that have the radius, mass, semimajor axis and host star mass including the errors listed. }

{In the panel depicting planets with semimajor axes between 0.01 and 0.05 AU, one sees that with the exception of one planet, there are only objects which do not possess a voluminous envelope.  This corresponds to the absence of close-in low-density planets of low and intermediate mass, i.e.  the (evaporation) desert. These very close-in planets therefore have a  small radii. The observed planets approximately following the theoretical mass-radius relation of bare rocky cores, which is given by the solid line (at 0.01 AU, in all simulation M0-M4, all planets with masses less than 20 $\mearth$ have completely lost their envelopes).}

{In the distance interval of 0.05-0.1 AU, there is already a much larger spread visible in the $M-R$ relation. At smaller masses $\lesssim 10 \mearth$, there are planets which continue to approximately fall on the envelope-free $M-R$ relation. A general trend of increasing radius with increasing mass is seen both in the theoretical and observed data, but with a large scatter. To be compatible with one single theoretical post-formation $M_{\rm e}(M_{\rm c})$ relation like for example the nominal relation of simulation M0, all actual exoplanets would have to fall between the dashed and solid red line. The plot shows that this is not the case. Instead it seems that the actual planets had a diversity of post-formation envelope mass that are covered by a combination of the simulation M0, M1, M2, and M3. The very high post-formation envelope masses for planets more massive than 10 $\mearth$ occurring in simulation M4 (which would result in planets with radii larger than 6 $\rearth$) are in contrast not occurring in the (small) observational sample. }

{The situation is similar for semimajor axes between 0.1 and 0.2 AU, but the number of planets without H/He decreases. There is also one planet which seems to have had a post-formation envelope mass that was higher than in simulation M3 but lower than in M4. This is compatible with a trend of increasing envelope mass fraction with increasing orbital distance which is predicted theoretically \citep[][Sect. \ref{sect:initialconditions}]{ikomahori2012,leechiang2015}. } 

{Moving even further out, we see that there is still a large spread in the observed $M-R$ relation. In contrast to smaller semimajor axes, however, there are now exoplanets that have very large radii for their mass, similar to or even larger than in simulation M4. In the context of the model, these planets would thus have started with envelope mass fractions even higher than in M4.}   

{In summary, Figure \ref{fig:MRdistances} shows that at distances of up to several 0.1 AU, it is necessary to take into account the reduction of the envelope mass during evolution when using the observed mass-radius relationship of sub-Neptunes to constrain the efficiency of  H/He envelope accretion during formation \citep{mordasiniklahr2014,lopezfortney2014}, as the envelope mass reduction may be significant also for planets above the valley (cf. \citealt{estrelaswain2020}). Only planets which are about 5-10 $\mearth$ more massive than  $M_{\rm bare}$ at their orbital distance still have envelope masses that differ only weakly from the primordial value. The comparison with the observed mass-radius relation shows  that the general trend of a post-formation envelope mass which increases with planet mass and orbital distance is present also in the observed population, but that the actual planets were born with a significant spread of envelope masses.}

\subsubsection{Normalization of  the initial envelope mass: N1, N2}\label{sect:N1N2}
In the context of the impact of the post-formation H/He mass on the locus of the valley, we have also explored the consequences of varying the normalization constant $M_{\rm e,1}$ when expressing the post-formation envelope mass as  $M_{\rm e,0}/\mearth=M_{\rm e,1} (\mcore/\mearth)^{p_{\rm e}}$  (Eq. \ref{eq:memcscaling}).  In simulation N1 and N2 $M_{\rm e,0}$ is set to 0.06 and 0.3, i.e., a factor two and ten increase relative to the  value of 0.03 $\mearth$ used in M0 to M4.

\begin{figure*}
\begin{center}
\begin{minipage}{0.48\textwidth}
	 \centering
        \includegraphics[width=0.95\textwidth]{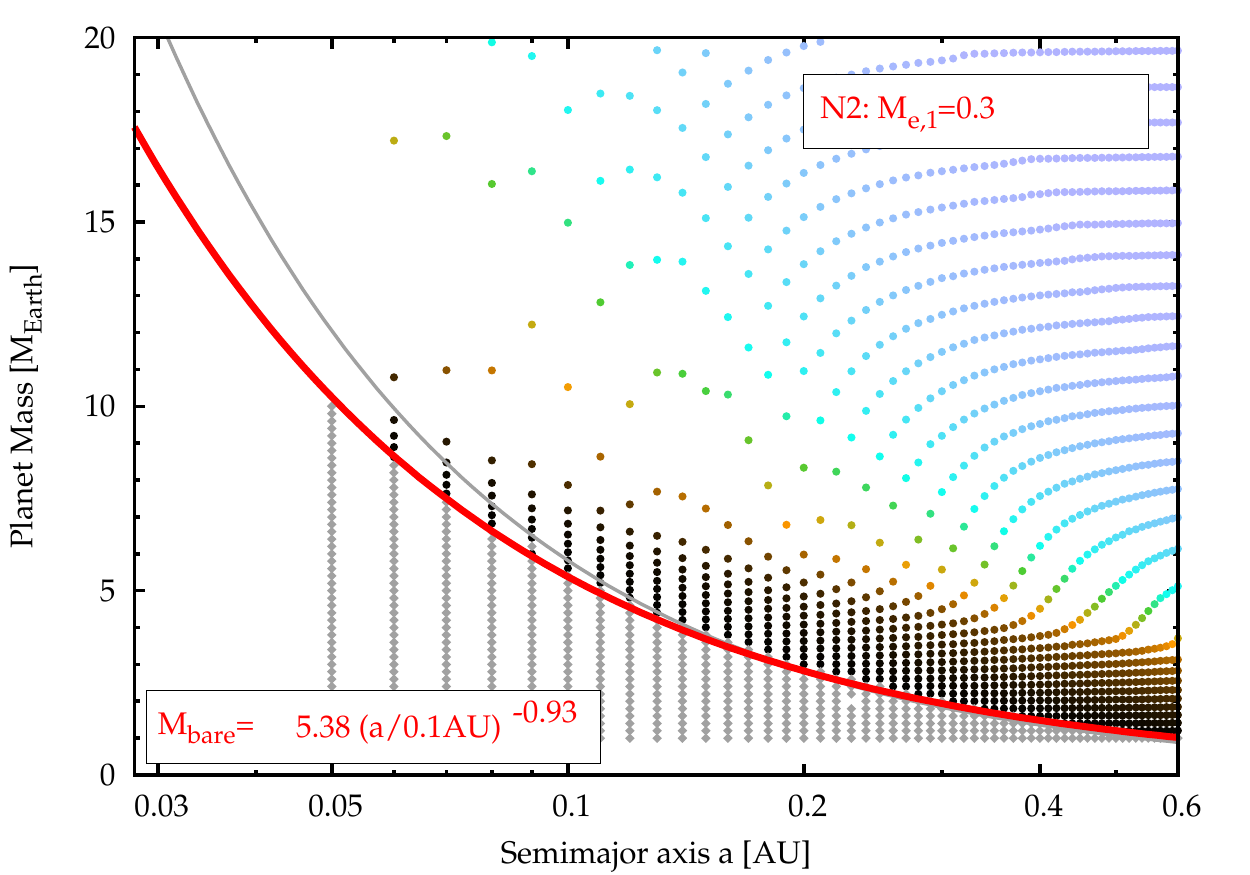}

     \end{minipage}%\hfill
     \begin{minipage}{0.48\textwidth}
      \centering
       \includegraphics[width=0.95\textwidth]{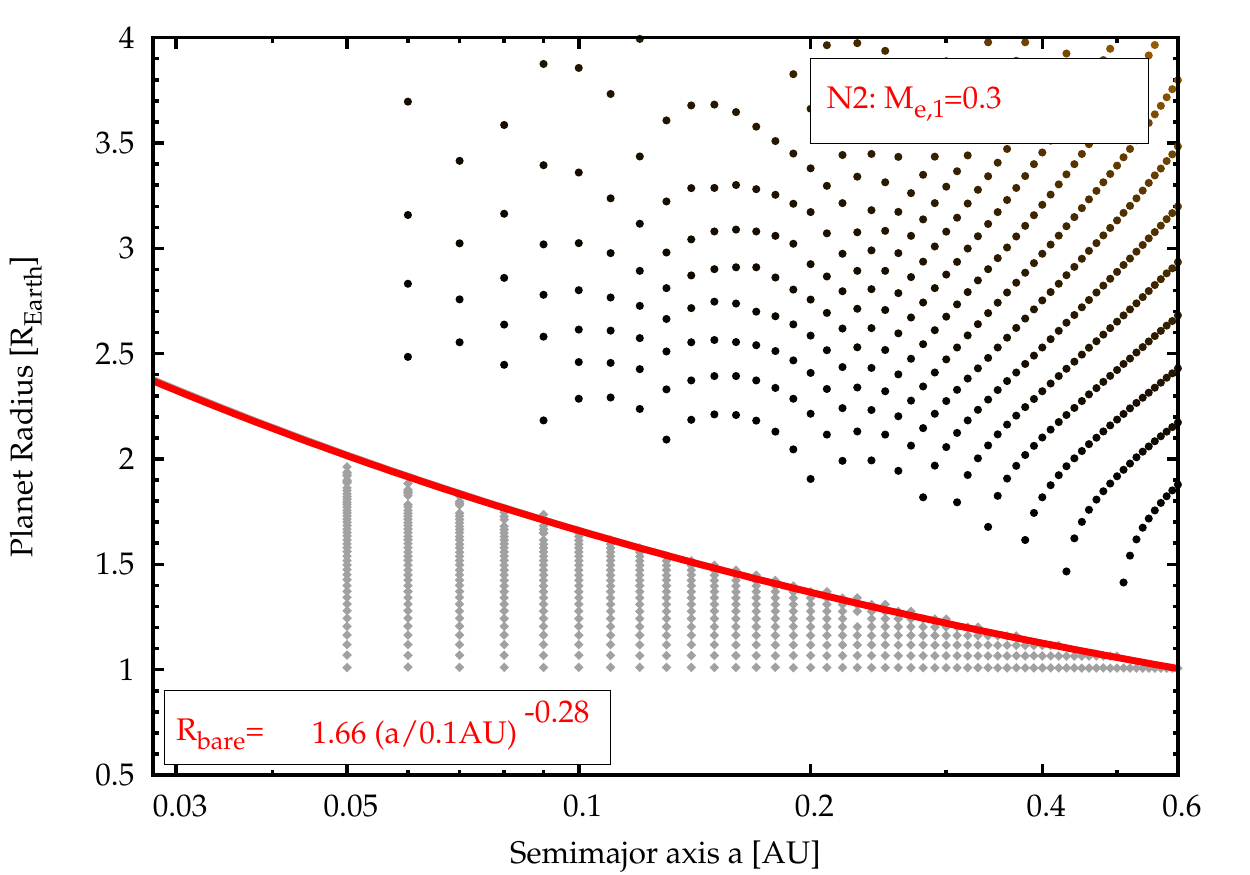}
     \end{minipage}
\caption{Impact of the normalization of  the initial envelope mass as found in simulation N2. The figure is analogous to Fig. \ref{fig:MRbareS0}. Compared to the reference case M3, the normalization constant of the envelope mass (i.e., the envelope mass for a 1 $\mearth$ planet) is here $M_{\rm e,1}=0.3$ instead of 0.03 $\mearth$, i.e. all the initial envelope masses are increased uniformly by a factor 10.}\label{fig:MRbareMenve}
\end{center}
\end{figure*}

Figure \ref{fig:MRbareMenve} shows the result for simulation N2 (tenfold increase of the envelope masses relative to M3). Despite this large increase, the locus of the transition again only changes marginally. This again shows that the post-formation envelope mass is not important for the final position of the evaporation valley. This is also predicted by the analytical model, where no dependency at all on $M_{\rm e,1}$ is found.
The simulation N1 where the normalization mass is twice as large as in the reference case is not shown as it is even closer to nominal simulation (but it is included in Table \ref{tab:sims}). 

\begin{figure*}
\begin{center}
\begin{minipage}{0.49\textwidth}
	 \centering
        \includegraphics[width=0.95\textwidth]{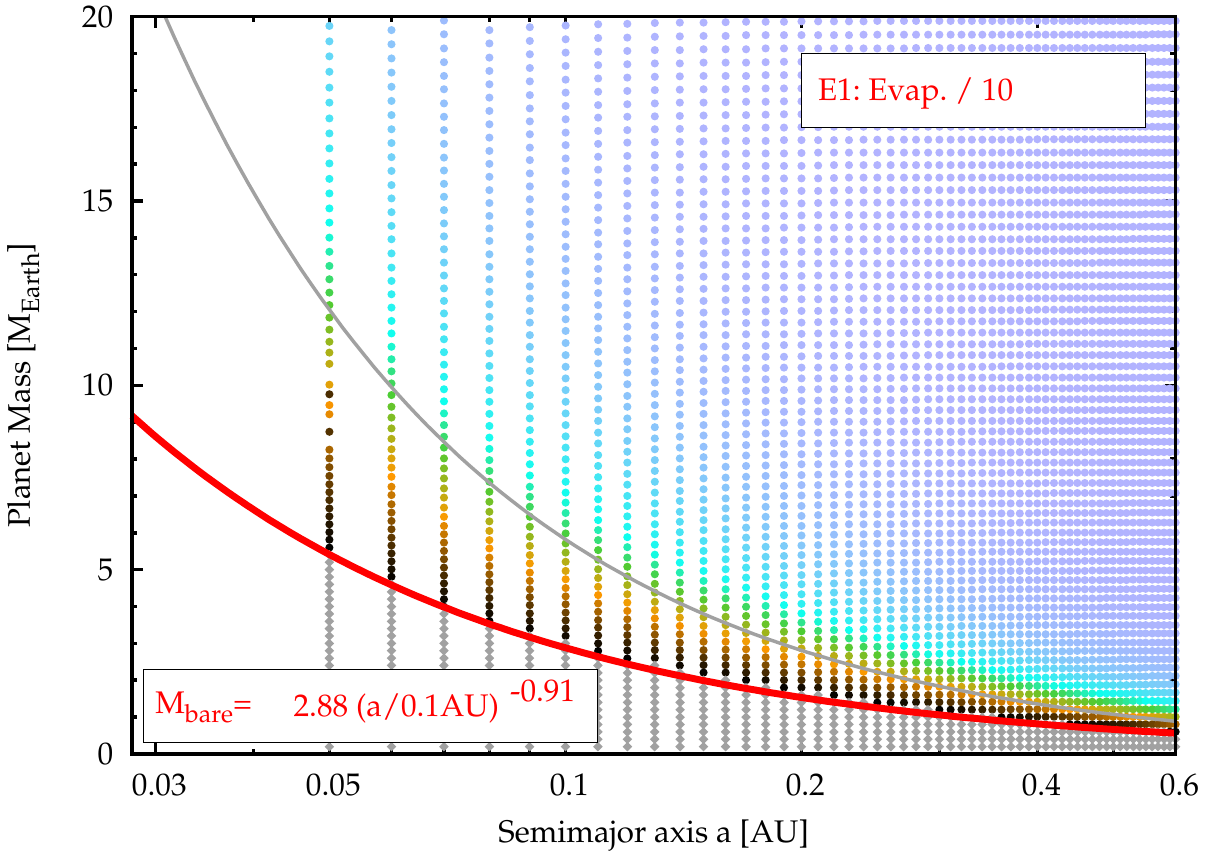}
        \includegraphics[width=0.95\textwidth]{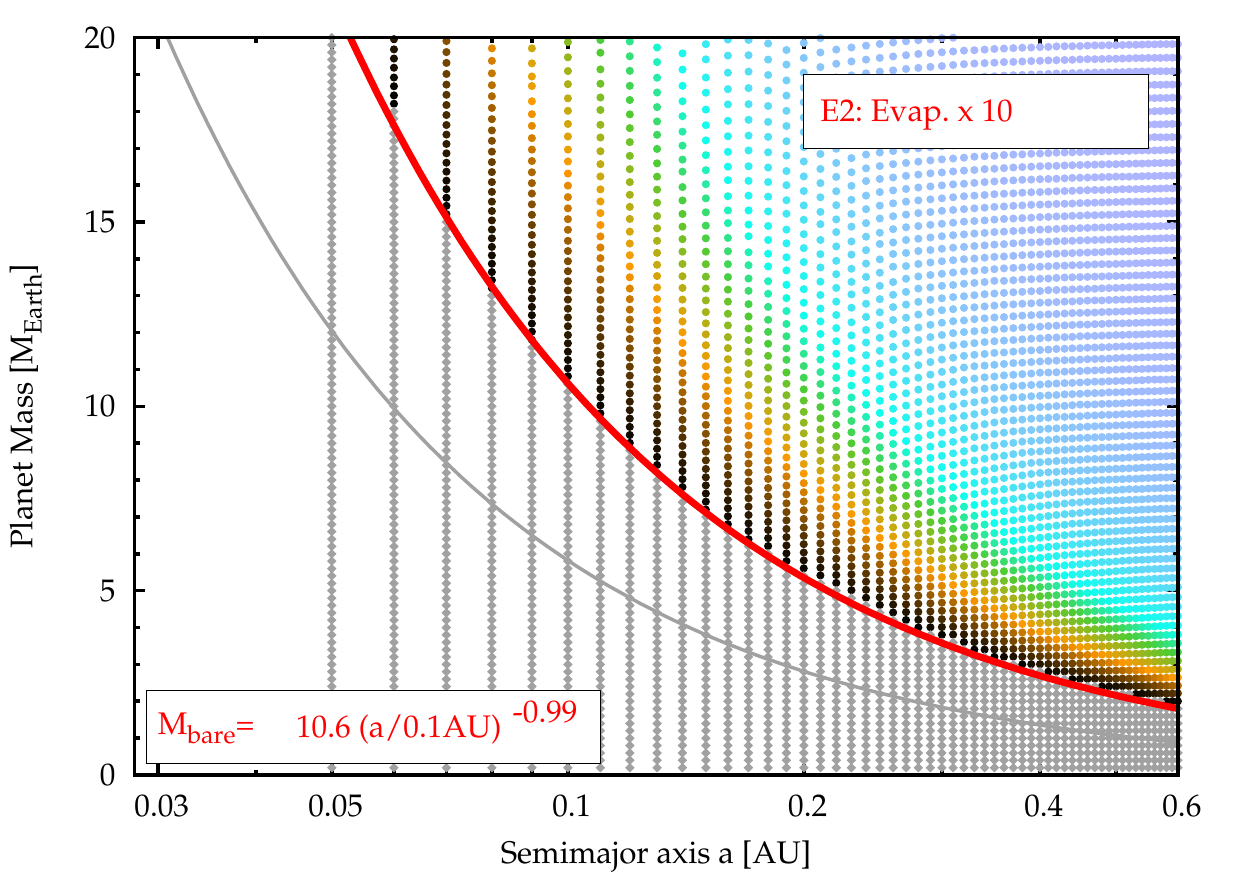}

         \end{minipage}
     \begin{minipage}{0.49\textwidth}
      \centering
       \includegraphics[width=0.95\textwidth]{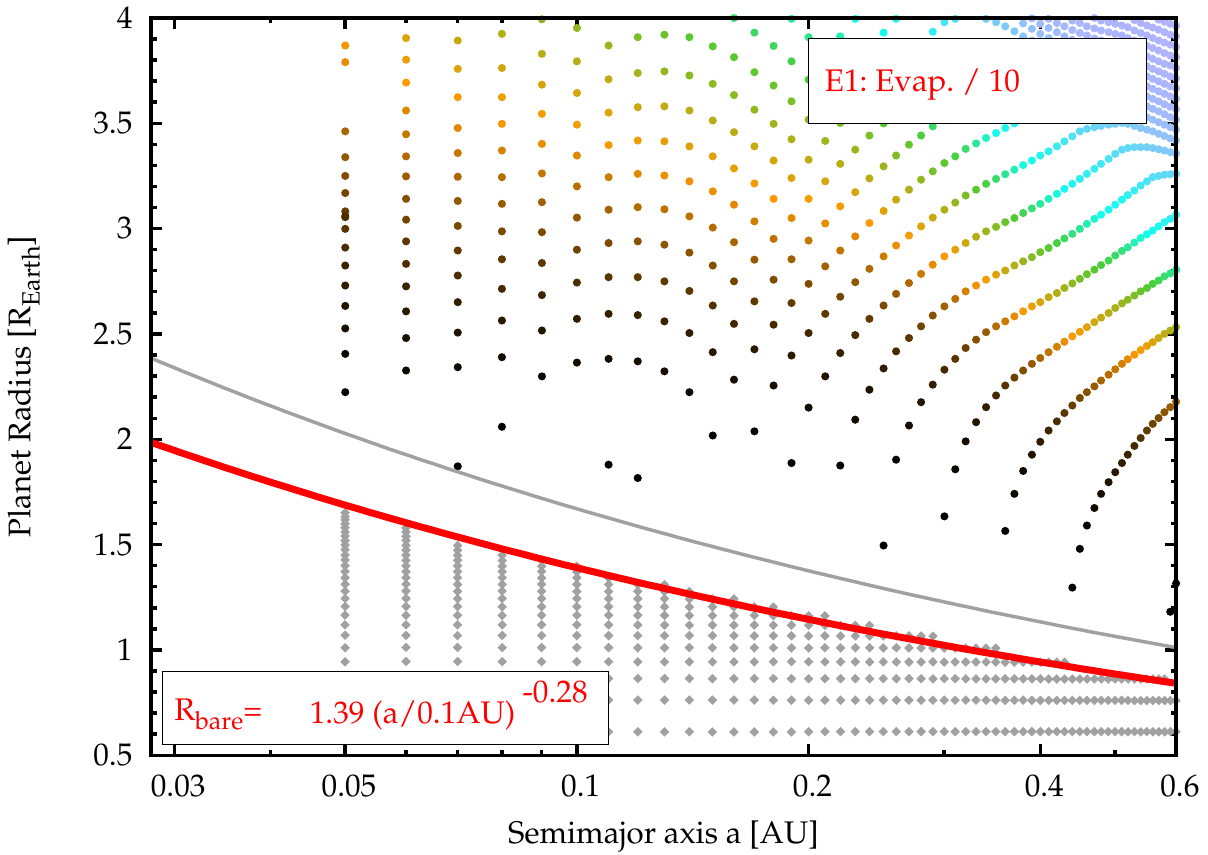}
       \includegraphics[width=0.95\textwidth]{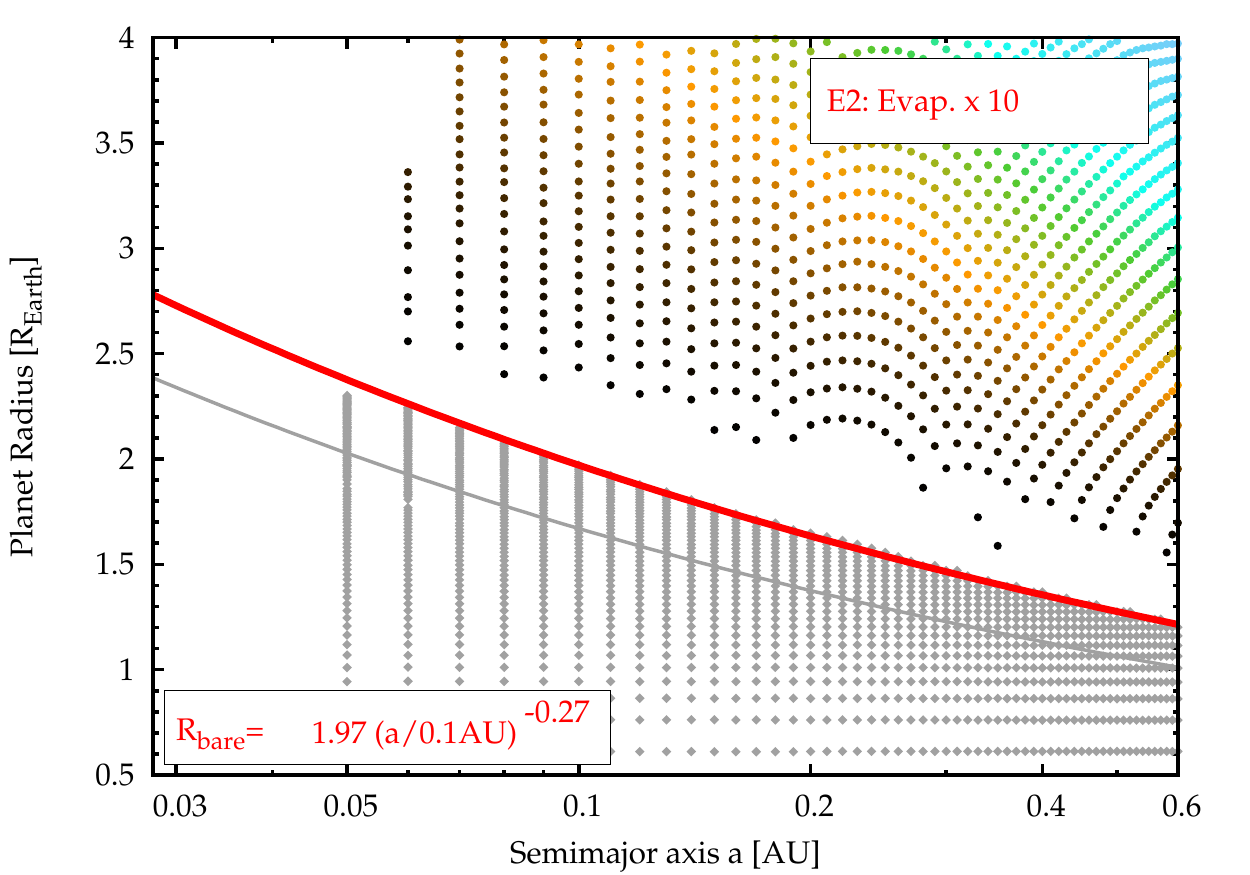}
     \end{minipage}
\caption{Impact of the strength of evaporation on the transition. The figure is analogous to Figure \ref{fig:MRbareS0} but for simulation E1 (top panels) and E2 (bottom panels) where the evaporation rate is uniformly reduced (E1) respectively increased (E2) by a factor 10 relative to the nominal case.  The transition mass and radius were again fitted with a power law for $\mbare$ and $\rbare$ as shown with the thick red line. The power law exponents for the dependency on the distance have remained  similar as in the reference case M3, but the absolute value of $\mbare$ is reduced  in E1 (weak evaporation) relative to M3  by a factor 2.02, and a factor 1.20 for  $\rbare$. In E2 (strong evaporation), $\mbare$ and $\rbare$ have increased by a factor 1.82 and  1.18, respectively. The thinner gray line shows the transition mass and radius in the nominal simulation M0 to allow direct comparison. M3 is in turn very similar to M0.}\label{fig:MRbareS15}
\end{center}
\end{figure*}

\subsubsection{Evaporation rate: E1, E2}\label{sect:evaprateE1E2}
The evaporation model used in this work is relatively simple, as it does not directly solve the conservation equations to obtain the escape rate as for example in \citet{murray-claychiang2009}, but instead uses the energy or radiation-recombination-limited formulae to calculate the loss rate, assuming constant efficiency factors. It also neglects the consequences of different atmospheric compositions or magnetic fields. It is clear that our escape rates are therefore only  rough estimates.

%=====================================

Furthermore, it is well known  that even at a fixed stellar mass of around 1 $\msun$,  stellar rotation rates \citep[e.g.,][]{johnstoneguedel2015} and associated $L_{\rm XUV}$ luminosities exhibit a spread of about a factor $\sim$30 at young ages \citep{tujohnstone2015} when most of the escape occurs. It is, however, also worth noting  that most stars still cluster around a typical rotation period of about 5 days \citep[][]{johnstoneguedel2015}, where  $L_{\rm XUV}\approx 10^{30}$ erg/s during the first $\sim$100 Myr \citep{tujohnstone2015}, and only a small fraction has markedly different rotation rates. Nevertheless, given this observed spread, it is therefore even  more important to investigate the impact of a variable strength of the evaporation.

To account for this, we reduce in simulation E1 the evaporation rate predicted by the model uniformly by a factor 10, while in simulation E2 we increase the rate uniformly by such a factor, so that the consequences of an escape rate varying by two orders of magnitudes can be assessed. Physically, this variation could be due to the (combined) effects introduced by different stellar $L_{\rm XUV}$, different durations of the saturated phase, or variations in the efficiency factors of escape. By simply scaling the evaporation rates, we do not need to specify this explicitly.

%young sun: large spread!
%Large spread in XUV flux: 10 x EUV, 30 x X rays, Tu, Johnstone, et al. 2015. Very different depending rotation of star.  But they peak at one value!

Figure \ref{fig:MRbareS15} shows the result of the E1 and E2 simulations. The higher (lower) evaporation rate leads to a higher (lower) transition mass and radius, as expected. For the low evaporation rate, relative to the reference simulation M3 (from which E1 and E2 only differ by the evaporation rate), the $M_{\rm b,0p1}$ has decreased from 5.52 to 2.88 $\mearth$, corresponding to a reduction by a factor 1.92. The radius $R_{\rm b,0p1}$ has decreased from 1.65 to 1.39 $\rearth$, a factor 1.19 change. 

For the high evaporation rate, $M_{\rm b,0p1}$ has increased from 5.52 to 10.6 $\mearth$, corresponding to decrease also by a factor 1.92. The radius $R_{\rm b,0p1}$ has increased from 1.65 to 1.97 $\rearth$, again a factor 1.19 change. So the total change in $M_{\rm b,0p1}$ by a factor 3.68 for a variation of a factor 100 in evaporation rate suggests a weak dependency on the factors determining the evaporation rate like $L_{\rm XUV}$ or $\varepsilon$ that scales only as approximately $M_{\rm b,0p1} \propto L_{\rm XUV}^{0.28}$. For comparison, the analytical model predicts a dependency like $L_{\rm XUV}^{0.5}$ (Eq. \ref{eq:mbarelaterocky}). Similarly, the total variation in radius by a factor 1.97/1.39=1.42 suggests a very weak power law dependency just like approximately $L_{\rm XUV}^{0.08}$. Analytically, we find for comparison $R_{\rm b,0p1} \propto L_{\rm XUV}^{0.135}$ (Eq. \ref{eq:rbarelate}). While the exact values of the exponents are certainly dependent on the details of the model used here, we consistently find weak dependencies, in particular for the transition radius.

This has a very important observational implication: From the location and width of the valley in the two simulations it becomes apparent that if we would overlay the two simulations, a depleted region would appear, but it would not be completely empty: regions completely devoid of planets in E1 would be partially populated by planets from E2, and vice versa. This is the case because the change in the valley's location (1.39 versus 1.97 $\rearth$ at 0.1 AU in E1 and E2, respectively, corresponding to a $\Delta R=0.58\rearth$) is comparable to the intrinsic width $\Delta W$ of the valley (about 0.5 $\rearth$, see also \citealt{owenwu2017}). If we would instead have $\Delta R\gg \Delta W$ (as it would be the case if the dependency on $L_{\rm XUV}$ would be stronger, as one could maybe naively expect from Eq. \ref{eq:mdotenergylim}), the valley would not be well visible (or even not at all), as it would scatter too much from star to star.  If we would have the other extreme, $\Delta R\ll \Delta W$ (for example if the strength of evaporation would be identical in all systems), then the valley would be completely empty. Both these things are not observed.
%==========================
But if in nature the location of the valley varies from star to star (because of different $L_{\rm XUV}$, but also different efficiency factors for example because of different atmospheric compositions) indeed by about 0.5 $\rearth$ around the mean value as we find here, while the intrinsic width $\Delta W$ has a similar magnitude (again as found here), then we deduce that we should still see a depleted valley, but not a completely empty one. This  corresponds to the observational result \citep{fultonpetigura2018}. Clearly, in these Kepler observations, we see the overlay of all individual system-specific valleys.

\subsubsection{Atmospheric opacity: O1}
\begin{figure*}
\begin{center}
\begin{minipage}{0.5\textwidth}
	 \centering
        \includegraphics[width=0.95\textwidth]{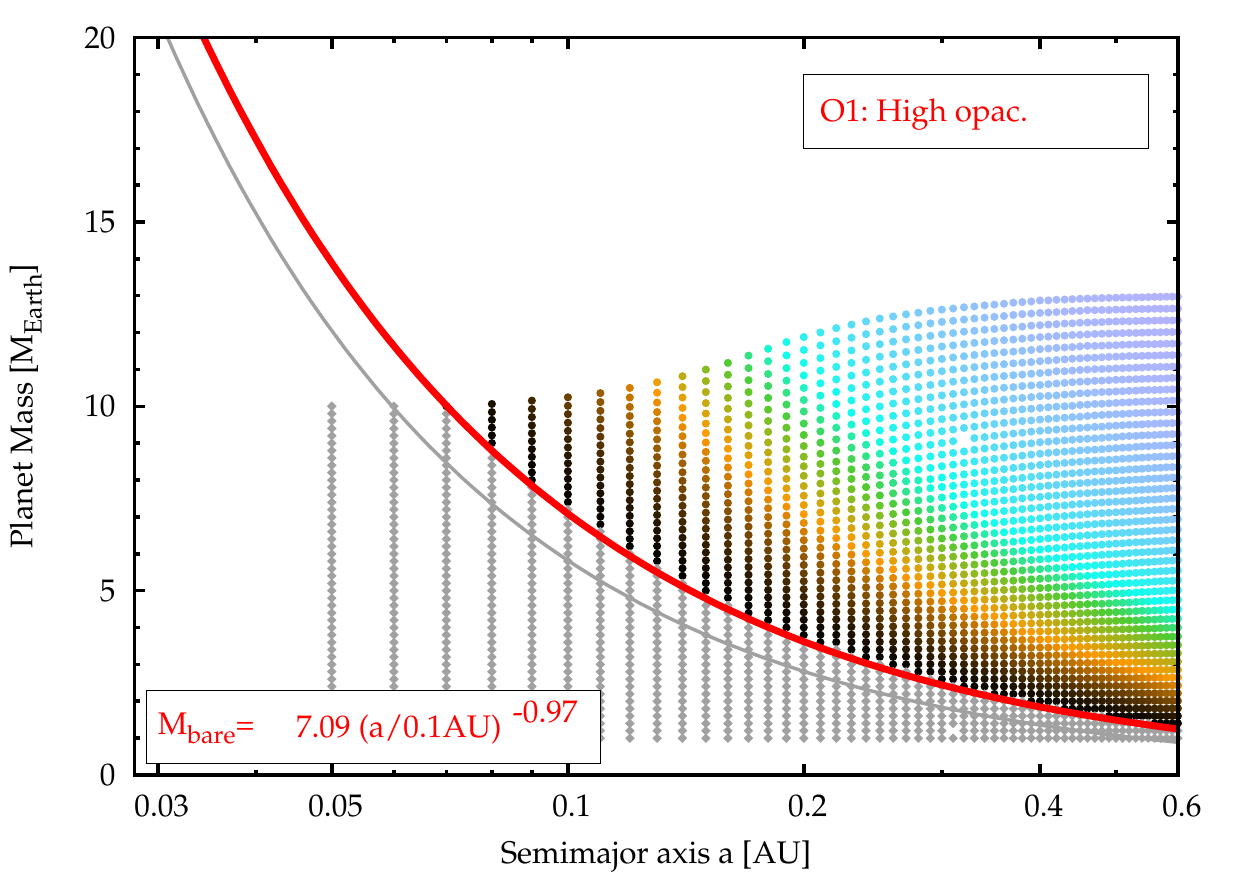}
         \end{minipage}\hfill
     \begin{minipage}{0.5\textwidth}
      \centering
       \includegraphics[width=0.95\textwidth]{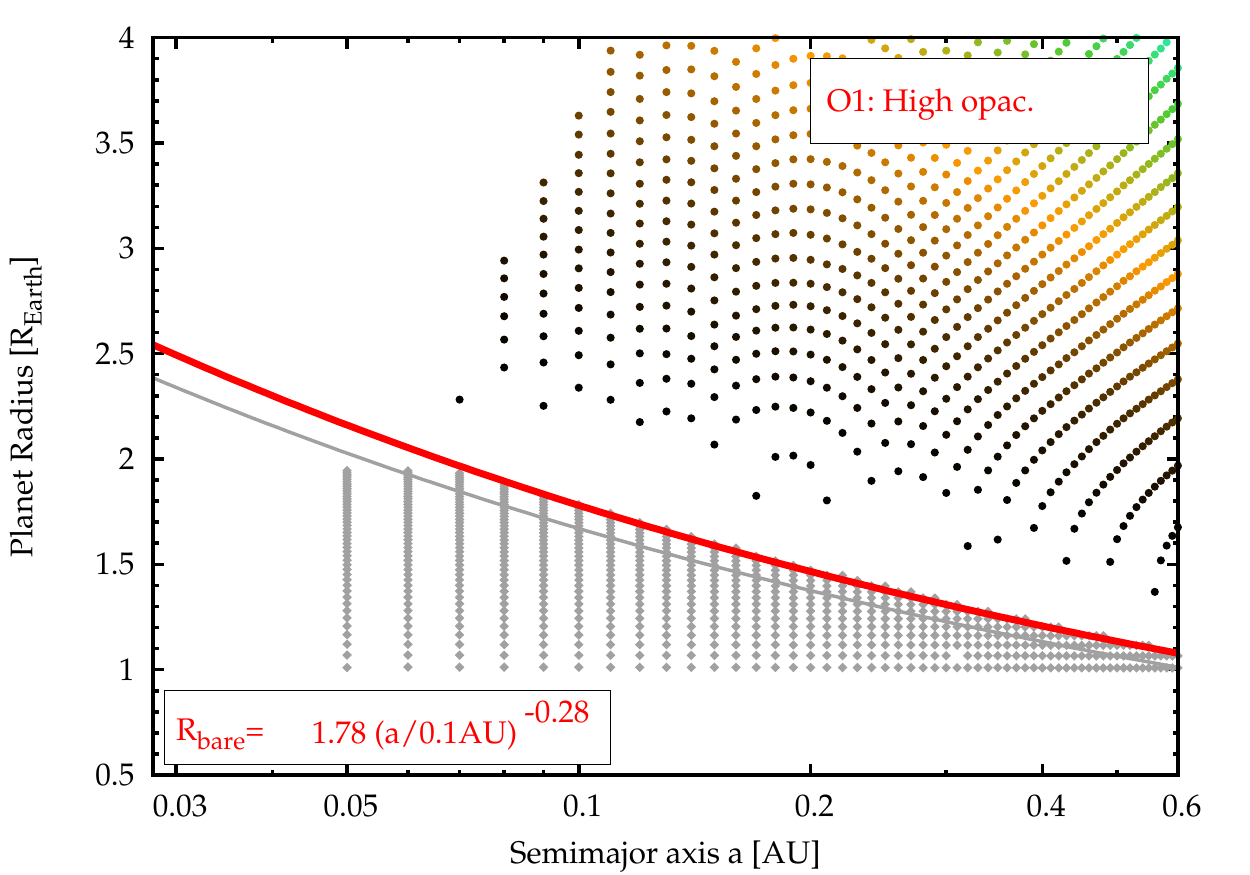}
     \end{minipage}
 \begin{minipage}{0.5\textwidth}
	 \centering
        \includegraphics[width=0.95\textwidth]{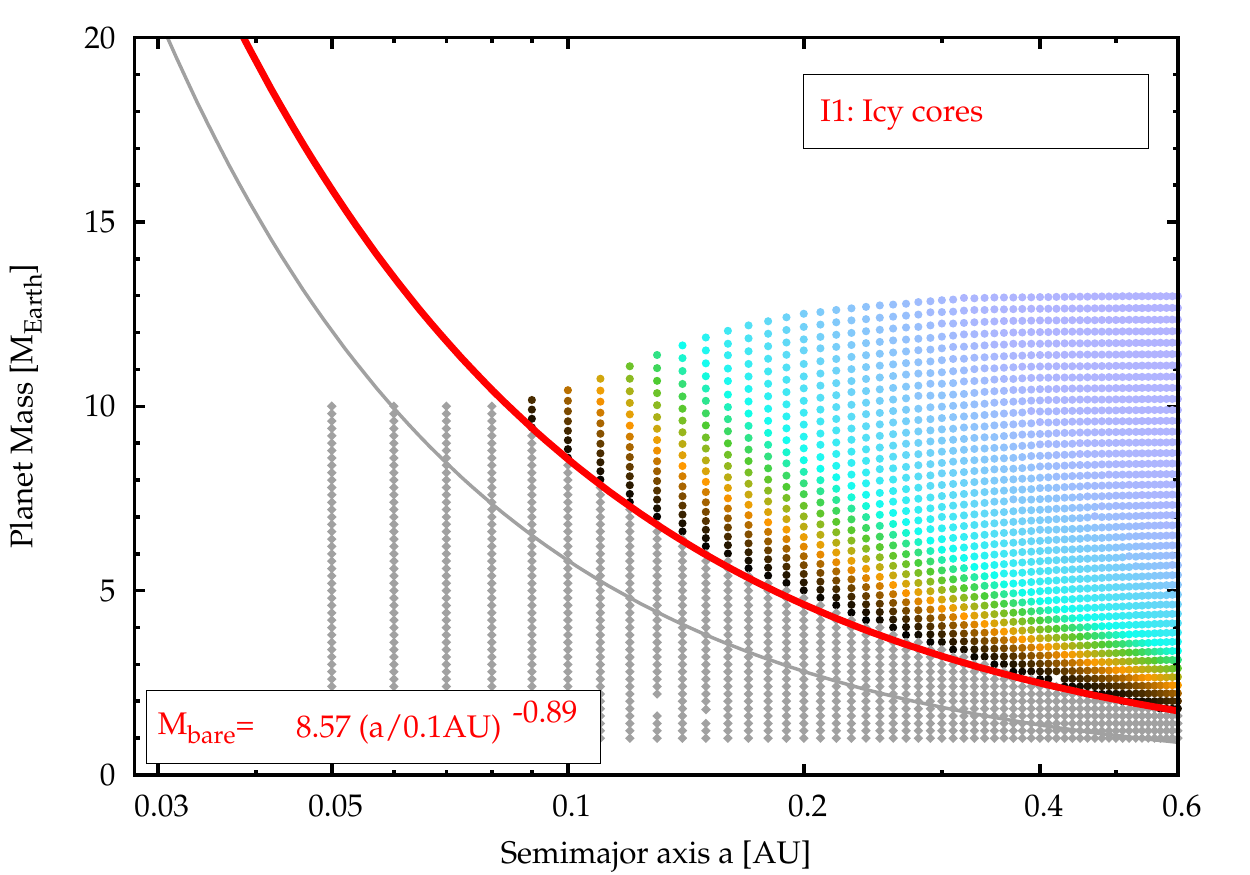}
     \end{minipage}%\hfill
     \begin{minipage}{0.5\textwidth}
      \centering
       \includegraphics[width=0.95\textwidth]{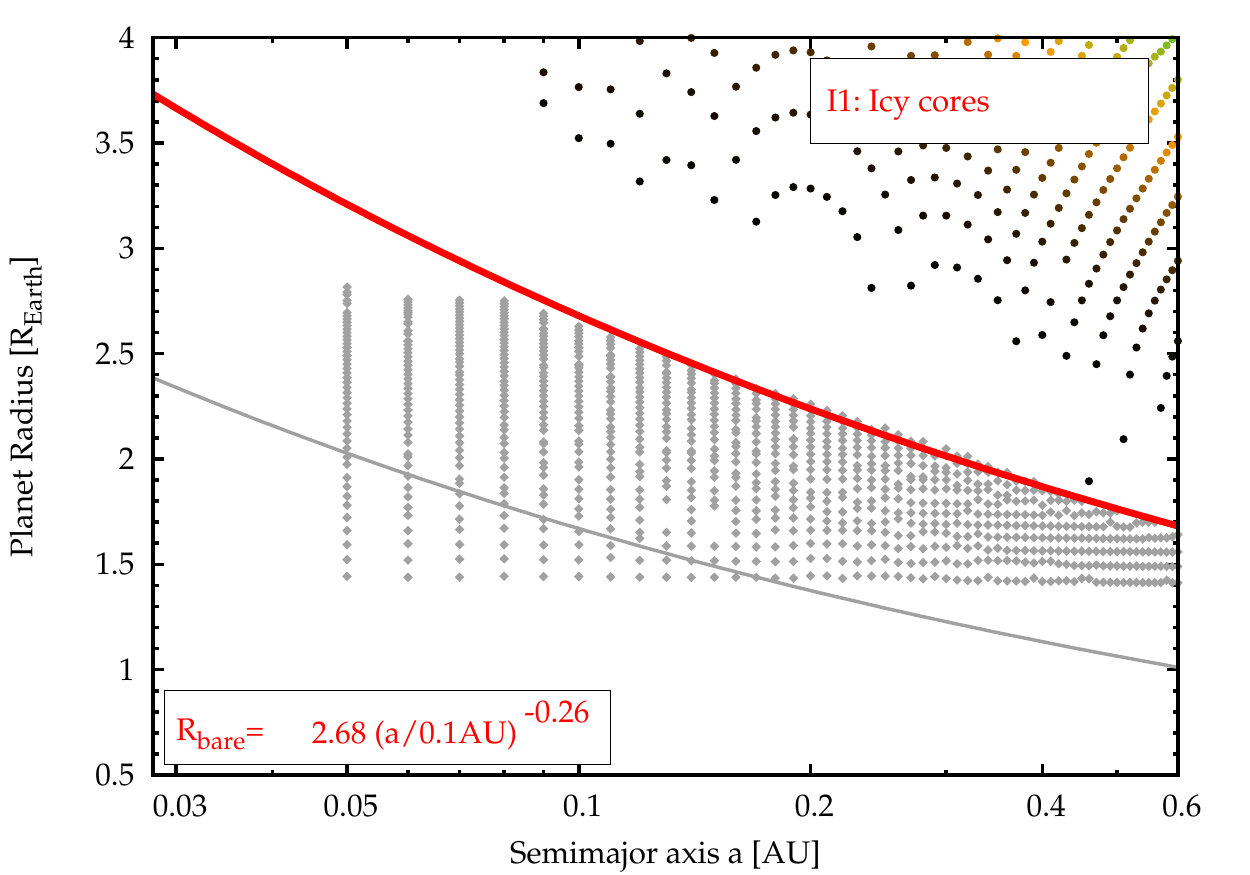}
     \end{minipage}    
\caption{{This figure is analogous to Figure \ref{fig:MRbareS0} but shows the i}mpact of the atmospheric opacity (O1, top panels) and of the composition of the solid core (I1, bottom panels). These simulations cover a smaller range in initial core masses than the other ones.  In {simulation O1} the {atmospheric} opacity {is uniformly increased} by a factor 10. {In simulation I1} an ice mass fraction in the core of unity instead of an Earth-like composition {was assumed.} The thin gray line shows the transition mass and radius in the nominal simulation M0 to allow direct comparison.}\label{fig:MRbareOpa}
\end{center}
\end{figure*}

The top panels of Fig. \ref{fig:MRbareOpa} show the consequences for the locus of the valley resulting from increasing  the Rossland mean opacity $\kappa_{\rm R}$ in all planets uniformly by a factor 10 relative to the nominal case. In the nominal case, we have used solar-composition grain-free opacities from \citet{freedmanlustig-yaeger2014} when calculating the cooling and contraction of the planets. {I}n the solar system, the atmospheric (and bulk) enrichment in metals is  increasing  with decreasing planetary mass \citep{mordasinivanboekel2016}. This is also predicted by planet formation models based on the core accretion paradigm both for the bulk \citep{fortneymordasini2013} and {(under the assumption of efficient mixing of atmosphere and envelope)} atmospheric  composition \citep{mordasinivanboekel2016}. 

{In extrasolar planets, the bulk metallicity also increases with decreasing planet mass, similarly  as in the solar system \citep{thorngrenfortney2016}. Regarding the atmospheric metallicity, the situation seems different:  \citet{fisherheng2018} find no trend in the retrieved atmospheric water abundances across nearly two orders of magnitude in exoplanet mass. \citet{wallackknutson2019} also see no evidence for a solar-system-like correlation between planet mass and atmospheric metallicity. Both studies do however find a large spread in atmospheric enrichments. It should also be noted that the large majority of the planets studied in these works are more massive giant planets far from the valley.} {In any case, these findings indicate that there  could  be a large diversity in atmospheric compositions, meaning} that the opacity in the low-mass planets we are studying here {could be} higher than solar.

A higher opacity in the atmospheres delays the cooling and contraction of the planets \citep[e.g.,][]{burrowshubeny2007}, reducing their mean density, which leads to a higher evaporation rate (Eq. \ref{eq:mdotenergylim}). We thus expect that at high opacity, the valley should move to higher $\mbare$ and $\rbare$ at fixed distance compared to the nominal case. Figure \ref{fig:MRbareOpa} shows that this is indeed the case. But we also see that the difference is small. Relative to M3, the reference simulation which differs from O1 only by the enhanced opacity, the transition mass grows by a factor 7.09/5.52=1.28, while for the radius forming the bottom of the valley, an increase by just 1.78/1.65=1.08 is observed.  These are small factors given the tenfold increase of $\kappa_{\rm R}$, which would correspond to a power law dependency approximately $\propto\kappa_{\rm R}^{0.03}$ for the radius. This weak impact could be related to a certain auto-regulation in the sense that at fixed mass, a larger radius (caused by higher $\kappa_{\rm R}$) leads to a higher evaporation rate, which reduces the envelope mass, which in turn reduces the radius, and eventually the evaporation rate.

A tenfold increase of the opacity approximately corresponds to an atmospheric metal enrichment of about 10-30 x solar \citep{freedmanlustig-yaeger2014}. In reality, the exact enrichment level corresponding to such an opacity increase depends on pressure and temperature \citep{freedmanlustig-yaeger2014} and is not uniform. For the 5-10 $\mearth$ planets we are mainly dealing with, the atmospheric enrichment could be even higher than 10-30 x solar \citep{mordasinivanboekel2016}. In the solar system for example, Uranus and Neptune are enriched in carbon (observed as CH$_{4}$ in the atmosphere) by a factor of about 80 relative to the Sun \citep{guillotgautier2014}. So we may expect enrichments level on the order of 100 x solar. But the weak dependency on $\kappa_{\rm R}$ found in O1 indicates that even for such a high enrichments, there would not be a very large shift of the valley.

\subsubsection{{Metal enrichment of the gaseous envelope: Z1-Z3}}\label{sect:metalenrichmentatmo}
{In simulation O1}, we have increased the opacity, but not modified two other quantities that in reality  also change for a higher amount of metals: first, the mean molecular weight {or more generally speaking, the equation of state}. At low enrichments levels ($\lesssim 10$ x solar), the opacity is already increased, but not yet significantly so the mean molecular weight. But if we go to more significant enrichment levels (several 10 x solar), the effect on the mean molecular weight becomes important as well. For example, if we approximate all the metals as water vapor, we  find a mean molecular weight of about 2.4, 2.6, and 5.3 for 1, 10, and 100 times solar. This clearly higher value for 100 x solar  lead{s} to a smaller planetary radius because of the increased density of the gas  \citep[e.g.,][]{baraffechabrier2008}. This effect counteract{s} the increase of the radius that {is} found when just increasing the opacity, but not self-consistently increasing the mean molecular weight also (which is what we {did in Simulation O1}, because the EOS of the gas is {still} the pure H/He EOS of \citealt{saumonchabrier1995} independently of $\kappa_{\rm R}$). 

Second, we have also not taken into account {in simulation O1} that the higher opacities are caused by higher quantities of heavy elements (i.e., the composition of the atmosphere), and composition influences the  {atmospheric}  evaporation rate, too \citep[e.g.,][]{johnstoneguedel2018}. Cooling via atomic lines of important metals like carbon or oxygen for example reduce{s} the temperature in the XUV heated gas which leads to lower evaporation rates \citep{owenmurray2018}. While the effect still needs to be {systematically} quantified in the context of strongly evaporating planetary atmospheres \citep{owenmurray2018}, it appears likely that this effect  also counteract{s} the stronger evaporation associated with higher opacities because of the aforementioned increase of the radius. 

Clearly, given these {opposing} effects, the role of the atmospheric composition {needs to be} treated {more} self-consistently  {than in simulation O1 by linking} the opacity, mean molecular weight (i.e., the EOS), the atmospheric and interior structure, and the efficiency of atmospheric escape {self-consistently. This is done in simulations Z1, Z2, and Z3, where the opacity, EOS, and evaporation rate were self-consistently coupled, as described in Sect. \ref{sect:simswithZ}.  The three simulations study an envelope heavy element mass fraction $Z_{\rm enve}$=0.1, 0.3, and 0.5, respectively.  }

{These simulations are also of interest because observationally, it is possible to study the valley position as a function of host star [Fe/H] \citep{owenmurray2018}.  As discussed below, the difficulty in connecting the simulations shown here with these constraints lies in the question whether (or to what extent) the atmospheric compositions of the low-mass planets near the valley are correlated with the host star [Fe/H].}
%leads to [M/H]=0.86, 1.43, and 1.78.

\begin{figure*}
\begin{center}
        \includegraphics[width=0.95\textwidth]{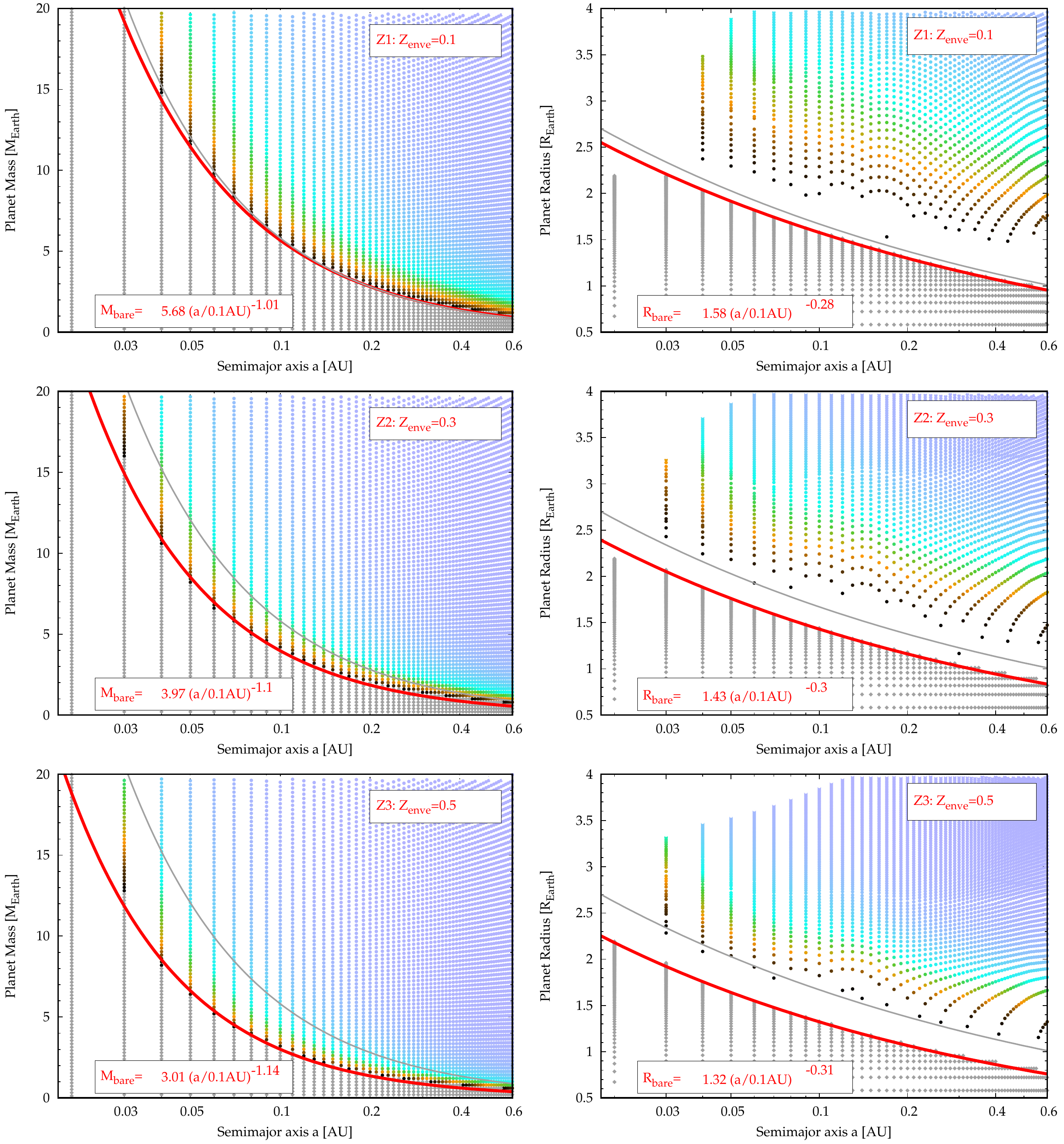}
\caption{{Impact of the envelope enrichment on the transition. The figure is analogous to Figure \ref{fig:MRbareS0} but for simulations Z1, Z2, and Z3 where envelope is enriched with H$_{2}$O with a mass fraction $Z_{\rm enve}$=0.1, 0.3 and 0.5, respectively.  Opacities and evaporation rates are modified accordingly, assuming a simple scaling law for the energy-limited evaporation rate $\propto Z_{\rm enve}^{-0.77}$. The transition mass and radius were again fitted with a power law for $\mbare$ and $\rbare$ as shown with the thick red line. The thinner gray line shows the transition mass and radius in the nominal simulation M0 to allow direct comparison. }}\label{fig:MRbareZenve}
\end{center}
\end{figure*}

{The resulting distance-mass and distance-radius plots are shown in Fig. \ref{fig:MRbareZenve}. We see that in contrast to simulation O1, where the enhanced opacity has led to a (slight) increase of $M_{\rm bare}$ and $R_{\rm bare}$, we now see that the higher $Z_{\rm enve}$, the lower $M_{\rm bare}$ and $R_{\rm bare}$. The valley thus shifts slowly downwards with increasing metallicity. The two effects by which a higher $Z_{\rm enve}$ reduces the evaporation rate (first by increasing the mean density of a planet at fixed core and envelope mass, and second by reducing the efficiency factor of evaporation) overwhelm thus the effect that the associated higher opacity leads via a reduce cooling to less dense planets and thus more evaporation. This downward shift was expected given that \citet{lopez2017} found that for pure water envelopes  ($Z_{\rm enve}$=1), the valley should be at about 1 $\rearth$ at an orbital distance of 0.1 AU. }

{One also sees in Fig. \ref{fig:MRbareZenve} that at $Z_{\rm enve}=0.1$, the innermost orbital distance where planets with an envelope survive is 0.04 AU. The same result holds for the nominal simulation M0. At higher  $Z_{\rm enve}$ in simulations Z2 and Z3, planets which have kept the envelope exist in contrast also  at 0.03 AU. This is not surprising: the same mechanisms which reduce the efficiency of evaporation at high $Z_{\rm enve}$ not only shift the valley downwards, but also push the limit of the sub-Neptune desert inwards to smaller orbital distances. If the envelope metallicities $Z_{\rm enve}$ of planets are indeed correlated with the stellar [Fe/H], this result is in agreement with the observational finding of \citet{owenmurray2018}  that  planets hosting H/He envelopes are more common around higher metallicity stars at small orbital distances.}

{\citet{owenmurray2018} also found that observationally, the locus of the valley is approximately independent of host star metallicity with a change of $\lesssim$15\% in radius over the observed range of metallicities. For a valley position in the nominal simulation at about 1.7 $\rearth$, 15 \% corresponds to a change of less than about 0.3 $\rearth$, bringing the valley on the lower limit down to about 1.4 $\rearth$. This value lies between the $R_{\rm bare}$ found in simulations Z2 and Z3 (1.43 and 1.32 $\rearth$). Again under the assumption of a correlation of host star metallicity and planetary $Z_{\rm enve}$, in order to be consistent with this limit, the envelopes in the valley region should not exhibit a systematic change in $Z_{\rm enve}$ with stellar [Fe/H] exceeding about 0.4. If we follow \citet{guptaschlichting2019} and assume that the stellar metallicities are directly representative of the metallicity of the planetary atmospheres, we typically expect a change in $Z_{\rm enve}$ because of the variation of stellar [Fe/H] (about -0.5 to 0.5 dex in the solar neighborhood) of $f_{\rm vol} Z_{\odot} 10^{-0.5}$ to $f_{\rm vol} Z_{\odot} 10^{0.5}$. Here, $f_{\rm ref} $ is the mass fraction of volatiles which are in the gas phase in the inner disk, and  $Z_{\odot}$ is the primordial solar heavy element mass fraction.  The value of $f_{\rm ref} $ is about 0.67 and $Z_{\odot}$=0.0149 \citep{lodders2003}.  Entering these number gives $Z_{\rm enve}$ ranging between 0.003 and 0.03. This is a very small range compared to the one needed to shift the valley significantly. Considering how $R_{\rm bare}$ varies across the simulations Z1 to Z3, variations between  0.003 and 0.03  shift the valley position only by about 0.02 $\rearth$. This is much less than the aforementioned observational limit of 0.3 $\rearth$, meaning that if the host star [Fe/H] indeed sets the planetary $Z_{\rm enve}$, the  weak dependency of  $R_{\rm bare}$ on $Z_{\rm enve}$ in the evaporation hypothesis for the valley is such that it is in agreement with the observed independency of the valley locus on host star metallicity.}

{An alternative explanation of the absence of a correlation of the valley position and the stellar [Fe/H] would be that there is no or only a  weak correlation of host star metallicity and planetary envelope/atmospheric metallicity for planets in the valley region. For (massive) giant planets where gas is the clearly dominant component  it seems likely that the disk gas metallicity, which should be correlated with the host star metallicity, is important for the atmospheric metal content \citep{mordasinivanboekel2016}. Observationally, \citet{wallackknutson2019} indeed find for giant planets a 1.9$\sigma$ trend of a correlation of planetary atmospheric metallicity  and stellar metallicity. Surprisingly, \citet{teskethorngren2019} find for the bulk composition in contrast no correlation between stellar [Fe/H] and planetary metallicity if one corrects for the effect that overall, planetary metallicity decreases with increasing planet mass. One should here keep in mind that there are limits to a situation where the atmospheric metallicity is correlated with host star [Fe/H] whereas the interior is not: at some point, such a configuration becomes unstable because density and pressure gradients would have opposite signs, which is  Raleigh-Taylor unstable \citep{thorngrenfortney2019}. }

{For the much lower mass planets near the valley, there are reasons to think why the envelope composition may not simply be set by the stellar metallicity: first, the pollution of the atmospheres by the accretion of planetesimals and pebbles  \citep[e.g.,][]{podolakpollack1988,mordasinialibert2006b,pinhasmadhusudhan2016,brouwersvazan2018,vallettahelled2019} is  proportionally more important as these planets contain in mass only a small gas fraction. The amount of metals contained in the accreted disk gas without pollution is likely low, roughly on a 0.3-3\% level, at least if the gas was accreted before the disk became chemically evolved \citep{guillothueso2006}. This means that the pollution by a relatively low amount of metals is sufficient to dominate the composition. When impactors starts to pollute the envelope is however rather intrinsic to a  planet (the envelope must be sufficiently massive) and not dependent on the host star [Fe/H]. Second, another form of  ``pollution'' is the outgassing of various chemical species from the solid core \citep[e.g.,][]{bowerkitzmann2019} which would again be a process intrinsic to a planet. For terrestrial planets, secondary outgassed atmospheres differ completely from the host star composition and in this sense the stellar [Fe/H]. The case of the planets near the valley with a few percents of their mass in H/He is obviously not as extreme as they still posses (parts of) their nebular atmosphere. But it could still be that the fact that these envelopes are over Gigayears in contact with a magna ocean influences the atmospheric composition more than the stellar [Fe/H].}

{The finding of \citet{owenmurray2018} that H/He atmosphere-hosting planets are more common around higher metallicity stars at short periods would then to be interpreted differently than that the high(er) atmospheric metallicity protects the atmosphere from photoevaporative loss. An alternative explanation could be the formation timescale: if planets around low [Fe/H] stars form later and therefore accrete less or even no H/He, this could potentially also explain this trend. }

{In the case that there is no correlation between stellar [Fe/H] and planetary atmospheric $Z_{\rm enve}$, the variation in $Z_{\rm enve}$ from planet to planet - if sufficiently large - would then simply induce a spread in the transition from solid to planets with H/He, as suggested by the results in simulations Z1-Z3. Observationally, this would make the valley less empty.  
}

{To finish this discussion it is important to stress  that we have adopted here a very simple and uncertain scaling of  the evaporation rate with $Z_{\rm enve}$. This should  be improved in future work. We have also used the same initial conditions for all $Z_{\rm enve}$.}  If possible it should {however} be taken into account that different compositions affect already a planet's formation, i.e., the initial conditions like for example the envelope mass fraction which depend on the envelope composition \citep{venturinialibert2016} {or the initial luminosity. }

\subsubsection{Rocky versus icy composition of the core: I1}\label{sect:corecomporockyvsicy}
The last simulation I1 is identical to M3 except for a mass fraction of (water) ice in the solid core that is equal to unity. In M3, an Earth-like composition without any ice was instead used. A core consisting exclusively of water ice like in I1 is certainly unlikely to occur in  nature (if ice can condensate in some part of the protoplanetary disk, then also more refractory element). However, it is interesting to compare the results with earlier studies \citep{owenwu2017,jinmordasini2018} and with the analytical model. 

\citet{owenwu2017,jinmordasini2018} had both independently found that the locus of the valley as observed by \citet{fultonpetigura2017} is consistent with a predominately Earth-like composition for the close-in low-mass planets probed by the valley, but not with a predominantly icy composition {(see also \citealt{owenadams2019} for the effects of magnetic fields)}. For an ice-rich composition, the valley would rather be located at about 2.3 $\rearth$ at 0.1 AU instead of the observed 1.6 $\rearth$ \citep{jinmordasini2018}. The reason for the larger $\rbare$ for icy cores is that first, an ice-rich composition reduces the mean density of the planet (when the planet still posses H/He), making planets of a higher core mass more vulnerable to escape. Second, once the envelope is lost, at fixed core mass, icy cores have a larger radius. 

A transition at 1.6 $\rearth$ as observed is instead consistent with an Earth-like iron-silicate composition. This indicates that these planets did not migrate to their current position from outside of the (water) iceline. As discussed extensively in \citet{jinmordasini2018}, this does, however, also not mean that these planets did not migrate at all, but that their migration was confined to inside of the water iceline. This is also what is predicted by planet formation simulations that couple planetary accretion, (full non-isothermal) type I (and type II) migration, and disk evolution \citep[e.g.,][]{mordasini2018}. In these simulations, only more massive Neptunian planets migrate from beyond the iceline close to the star, as the type I migration timescale decreases with increasing mass. But these more massive planets can keep their H/He, except if they are very close to the star, like 55 Cnc e \citep{jinmordasini2018}. 

It is also clear that the some individual planets can still have a high ice content as the observations only show the mean locus of the valley, but ice-rich should not be the dominant composition. It is interesting to note that thanks to evaporation and the resulting valley, it was possible for \citet{owenwu2017,jinmordasini2018} to infer Earth-like compositions based on the location of the valley alone, i.e., based on radii only. The density of the planets which is usually used to infer such informations is not needed (and also not available for most Kepler planets). 

As shown in \citet{jinmordasini2018}, an Earth-like composition is also consistent with the (measured) densities of the (few) planets in the ``triangle of evaporation'' with a relatively well-constrained density (i.e., planets for which both radius and mass were relatively precisely measured). The ``triangle of evaporation'' is the approximately triangular-shaped region (in a log-log plot) in the planet radius-orbital distance plane where planets completely lose the H/He, i.e., planets with $R<\rbare(a)$. For such planets without H/He envelopes, the degeneracy of the mass-radius relation \citep[][]{rogersseager2010} is strongly reduced, and the density can be used to constrain the ice content. Unfortunately, only very few planets are currently known that have a density that is observationally constrained well enough to really infer the ice mass fraction (see also \citealt{dornventurini2017}). 

To improve this  situation, several tens of planets with masses below 10 $\mearth$, radii below 4 $\rearth$, and inside of about 0.5 AU with very accurately measured masses and radii, for example from TESS \citep{Rickerwinn2014} and CHEOPS  \citep{fortierbeck2014} observations, would be very helpful.

Coming back to the two bottom panels of Figure \ref{fig:MRbareOpa},  we see that relative to the reference simulation M3, $\mbare$ increases from 5.52 to 8.57 $\mearth$ (factor 1.55 increase) for a fully icy composition, while $\rbare$ grows from 1.65 to 2.68 $\rearth$, i.e., by a factor 1.62. We thus see that especially for the radius, the ice mass fraction has a strong influence on the transition locus. It leads, with a large margin, to the strongest increase of $\rbare$ relative to the reference case of all simulations conducted in this work as can be seen from Table \ref{tab:sims}. The impact of an icy core is in particular also larger then the one of varying the evaporation rate by a factor of 10. 

The shift from about 1.6 to 2.7 $\rearth$ is also more than twice as large as the intrinsic width of valley of about 0.5 $\rearth$. This remains true also for a more realistic ice mass content of ice of 50 to 75\% \citep{jinmordasini2018}. This is very different from the results for the impact of various initial envelope masses and strengths of evaporation presented in the previous sections. This strong impact of the composition explains why the valley is such a good probe for composition of the cores, as first suggested by \citet{lopezfortney2013}.

While the (double) physical effect leading to a shift of $\rbare$ to higher radii for an icy core composition was qualitatively explained earlier in this section, it will become possible to more quantitatively assess the impact of the ice mass fraction in the context of the analytical model (Sect. \ref{sect:rbare}). There, it is found that the $\rbare$ radius depends linearly on the ice mass fraction, whereas it depends only weakly on other quantities, for example  only as $L_{\rm XUV}^{0.135}$ on the stellar XUV luminosity.

\section{Analytical study}\label{sect:analyticalstudy}
We now develop  a simple analytical criterion for the evaporative transition from close-in super-Earth to sub-Neptunes that can explain the main results found in the numerical simulations.  It explains in particular (i) the decrease of the transition mass from solid planets to planets with H/He $M_{\rm bare}$ with approximately $a^{-1}$ or equivalently $R_{\rm bare}$ with $a^{-0.27}$, (ii) the weak dependency on the efficiency of atmospheric escape \citep{jinmordasini2014}, and (iii) the independency of $R_{\rm bare}$ on the (power law) scaling of the initial envelope mass with core mass. The dependency of the transition on these quantities were previously investigated numerically in \citet{lopezfortney2013}.  %lopezfortney2013apj776

\subsection{Energy comparison principle}
The analytical model is based on the principle that the ability to keep an envelope is controlled by the comparison of the relevant energies: it compares the binding energy of the envelope in the potential of the core with the energy deposited by the time-integrated XUV flux received by the upper atmosphere due to stellar XUV irradiation {(the lifetime-integrated X-ray and extreme ultraviolet flux in the words of \citealt{mcdonaldkreidberg2019})}. Then, the mass of H/He envelope that is evaporated at some moment in time is given as the mass which has the same binding energy as the time integrated absorbed stellar XUV flux received  up to that time by the planet (modulo the efficiency factor). 

The same basic idea was first used by \citet{lecavelier2007} to study the evaporation of gaseous giant planets (for other energy-based studies for giant (i.e., H/He gas-dominated) planets, see also \citealt{ehrenreichdesert2011} and \citealt{jacksondavis2012}). In this work, we deal in contrast with low-mass, core-dominated planets. Therefore, we need to modify Lecavelier's approach in two crucial points: 

(i) instead of the total binding energy of the planet (including all mass), the relevant quantity is in our case the binding energy of the gaseous envelope in the gravitational potential of the core (the core itself does not evaporate, but is much more massive than the envelope and thus dominates the gravitational potential). 

(ii) the mass-radius relation is different from giant planets and more complex, as it depends on both the core and envelope mass. The radius of a low-mass planet depends in particular strongly on the envelope mass fraction \citep[e.g.,][]{lopezfortney2013b}. This is very different from giant planets were the radius is always around 1 Jovian radius and only weakly dependent on the (total) mass over a large mass range because of increasing degeneracy in the interior \citep[e.g.,][]{baruteaubai2016}. 

These two differences  lead to a more complex behavior than in the three aforementioned energy-based studies that addressed the ``evaporation desert'' important for (at least initially) gas-dominated planet, and not the evaporation valley of \citet{fultonpetigura2018,vaneylen2018} that is relevant for low-mass planets starting only with a low-mass H/He envelope.

\subsection{Derivation of $M_{\rm bare}$ and $R_{\rm bare}$}\label{sect:derivationmbare}
Based on this idea, we now derive the analytical expressions  for $M_{\rm bare}$ and $R_{\rm bare}$ which is the mass and radius of the most massive planet losing all its primordial H/He envelope at a given orbital distance. This corresponds to the bottom of the valley.  The binding energy $U_{\rm e}$ of the gaseous envelope with mass $M_{\rm e}$ in the gravitational potential of the core with mass $M_{\rm c}$ can be approximated as 
\beq\label{eq:ue}
U_{\rm e}=-\frac{k_{\rm pot} G M_{\rm c} M_{\rm e}}{R}
\eeq
where $G$ is the gravitational constant and $k_{\rm pot}$ a number  on the order of unity which depends on the density structure. Since we are dealing with low-mass, core-dominated planets where $M_{\rm e} \ll M_{\rm c}$, we neglect the self-gravity of the envelope, but we do take into account that the (outer) radius of the planet can be much larger than the core radius even for $M_{\rm e} \ll M_{\rm c}$. Therefore, we use in the equation the total planetary radius $R$ and not just the core radius $R_{\rm c}$.

The cumulative (time integrated) XUV energy input that drives evaporation - assuming energy-limited escape -  since the formation of the planet at $t_{\rm form}$ up to some time $t$ is given as 
\beq
V_{\rm XUV}=\int_{t_{\rm form}}^{t} \varepsilon \pi R^{2} \frac{L_{\rm XUV}}{4 \pi a^{2}} dt.
\eeq
In this equation, $\varepsilon$ is the fraction of the incoming stellar XUV flux that drives evaporation (the efficiency factor), $a$ the planetary semimajor axis, $L_{\rm XUV}$ the stellar XUV luminosity, and $R$ is again the planetary radius. For the analytical model, we assume that the radius where the XUV radiation is absorbed is the same as the radius in Eq. \ref{eq:ue}, neglecting a potential difference \citep{murray-claychiang2009}.  All these quantities are variable in time:  The stellar  $L_{\rm XUV}$ decreases in time \citep[e.g.,][]{ribasguinan2005,tujohnstone2015},  $a$ could potentially change due to tides or N-body interactions, and $R$ decreases due to the cooling and contraction of the planet, and because of the decrease of the envelope mass due to escape. For the analytical model, we retain only the temporal dependency of $L_{\rm XUV}$, and assume that $a$ is constant, while for $R$ we then need to specify a representative mean radius (during the relevant time interval) , as described below. For this temporal mean, we must keep in mind  that early after formation, the planetary radii are much larger \citep{mordasinialibert2012b}. 

Using the observational results for the EUV flux as a function of time of G and K stars \citep{ribasguinan2005}, we can evaluate the integral that now only runs over $L_{\rm XUV}$. These studies have shown that the flux decreases in time as a power law $L \propto t^{-\alpha}$, where $\alpha\sim1.23$ for G type stars and 1-1200 \AA, with a saturated (constant) EUV luminosity $L_{\rm sat}$ at ages lower than $t_{\rm sat}\sim 100$ Myr. Taking $\alpha=5/4$, the integral then yields for $t>t_{\rm sat}$ (and thus also $t\gg t_{\rm form}$)  
\beq
V_{\rm XUV}= \varepsilon \pi R^{2} \frac{L_{\rm sat} t_{\rm sat}}{4 \pi a^{2}} \left(5-4\left(\frac{t_{\rm sat}}{t}\right)^{1/4}\right)
\eeq
It is convenient to write this in the following form
\beq
V_{\rm XUV}= \frac{ \varepsilon \pi R^{2} } {4 \pi a^{2}}  E_{\rm XUV}
\eeq
with the energy of the star emitted in the XUV since the formation of the planet given as \citep[see also][]{jacksondavis2012}
\beq
E_{\rm XUV}=\int_{t_{\rm form}}^{t}  L_{\rm XUV} dt = L_{\rm sat} t_{\rm sat}\left(5-4\left(\frac{t_{\rm sat}}{t}\right)^{1/4}\right)
\eeq
At this point, we could in principle set the binding energy and the integrated energy flux equal and solve for $M_{\rm c}$. But this would not lead to a conclusive result, as $R$ is itself a function of $M_{\rm c}$ and $M_{\rm e}$.

\subsubsection{Envelope and core mass}\label{sect:enveandcoremass}
Therefore, we next need to establish a relation of the post-formation envelope mass and the core mass. If the Kelvin-Helmholtz cooling timescale of the gaseous envelope regulates the envelope accretion rate during the nebular phase, and if $\tau_{\rm KH}\propto M_{\rm c}^{q_{\rm KH}}$ as suggested by various past studied \citep[e.g.,][]{ikomanakazawa2000}, we expect that the final envelope masses will also follow a power law, with $M_{\rm e} \propto M_{\rm c}^{-q_{\rm KH}+1}$ (see Appendix A in \citealt{mordasiniklahr2014}). Indeed, population synthesis simulations of the formation of planets where the gas accretion rate is found by explicitly solving the planetary structure equations find roughly speaking such a power law dependency for low-mass, subcritical cores \citep[Sect. \ref{sect:initialconditions} and][]{mordasiniklahr2014}. Thus, we write
\beq\label{eq:memcscaling}
M_{\rm e}= M_{\rm e,1} \left(\frac{M_{\rm c}}{\mearth}\right)^{p_{\rm e}}.
\eeq  
In this equation, $M_{\rm e,1}$ is the primordial (i.e., post-formation) H/He envelope mass of a 1 $\mearth$ core. Formation models  \citep[e.g.,][]{ikomanakazawa2000,ikomahori2012} show that its value depends on the planet's orbital distance, the disk properties, and on the (grain) opacity in the protoplanetary atmosphere \citep{ormel2014,mordasini2014}. The latter can be represented by choosing different values of $M_{\rm e,1}$. As the reference value, we use $M_{\rm e,1}=M_{\rm e,1,r}=0.03 \mearth$, i.e., a three percent envelope mass. 

The power law exponent   $p_{\rm e}$ gives the increase of the envelope mass with increasing core mass, that can be related to the KH exponent, as seen above. From the numerical results of Sect. \ref{sect:initialconditions}, we use as the nominal value $p_{\rm e}=2$, but we also study $p_{\rm e}=0$, 1, and 3. 

The exponent $p_{\rm e}$ transports important informations on the formation process. From a formation point of view, it would be very interesting if we could obtain it from the evaporative imprints. As we suspect from the numerical results exhibiting a very weak dependency on the post-formation envelope mass, this is unfortunately not the case: as we will see, in the analytical model, $M_{\rm bare}$ is even completely independent of $p_{\rm e}$ (or the power law normalization constant). From an evolution point of view, this is in contrast positive, as it means that the initial conditions are not important. 

\subsubsection{Mass-radius relation}\label{sect:analyticmrrelation}
The final step before we can calculate $M_{\rm bare}$ is to specify the mass-radius relation. This relation is crucial to understand why $R_{\rm bare}$ is independent of  $p_{\rm e}$. We first treat the core radius and the envelope radius separately. For the core radius we use the result of \citet{valenciaoconnell2006} or \citet{lopezfortney2014} who found that in the Super-Earth mass domain, planets with an Earth-like rocky  composition (i.e., 1:2 iron-silicate by mass) closely follow a power law with
\beq\label{eq:rcoremcoreearthlike}
\frac{R_{\rm c,r}}{\rearth} =  \left( \frac{M_{\rm c}}{\mearth} \right)^{p_{\rm c}}.
\eeq
with $p_{\rm c}$ between 0.25 and 0.27. 

\begin{figure}[tb]
    \centering
    \includegraphics[width=0.49\textwidth]{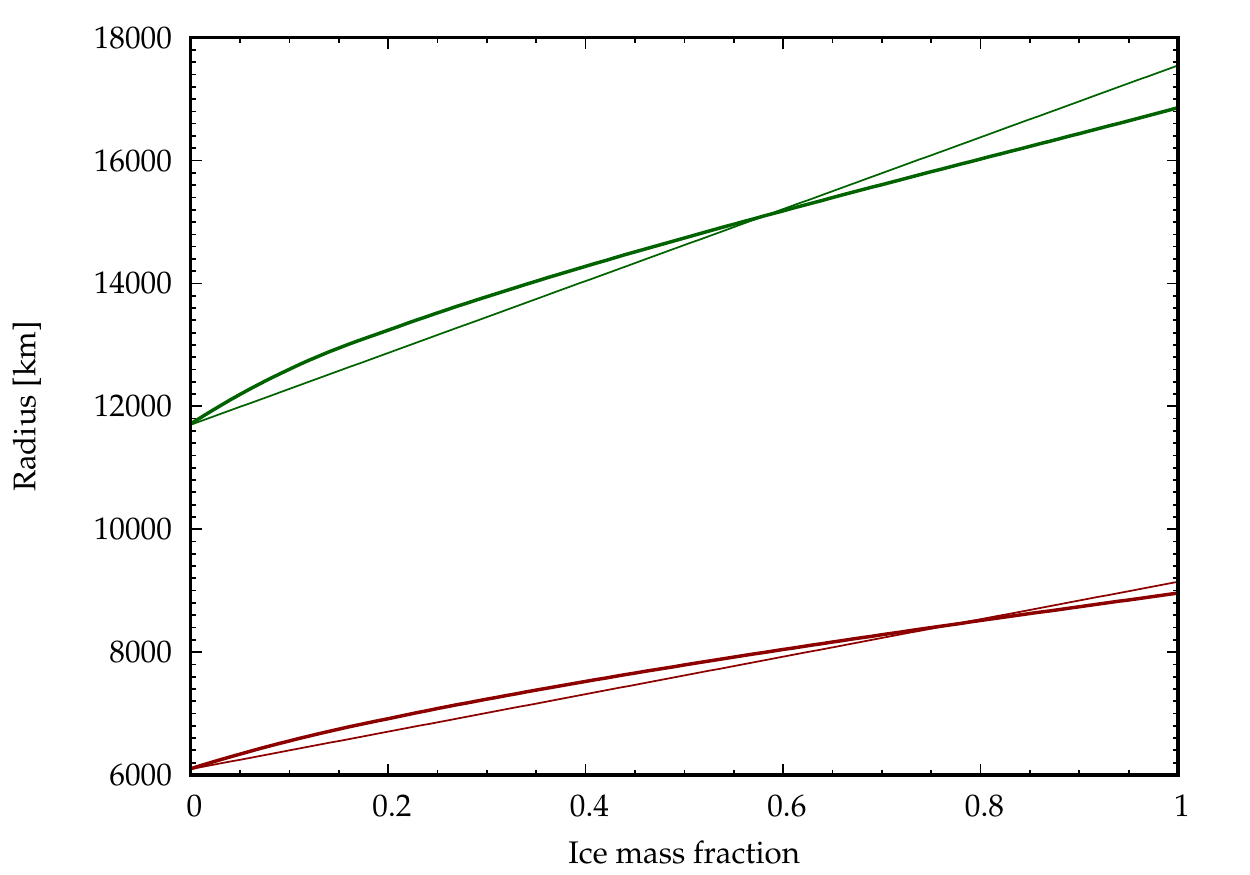}
   \caption{Radius of the solid core as a function of ice mass fraction (the rest has an Earth-like iron and silicate composition). The top and bottom  pairs of lines are for a mass of 10 and 1 $\mearth$, respectively. The thick lines show the radius as found from solving the interior structure equations, while the thin lines show the simple approximation from  Eq. \ref{eq:rcoreice} with the linear dependency.  }\label{SimpleIncreaseRadiusFice2}
\end{figure}

Below we will study how the location of the evaporation valley depends on the ice content of the  core. We thus also need to specify how the presence of (water) ice increases the core radius. Considering the $M-R$ relation of rocky and icy low-mass planets  (e.g., \citealt{mordasinialibert2012c}) as predicted from interior structure models suggest that to first order, we can write the following expression for the radius as a function the ice mass fraction $f_{\rm ice}$ {(see also \citealt{zengjacobsen2019} for a similar expression)} 
\beq\label{eq:rcoreice}
R_{\rm c}\approx R_{\rm c,r}\left(1+\frac{1}{2} f_{\rm ice}\right)
\eeq
for masses between about 1 and 10 $\mearth$. It is clear that this is not meant as an accurate expression for the impact of the ice mass fraction (see for example \citealt{fortneymarley2007}), but for our purpose it is sufficient. The increase of the core radius with increasing ice mass fraction as found from solving the interior structure equations is compared to this simple linear approximation in Fig. \ref{SimpleIncreaseRadiusFice2}. The differences to the numerically obtained radii are $\lesssim$ 4 \% in this mass interval. 

For the extent of the gaseous  envelope we use the results of \citet{lopezfortney2014} %ApJ vol. 792 pp. 1
 who derived a fit to the thickness of the H/He envelope $W_{\rm e}$ as found in their evolutionary calculations. Their fit gives $W_{\rm e}$ as a function of core mass, envelope mass, semimajor axis, and time, where the latter two dependency are weaker. As mentioned above, for the analytical model we do not take into account explicitly the temporal dependency of the radius, but rather evaluate their expression at $t=$100 Myr as the representative timescale on which most of the evaporation occurs (e.g., Fig. 4 in \citealt{jinmordasini2014}). Furthermore, we also neglect the weak dependency on the semimajor axis, but take again a representative distance, namely 0.1 AU. In this case, the envelope thickness $W_{\rm e}$ as a function of core and envelope mass given by \citet{lopezfortney2014} is
 \beq\label{eq:widthenve}
\frac{W_{\rm e}}{\rearth}=4.2 \left(\frac{M_{\rm c}+M_{\rm e}}{\mearth}\right)^{-0.21}\left(\frac{M_{\rm e}/(M_{\rm c}+M_{\rm e})}{0.05}\right)^{0.59} 
\eeq
We see that the radius decreases with increasing total mass (approximately equal to $M_{\rm c}$ for the planets studied here),  but increases with the envelope mass fraction. The total radius is then $R=R_{\rm c}+ W_{\rm e}$, neglecting the usually small contribution of the radiative atmosphere  \citep{lopezfortney2014}. 

\begin{figure*}[]
    \centering
    \includegraphics[width=0.45\textwidth]{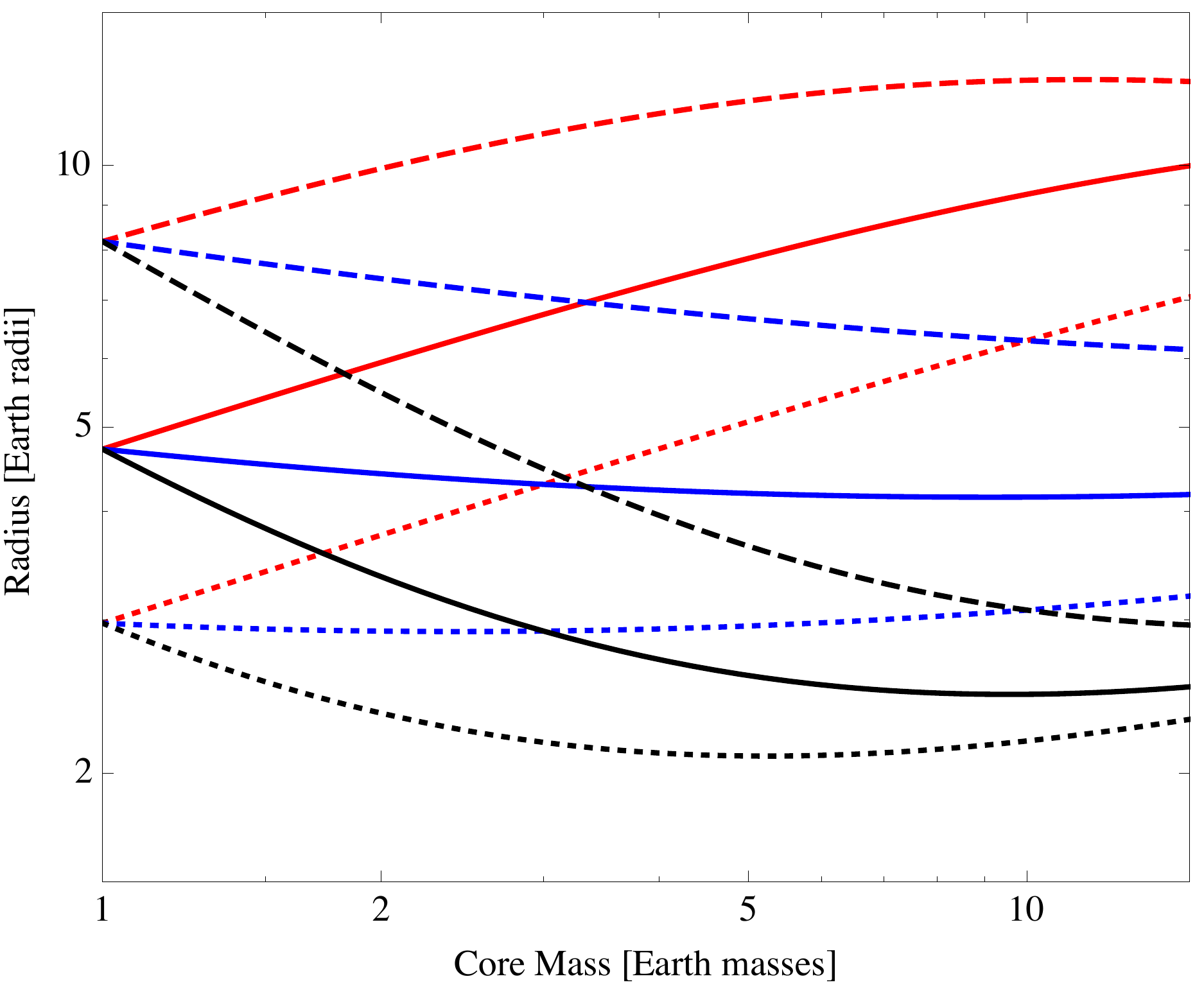} 
   %\hspace{0.5cm}%  \hfill
    \includegraphics[width=0.45\textwidth]{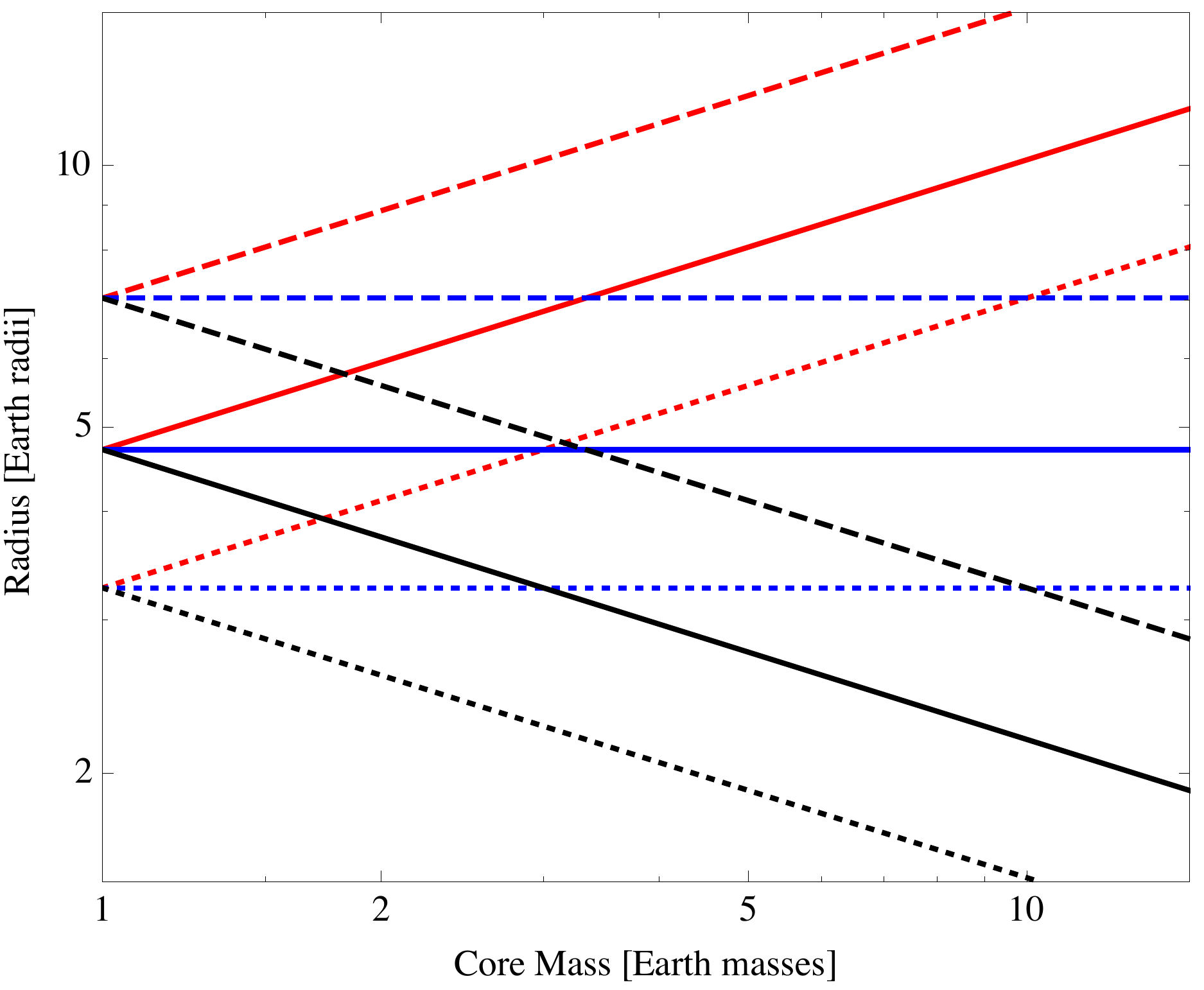}
   \caption{Radius as a function of core and envelope mass for planets at 100 Myr and 0.1 AU. The left panel shows the radius as found from the full expression (Eqs. \ref{eq:rcoremcoreearthlike}, \ref{eq:widthenve}) while the right one uses the simple approximation of Eq. \ref{eq:rapprox}. The dotted, solid, and dashed lines {show the radius for} envelope masses of a 1 $\mearth$ core of $M_{\rm e,1}$=0.01, 0.03, and 0.1 $\mearth$. Black, blue, and red lines show $p_{\rm e}$=0, 1, and 2, the power law exponent representing how H/He envelope mass $M_{\rm e}$ scales with core mass, $M_{\rm e}\propto M_{\rm e,1} M_{\rm c}^{p_{\rm e}}$.}\label{fig:mrfullandapprox}
\end{figure*}

In Figure \ref{fig:mrfullandapprox}, left panel, we plot the total radius $R$ as a function of core mass. The envelope mass is given according to Eq. \ref{eq:memcscaling} with $M_{\rm e,1}$=0.01 (less efficient gas accretion for example because of higher opacities), 0.03 ($=M_{\rm e,1,r}$, the nominal value), and 0.1 (efficient H/He accretion) and for power law exponents of the increase of the envelope mass with core mass of $p_{\rm e}=0$ (envelope mass independent of core mass), 1 (linear increase), and 2 (nominal case).   

Despite being analytical, the expression for the total radius as the sum of core radius plus  envelope thickness (Eqs. \ref{eq:rcoremcoreearthlike} and \ref{eq:widthenve}), is unhandy for the present application. Fortunately, the plot suggest that for a large part of the parameter space, one can write the total radius with a simpler functional form that contains in particular no more an explicit dependency on $M_{\rm e}$, but expresses it with $M_{\rm c}$ and $p_{\rm e}$. Examining various  expression eventually leads to the following form, shown in the right panel of Fig. \ref{fig:mrfullandapprox}:
\beq\label{eq:rapprox}
R \approx R_{\rm 1,r} \left(\frac{M_{\rm e,1}}{M_{\rm e,1,r}}\right)^{\frac{1}{3}}  \left(\frac{M_{\rm c}}{\mearth}\right)^{\frac{p_{\rm e}-1}{3}}
\eeq
In this expression, the normalization constant $R_{\rm 1,r}\approx4.6 \rearth$ is the total radius of a young 1 $\mearth$ planet at 0.1 AU with an envelope mass of $M_{\rm e,1,r}$. Note that such a planet loses its envelope on a timescale of less than 100 Myr when including evaporation. This is however not a problem, as the specific mass of 1 $\mearth$ is just the arbitrary normalization mass. Therefore, Eq. \ref{eq:rapprox} yields the relationship of core mass, envelope mass, and total radius for young low-mass planets as long as they still have an envelope, which is the quantity we need. The validity of the expression was checked by comparing it to the  radii  in the numerical simulations. A typical agreement on a $\sim20$\% level (or better) was found.

The second term in Eq. \ref{eq:rapprox} shows how the  radius depends on the envelope mass at a fixed mass of 1 $\mearth$. The third term finally shows the dependency on the core mass. From the left of Fig. \ref{fig:mrfullandapprox}, we  see that we can write the dependency approximately also as a power law. Crucially, the exponent is itself a function of  $p_{\rm e}$. One finds a dependency as $(p_{\rm e}-1)/3$. This exponent shows an interesting behavior: if the envelope mass is independent of core mass ($p_{\rm e}=0$), the total radius decreases with increasing core mass as approximately $M_{\rm c}^{-1/3}$. The increasing core gravity compresses the envelope, and no additional gas must be accommodated with increasing core masses. For a linear increase of the envelope mass with core mass ($p_{\rm e}=1$), the total radius is independent of core mass (see the approximately horizontal middle lines in the left panel of Fig. \ref{fig:mrfullandapprox}). Here the effect of the increasing gravity of the core leading to a smaller radius and the increase of the envelope mass and thus the necessary volume cancel each other. Finally, for a quadratically increasing envelope, the increase of the volume because of more gas wins over the increasing gravity of the core; the radius increases as $M_{\rm c}^{1/3}$.

We comment that it appears possible that the (approximative) core mass-envelope mass-radius relation of Eq. \ref{eq:rapprox} could be derived also analytically from (a simplified) solution of the structure equations. We have not tried to do this here.
If such an analytical solution exists, it would  make the model presented in this paper  fully analytical.

\subsubsection{Limits of the approximation}\label{sect:limitationsapprox}
The approximation of Eq. \ref{eq:rapprox} works well for core-envelope mass ratios suggested by the formation simulations (Sect. \ref{sect:initialconditions}), i.e., a few percent of envelope in mass for cores of a few Earth masses, increasing with core mass to $\sim$10 \% in mass for cores approaching 10 $\mearth$. As visible from comparing the left and right panel of Fig. \ref{fig:mrfullandapprox},  our approximation breaks down in the limit both of very low and high core-envelope mass ratios: 

In the former case, this is due to the fact that nothing prevents the approximation for the total radius from becoming smaller than the core radius, which is of course physically impossible. According to Eq. \ref{eq:rcoremcoreearthlike}, the core radius of a 10 $\mearth$ planet of terrestrial composition is for example about 1.9 $\rearth$, more than the value given by the black dotted line in the right panel of the figure. The approximation is thus not usable for cases where the core radius is not clearly smaller than the total radius, as the approximation would predict too small radii. 

On the other end of the spectrum of envelope masses, it is also not applicable once the self-gravity of the (massive) envelope starts to be significant, compressing the planet. Here the approximation predicts too large radii.  These two regimes manifested themselves in the largest differences between the full expression (left panel) and the approximation (right panel) occurring in the bottom and top right corners.

\subsection{Locus of the valley of evaporation in the distance-mass and distance-radius plane}\label{sect:finalresultanalytical}

\subsubsection{Maximum bare mass $M_{\rm bare}$}\label{sect:mbare}
With the expression for $M_{e}(M_{\rm c})$ and $R(M_{c},M_{\rm e})$ we can finally set $-U_{\rm e}=V_{\rm XUV}$ and solve for the largest core mass that loses it{s} entire envelope. We follow here \citet{lecavelier2007} in neglecting  the small contributions of the kinetic and thermal energy in the energy budget. We then have
\beq
\frac{k_{\rm pot} G M_{\rm c} M_{\rm e}}{R}=\frac{ \varepsilon  \pi R^{2}}{4 \pi a^{2}} E_{\rm XUV}. 
\eeq
Rearranging and inserting the equations for  $M_{\rm e}$ (Eq. \ref{eq:memcscaling}) and $R$ (Eq. \ref{eq:rapprox}) yields
\beq\label{eq:lhseqrhs}
k_{\rm pot} G M_{\rm c} M_{\rm e,1}\left(\frac{M_{\rm c}}{\mearth}\right)^{p_{\rm e}}=\frac{ \varepsilon E_{\rm XUV } }{4 a^{2}} R_{\rm 1,r}^{3} \left(\frac{M_{\rm e,1}}{M_{\rm e,1,r}}\right) \left(\frac{M_{\rm c}}{\mearth}\right)^{{p_{\rm e}-1}}
\eeq
Solving for the $M_{\rm c}$ which now becomes the maximal core mass that evaporates to a bare core  $M_{\rm bare}$, we obtain the final result
\beq
M_{\rm bare}=\frac{1}{2 a}\left(\frac{\varepsilon E_{\rm XUV} R_{\rm 1,r}^{3}}{k_{\rm pot} G}\right)^{\frac{1}{2}}\left(\frac{M_{\rm e,1,r}}{\mearth}\right)^{-\frac{1}{2}}.
\eeq
Together with the associated radius (Eq. \ref{eq:rbarelate}), this is the main result of the analytical part of this paper. 

Note that in the equation, $R_{\rm 1,r}$ and $M_{\rm e,1,r}$ are constants from the mass-radius relation equivalent to the statement that a 1 $\mearth$ planet has a 1 $\rearth$ radius for an Earth-like composition, and not variables. There is in particular no dependency on $p_{\rm e}$ and $M_{\rm e,1}$ (i.e., the initial envelope mass) as the term on the RHS and LHS in Eq. \ref{eq:lhseqrhs} have cancelled.  This independency is in excellent agreement with the numerical experiments presented above in Sect. \ref{sect:numresults} and explains why independently of the primordial H/He content, an always similar evaporation valley is found. It appears likely that it is actually because of this independency on the post-formation envelope mass that the valley is so clearly appearing observationally: it means that the (likely) significant spread in the post-formation envelope masses resulting from the formation epoch does not  matter. 

The reason behind this cancelation becomes clear with the analytical model: a more massive envelope (high $p_{\rm e}$ or $M_{\rm e,1,r}$) means more mass needs to be  evaporated (on the LHS in Eq. \ref{eq:lhseqrhs}), but it also means that the radius is larger (at nearly fixed total ($\approx$core) mass) and thus the evaporation stronger, which appears on the RHS. Because of the particular functional form of the mass-radius relation of low-mass planets with H/He, the two cancel.  We also see that $M_{\rm bare}$ scales as $1/a$, again as in the numerical simulations. From Kepler's law it follows that in terms of orbital period $P$, $M_{\rm bare}\propto P^{-2/3}$. 

It is important to note that there is only a square root dependency on $\varepsilon$ and $E_{\rm XUV}$. This explains now analytically  why, first, on the theoretical side various  models of evaporation (which can be see as different effective $\varepsilon$) found similar locations of the valley, i.e., why the details of these models do not matter so much. Second, on the observational side, the rather weak  dependency on $E_{\rm XUV}$  explains why we see a clear valley at all: the  large observed spread in the $L_{\rm XUV}$ of young stars does not so strongly alter the location of the valley around an individual star. This means that the valley is not blurred away when we consider the entire population of stars/planets. As we will in the following section, for $R_{\rm bare}$ the dependency is even much weaker than for $M_{\rm bare}$.
 
Inserting the parameters $M_{\rm e,1,r}=0.03 \mearth$ and  $R_{\rm 1,r}=4.6\rearth$, setting $k_{\rm pot}=1$,  and evaluating at 4.5 Gyr $\gg t_{\rm sat}$, we get a value of 
\begin{align}\label{eq:mbarelaterocky}
M_{\rm bare,late,rocky}&\approx6.5\left(\frac{a}{0.1 \, \mathrm{AU}}\right)^{-1}\left(\frac{\varepsilon}{0.1}\right)^{\frac{1}{2}} \nonumber \\ 
&\times \ \ 
\left(\frac{L_{\rm sat}}{10^{30} \,  \mathrm{erg/s}}\right)^{\frac{1}{2}} \left(\frac{t_{\rm sat}}{100 \,  \mathrm{Myr}}\right)^{\frac{1}{2}} \mearth.
\end{align}
This is also quantitatively in quite good agreement with the numerical results (Table \ref{tab:sims}).  The numerical  value of 6.5 $\mearth$ is for an Earth-like rocky core composition. For ice-rich compositions, if we assume that the increase in the core radius (Eq. \ref{eq:rcoreice}) translates into the same increase of $R_{\rm 1,r}$, then an ice mass fraction of for example 1/2 should lead to an increase of $M_{\rm bare,late}$ by a factor $5/4^{3/2}\approx1.4$ larger, i.e., 9.1 $\mearth$.  
\subsubsection{Maximum bare radius $R_{\rm bare}$}\label{sect:rbare}
With the relations of Eq. \ref{eq:rcoremcoreearthlike} and \ref{eq:rcoreice} we convert $M_{\rm bare}$ trivially into the radius of the largest planet that becomes a bare core, $R_{\rm bare}$. One finds
\begin{align}
R_{\rm bare}&=\left(1+\frac{1}{2}f_{\rm ice}\right)\left(\frac{1}{2 a}\right)^{p_{\rm c}}\left(\frac{\varepsilon E_{\rm XUV} R_{\rm 1,r}^{3}}{k_{\rm pot} G}\right)^{\frac{p_{\rm c}}{2}} \nonumber \\ 
 &\times   \ 
 \left(\frac{M_{\rm e,1,r}}{\mearth}\right)^{-\frac{p_{\rm c}}{2}}\left(\frac{1}{\mearth}\right)^{p_{\rm c}} \rearth
\end{align}
{The equation makes clear that the time-integrated X-ray and extreme ultraviolet luminosity of a star is key in determining the impact of atmospheric photoevaporation. This picture is supported by the finding of   \citet{mcdonaldkreidberg2019} that the (evaporation) desert exhibits much greater variability in the desert onset in the bolometric flux space compared to the integrated X-ray flux space.} 

Inserting the values of $M_{\rm e,1,r}$ and  $R_{\rm 1,r}$, setting $k_{\rm pot}$=1 and $p_{\rm c}$=0.27 and evaluating  $E_{\rm XUV}$ again at 4.5 Gyr yields the main result of the paper in terms of the location of the bottom of the evaporation valley as a function of the determining quantities
\begin{align}\label{eq:rbarelate}
R_{\rm bare,late}\approx 1.6 \left(1+\frac{1}{2}f_{\rm ice}\right) \left(\frac{a}{0.1 \, \mathrm{AU} }\right)^{-0.27}    \left(\frac{\varepsilon}{0.1}\right)^{0.135}   \nonumber \\   \times \ \ 
 \left(\frac{L_{\rm sat}}{10^{30}\, {\rm erg/s^{-1}}}\right)^{0.135} 
\left(\frac{t_{\rm sat}}{{\rm 100 \, Myr}}\right)^{0.135}   \rearth
\end{align}
As expected, in Table \ref{tab:sims} we again find good agreement with the numerically obtained results, in particular for the dependency on the orbital distance $\propto a^{-p_{\rm c}}=a^{-0.27}$  (compare with $e_{\rm r}$ in Table \ref{tab:sims}). With Kepler's law, this corresponds in terms of orbital period $P$ to $R_{\rm bare}\propto P^{-2 p_{\rm c}/3}\approx P^{-0.18}$, where the normalization distance of 0.1 AU corresponds for a 1 $\msun$ star to $P$=11.6 d. 

It is certainly remarkable that two fundamental properties of solid planets are imprinted into the valley's locus: first their mass-radius scaling in terms of $d \log R / d \log M = p_{\rm c}$ (Eq. \ref{eq:rcoremcoreearthlike}) which determines  the valley's distance dependency. Second, the ice mass fraction (or more generally, the bulk composition), which determines the absolute position at a fixed distance.  These dependencies can be compared with observational constraints (Sect. \ref{sect:compobs}, see also \citealt{guptaschlichting2018}). 

In this expression it becomes apparent that the transition radius has a very weak dependency on the strength of evaporation encapsulated in $\varepsilon$, $L_{\rm sat}$, and $t_{\rm sat}$, even more so than the transition mass $M_{\rm bare}$, which has important implications, as discussed in the last section and in  Sect. \ref{sect:conclusions}.

The strongest dependency is the linear dependency on the ice mass fraction $f_{\rm ice}$ (or the composition of the core in a more general sense). This is the reason why the valley of evaporation is such a good diagnostic of the core composition. For an ice mass fraction of 0.75 as used in \citet{jinmordasini2018}, Eq. \ref{eq:rbarelate} predicts that the transition radius at 0.1 AU increases from 1.6 to 2.2 $\rearth$.  This estimate neglects that $M_{\rm bare}$ is itself already larger as icy cores are more vulnerable to loss already when the envelope is still present, and not only that they are bigger once the envelope is lost. The numerically found value is thus somewhat larger (2.3 $\rearth$ for 75\% ice).  Above, in Sect.  \ref{sect:corecomporockyvsicy}  for 100\% ice, numerically a $\rbare=2.7 \rearth$ at 0.1 AU was found (Simulation I1), while  Eq. \ref{eq:rbarelate}  predicts about 2.4 $\rearth$.

\subsubsection{Incorporating a distance dependent efficiency factor}\label{sect:distancedepeff}
Up to now, we have assumed that the efficiency factor in the energy-limited escape formula is a constant. In reality, it depends on several factors \citep[e.g.,][]{kubyshkinafossati2018,wu2018}, like the composition and temperature in the atmosphere which in turn depends on stellar irradiation  and thus the orbital distance. Assuming that the latter dependency is the most important (and the one we can most easily study, given that Kepler planets forming the evaporation valley are found at a range of orbital distances), we may approximate the efficiency factor at least locally with a power law in orbital distance as 
\beq
\varepsilon=\varepsilon_{0} \left(\frac{a}{a_0}\right)^{q}.
\eeq
where $\varepsilon_{0} $ is the value of the efficiency factor at some normalization distance $a_{0}$ which we set to 0.1 AU.

Repeating the analysis as before for the constant $\varepsilon$, we now find a transition mass at late times of
\begin{align}
M_{\rm bare,late,rocky}&\approx6.5\left(\frac{a}{0.1 \, \mathrm{AU}}\right)^{-1+q/2}\left(\frac{\varepsilon_{0}}{0.1}\right)^{\frac{1}{2}} \nonumber \\ 
&\times \ \ 
\left(\frac{L_{\rm sat}}{10^{30} \,  \mathrm{erg/s}}\right)^{\frac{1}{2}} \left(\frac{t_{\rm sat}}{100 \,  \mathrm{Myr}}\right)^{\frac{1}{2}} \mearth.
\end{align}
The corresponding transition radius becomes 
\begin{align}
R_{\rm bare,late}\approx 1.6 \left(1+\frac{1}{2}f_{\rm ice}\right) \left(\frac{a}{0.1 \, \mathrm{AU} }\right)^{-0.27(1-q/2)}    \left(\frac{\varepsilon_{0}}{0.1}\right)^{0.135}   \nonumber \\   \times \ \ 
 \left(\frac{L_{\rm sat}}{10^{30}\, {\rm erg/s^{-1}}}\right)^{0.135} 
\left(\frac{t_{\rm sat}}{{\rm 100 \, Myr}}\right)^{0.135}   \rearth
\end{align}
i.e., $R_{\rm bare,late}\propto a^{-p_{\rm c}(1-q/2)}$. The idea behind this approach is the following: given the recent observational determination of the power law slope of the valley  \citep{vaneylen2018} as a function of orbital period which corresponds in the model to $-(2/3) p_{\rm c} (1 - q/2)$, we can  determine from the observations $q$ (assuming that $p_{\rm c}\approx 0.27$ is known) i.e., how the efficiency factor varies a function of orbital distance (or in a more general sense, depends on planet properties like the escape velocity, see \citealt{wu2018}). This is an important constraint for evaporation models.

The slope of the middle of the valley  varies as $a^{-0.15\pm 0.05}$ as found by \citet{vaneylen2018}, which translates to $q=0.89\pm0.26$. This means that taken at face value, we would find that the efficiency factor varies roughly speaking a bit weaker than linearly with orbital distance, which is caused by the more horizontal valley in the observations compared to the predictions for a constant $\varepsilon$. At small orbital distances, such a lower (effective) $\varepsilon$ could be caused by stronger radiative cooling, corresponding to a mass loss rate which is (partially) radiation-recombination-limited instead of energy-limited \citep{murray-claychiang2009}. As shown by \citet{owenwu2017}, simple energy-limited evaporation models with a fixed $\varepsilon$ indeed predict a steeper slope of the valley compared to more realistic models which directly calculate the evaporation rate from first principles \citep{owenjackson2012}.

At large distance, this inferred dependency  formally predict very high values for  $\varepsilon$. But here we  have to take into account  that at distances much outside of about 0.2 AU \citep{owenjackson2012}, the mass loss rate may in any case be much different (smaller) than assumed here: outside of such a distance, the evaporation should occur in the Jeans' regime, and no more in the hydrodynamic regime as assumed in the model, which would strongly reduce the escape rate. This should be tested in future with more detailed evaporation models \citep[e.g.,][]{kubyshkinafossati2018,kubyshkinafossati2018b}.

\subsubsection{Comparison with the numerical simulations}\label{sect:companalyticalnumerical}
Comparison of the locus of the valley as predicted by the analytical model with the numerically found values in Table \ref{tab:sims} shows that the analytical model can very well explain the main results found with the simulations. The analytical model explains in particular:
\begin{enumerate}
\item The dependency of the transition mass from solid to planets with H/He $M_{\rm bare}$ with orbital distance $a$ as $a^{-1}$ or equivalently $R_{\rm bare}$ with $a^{-0.27}$. This was numerically found to good approximation in all simulations except M1.
\item The (at first) surprising independency or very weak dependency of $R_{\rm bare}$ on the scaling of the envelope mass with core mass, i.e., the post-formation envelope mass (simulations M0-M4, N1 and N2).
\item The weak dependency on the strength of atmospheric escape encapsulated in the efficiency factor $\varepsilon$ and $L_{\rm XUV}$ (simulations E1 and E2).
\item The comparatively strong dependency on the ice mass fraction (simulation I1) or, more generally speaking, the composition of the solid core.
\end{enumerate}

The largest discrepancy regarding the radial slope of the valley occurs for simulation M1, where a clearly steeper slope was found numerically with $\rbare\propto a^{-0.36}$, whereas the analytical model always predicts $\rbare\propto a^{-0.27}$. In this simulation, $p_{\rm e}=0$ which means that the envelope mass is equal to 0.03 $\mearth$ independent of core mass. This leads to much lower core-envelope mass ratios for massive cores $\gtrsim 5-10 \mearth$ than in the other simulations. They are also much lower than the ones suggested by the formation model. The discrepancy can be explained when  we recall that the analytical model uses an approximation for $R(\mcore,\menve)$ (Eq. \ref{eq:rapprox}) which breaks down in the limit of both very low and high core-envelope mass ratios as discussed in Sect. \ref{sect:analyticmrrelation}. It is thus not surprising that the analytical model does not  well recover the numerical results in M1: for massive cores with very low mass envelopes, $R(\mcore,\menve)$ predicted by Eq. \ref{eq:rapprox} can become smaller than the core radius, which is obviously impossible.

Interestingly, however, it is possible to construct a more complex analytical model which uses the full expression for $R(\mcore,\menve)$ (Eqs. \ref{eq:rcoreice} and \ref{eq:widthenve}) instead of the approximation of Eq. \ref{eq:rapprox}. For this model, it is no more possible to analytically solve for $\mbare$ (and $\rbare$), but instead one has to find the root of a more complex version of Eq. \ref{eq:lhseqrhs} numerically. But with this more complex analytical model, one finds excellent agreement with the locus of the valley also for simulation M1. This is a strong indication that the principle underlying the analytical model, namely that envelopes are lost when the temporal integral of the stellar XUV irradiation absorbed by the planet is larger than binding energy of the envelope in the potential of the core, indeed captures  the governing physics.

\subsection{Comparison to Owen \& Wu (2017)}\label{sect:compOW17}

\begin{figure}[]
    \centering
    \includegraphics[width=0.47\textwidth]{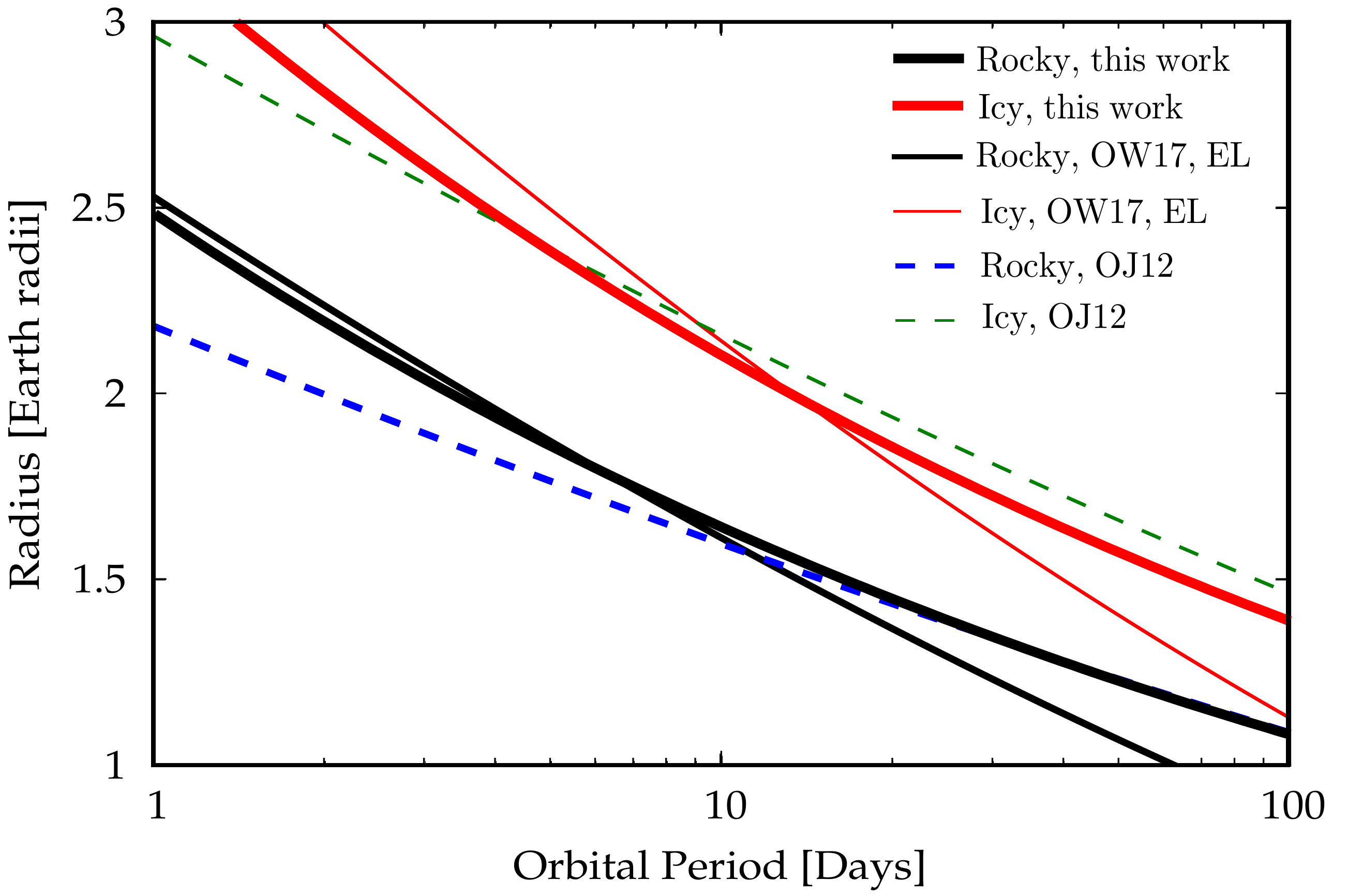}
   \caption{Comparison of the radius $R_{\rm bare}$, i.e., the radius the largest stripped core as a function of orbital period as predicted by the analytical model of this paper (thick lines) and the model of \citet{owenwu2017}. For the latter, the results for both the simpler energy-limited model with a constant efficiency factor and for the model of variable efficiency \citep[based on][]{owenjackson2012} are shown. The upper and lower three lines are for an Earth-like and ice-rich (1/3 in mass) core composition, respectively.  }\label{fig:compow17}
\end{figure}

The analytical model presented here differs in a number of aspects from the one presented by \citet{owenwu2017}. First, their model is fully analytical, while we use the fits to the numerically obtained envelope mass-core mass-radius relation from \citet{lopezfortney2013}. Second, we only consider the regime of more massive envelopes (high $X$ case in the terminology of \citealt{owenwu2017}), which is inspired by the post-formation envelope masses found in our core accretion formation simulations (Sect. \ref{sect:initialconditions}). Our mass-radius relation is not applicable for the case of massive cores with tiny envelopes: in this case, Eq. \ref{eq:rapprox} can predict total radii that are smaller than the core radius, which is of course not possible. Third, the \citet{owenwu2017} model is first timescale-based, and uses binding vs. irradiated XUV energies in the final step only. The approach here is directly based on comparing these energies. Finally, the full model of \citet{owenwu2017}  uses variable evaporation efficiencies \citep{owenjackson2012} while we normally consider a constant efficiency factor, as \citet{owenwu2017} also do in their simplified model. The case of an efficiency factor changing with orbital distance is addressed in Sect. \ref{sect:distancedepeff}.
In summary, the \citet{owenwu2017} model provides a more general and fully analytical view of the mechanisms controlling envelope loss. The model presented here on the other hand makes the physical mechanism determining the boundary between Super-Earth and Sub-Neptunes very directly comprehensible, and illustrates well the dependencies of important parameters like the initial envelope mass or the stellar XUV-luminosity. 

It is interesting to quantify the differences introduced by these point for the main result, which is the locus of the valley. \citet{owenwu2017} find a dependency of the radius of the largest stripped core $\rbare$ as a function of period $P$ of $\rbare \propto P^{-0.25}$ for the simple energy-limited model with a constant efficiency factor, and of $\propto P^{-0.16}$ in a model with a variable efficiency according to the \citet{owenjackson2012} evaporation models. The analytical model derived here leads for a constant $\varepsilon$ to a similar dependency with $\rbare \propto P^{-0.18}$, i.e., between the two exponents found by \citet{owenwu2017}.

Figure \ref{fig:compow17} compares the radius of the largest stripped planet $\rbare$ (i.e., the bottom of the valley) as a function of orbital period $P$ as found in \citet{owenwu2017} and  analytically in this work. Two compositions of the core are shown: an Earth-like composition with a 2:1 silicate:iron mass ratio, and a planet consisting of silicates and 1/3 ice. Thin lines show the results of \citet{owenwu2017}. Thick lines show the analytical results of the present paper using Eq. \ref{eq:rbarelate} with the nominal parameters, i.e., $\rbare/\rearth\propto 1.6 p^{-0.18}$ for the Earth-like composition, and $\propto 2.05 p^{-0.18}$ for the case with ice. The latter normalization radius of 2.05 $\rearth$ is obtained with the aforementioned equation with $f_{\rm ice}$=1/3, and taking into account that the assumption of a pure silicate composition (without iron - in contrast to the assumption underlying our Eq. \ref{eq:rcoremcoreearthlike})  leads to a further increase of the radius by about 10\% \citep{fortneymarley2007}.

Figure \ref{fig:compow17} shows that despite the aforementioned differences, the two models yield comparable results, regarding both the distance dependency of the valley and the impact of varying the core composition. We see that the differences between the analytical model presented here and the \citet{owenwu2017} results are comparable to the differences between the two approaches discussed by \citet{owenwu2017}, namely the simple energy-limited approach with a constant efficiency factor, and the more realistic model with a variable efficiency (which is based on \citealt[][]{owenjackson2012}). This leads to a shallower slope \citet{owenwu2017,wu2018}.

One also sees that at a period of around 10 days, which approximately corresponds to an orbital distance of about 0.1 AU around a solar-like stars, the results of the various models are very similar. This means that our results for the valley locus at the place where we normalize them (period of 10 days, resp. 0.1 AU), should not be strongly affected by our assumption of a constant efficiency factor. The dependency on orbital distance will in contrast be more affected.

\section{Comparison with observations}\label{sect:compobs}
\subsection{Comparison with Kepler data}
\begin{figure*}
\begin{minipage}{0.69\textwidth}
	      %\left
       \includegraphics[width=\textwidth]{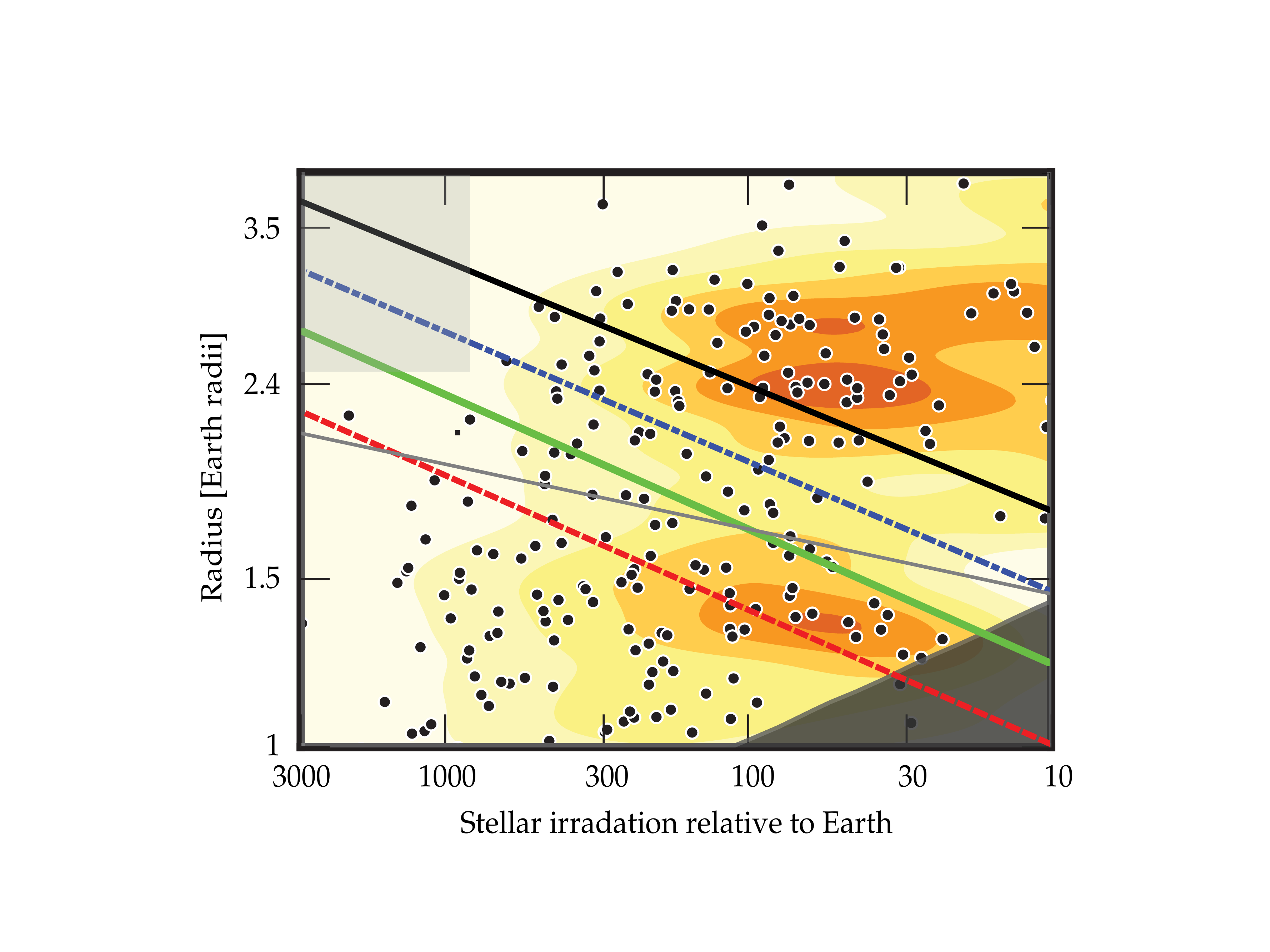}%
      \end{minipage}
     \begin{minipage}{0.30\textwidth}  
     \vfill
     \hfill
     \caption{Comparison of the locus of the valley as observed for Kepler planets around solar-like stars (points and yellow-brown occurrence  maps, \citealt{fultonpetigura2018}, with permission) and as theoretically found in the simulations in Table \ref{tab:sims} (lines). Note that all the theoretical lines show the \textit{bottom} of the valley, not its middle. The green solid line is the nominal simulation M0. The red dashed and blue dashed-dotted lines are E1 and E2 (strong and weak evaporation).  The black line is  I1 (ice cores). The thin gray line  assumes a distance-dependent  evaporation efficiency factor. The gray rectangle in the  top  left corner is the empty evaporation desert as found in the nominal simulation M0. The black triangle is observationally unconstrained.}\label{fig:compfp18} 
             	\end{minipage}
\end{figure*}

Figure \ref{fig:compfp18} shows the location of the valley as observed for Kepler planets around solar-like stars (0.96$\leq M/M_{\odot}\leq$ 1.11) from \citet{fultonpetigura2018} with points and yellow-brown completeness-corrected occurrence rate maps (dark colors indicate a high occurrence). This data is obtained with new Gaia parallaxes, Kepler photometry, and spectroscopically determined stellar temperatures from the California-Kepler Survey, resulting in a typical precision in the planetary radii of 5\%. {A similar study of \citet{martinezcunha2019} with a  median internal uncertainty of 3.7\% in the planetary radii finds results that are in overall agreement with   \citet{fultonpetigura2018}.}

The lines show the valley's location predicted by the simulations (using $R_{\rm b,0p1}$ and $e_{\rm r}$ from Table \ref{tab:sims}). The  theoretical lines thus show the \textit{bottom} of the valley, not its middle. The nominal simulation M0 shown by the green solid lines agrees quite well with the observed bottom of the valley, as noted previously \citep{owenwu2017,jinmordasini2018}. The red dashed and blue dashed-dotted lines show the  simulations E1 and E2 (strong and weak evaporation). They give an impression of how much the valley's location shifts if the evaporation rate is varied by a factor 10 and 0.1 relative to the nominal case, for example because of the observed spread in stellar $L_{\rm XUV}$. As discussed in Sect. \ref{sect:evaprateE1E2}, the valley is found to shift by a similar extent as its intrinsic width (about 0.5 $\rearth$), which explain why the valley is visible, but still not completely empty.

The black line is simulation I1 (ice cores) which is ruled out by the observations, again as already noted in \citet{owenwu2017,jinmordasini2018}. The thin gray line finally assumes the linearly distance-dependent evaporation efficiency factor derived in Sect. \ref{sect:distancedepeff} based on the slope of the valley determined observationally with astroseismology by \citet{vaneylen2018}. Comparing (by eye) with the radial dependency of the observed super-Earth over-density, it seems to provide a better reproduction of the observed slope also in this data set, but a quantitative statistical assessment using the exact occurrent maps would be required to really address this. Observations of the frequency of planets in the black triangle which is currently observationally unconstrained would obviously also be helpful for a better determination of the slope. Regarding the morphology of the valley at such larger distances, we recall that our theoretical model does not include the effect of the transition from hydrodynamic escape to the much slower Jeans escape. This could effectively put an end to the valley at around a stellar irradiation relative to Earth $F_{\oplus}$ of roughly  25 \citep{owenjackson2012} because the planets could keep the H/He. This {o}f course only holds if also at such larger distances, most planets grow to their final mass during the presence of the gas disk. This becomes however increasingly unlikely, as the formation timescale via collisional growth increases with orbital distance \citep[e.g.,][]{mordasinialibert2009a}, as exemplified by Earth's last giant (moon-forming) impact at an age clearly higher than protoplanetary disk lifetimes \citep{jacobsonmorbidelli2014,haischlada2001}. As discussed in detail in \citet{owenmurray2018}, it could therefore be that at such distance, terrestrial planets could become dominant that formed after disk dispersal, with a limiting radius that increases with distance \citep{lopezrice2018}. Simulations of combined planetary growth and self-consistently coupled evolution which includes the accretion of solids and of H/He, and the loss thereof via atmospheric escape and impacts along the lines of the population syntheses by \citet{mordasini2018} should help to clarify this picture.

Coming back to Figure \ref{fig:compfp18}, the gray squared region in the  top  left corner is the empty ``evaporation desert'' as found in the nominal simulation M0. It is another characteristic imprint of atmospheric escape \citep[e.g.,][]{kurokawanakamoto2014}. While there are smaller planets observed at such very close-in orbits, it is empty also in the observational data \citep[cf.][]{lundkvistkjeldsen2016,mazehholczer2016}. It is certainly a strong point of the evaporation hypothesis for the valley that evaporation naturally explains two a priori unrelated observed features, the  valley \textit{and} the desert.

\subsection{Kepler-36}\label{sect:kepler36}
Planets in multiple systems with dissimilar densities like the planets around Kepler-36 \citep{carteragol2012},  K2-106 \citep{guentherbarragan2017} or HD3167 \citep{gandolfiarragan2017} are of particular interest as these planets  have evolved in the same stellar XUV environment, allowing to study the differential history of the planets under the action of evaporation with one variable less {\citep{owencampos2019}}. A prime example in this context are the two planets around Kepler-36. They are characterized \citep{carteragol2012} by an extreme density difference (about 6.8 and 0.8 g/cm$^{3}$ for planets b and c, respectively) despite the very similar orbital distances (0.115 and 0.128 AU). But planet c is with 7.7 $\mearth$ clearly more massive than b (4.3 $\mearth$). 

As shown in \citet{lopezfortney2013} (see also \citealt{owenmorton2015}), %776 pp. 2
this mass difference means that planet b with its weaker gravity cannot keep its H/He envelope, while c can, meaning that models of atmospheric escape can successfully explain this surprising system. This is also the case with the calculations presented here: The analytical model predicts a transition mass $\mbare=5.65 \mearth$ at b's position, and $\mbare=5.08 \mearth$ at c's position (Eq. \ref{eq:mbarelaterocky}), i.e., between the masses of b and c. So planet b with a mass inferior to the local $\mbare$ should lose its envelope, whereas c should keep it - exactly as observed. In this sense, the simple analytical model passes the Kepler-36 acid test very well, without any model tuning. Kepler-36 is thus a very interesting system that is at the border of the triangle of evaporation, and it should serve as a test case for any model trying to explain the observed properties of close-in low mass planets (see for example \citet{liuhori2015} for a scenario of devolatilized by giant impacts).
 
Comparing with the numerically obtained results (Table \ref{tab:sims}), we see that the minimum core masses $\mbare$ for the  planets to retain their envelopes is also here between 5 and 6 $\mearth$ (i.e., between the masses of b and c) for most simulations, including the nominal one, again in agreement with observations. The same result  was found already also by \citet{owenwu2013,owenmorton2015}.

In the context of these numerical models it is possible to go a bit further with the analysis of  Kepler-36, as simulations M1, E1, E2, and I1 predict - taken at face value - in contrast an incompatible $\mbare$. From this, we can infer the following: 

First, from the incompatibility with simulation M1 (low envelope masses) we can deduce that the planet c initially had an envelope more massive than 0.03 $\mearth$ (corresponding to an envelope mass fraction of about 0.4\%). This is consistent with the detailed analysis of  \citet{lopezfortney2013} who demonstrated that both planets could have been born with H/He envelopes of approximately 22\% of the planets' initial mass. Similarly, \citet{owenmorton2015} also find a higher initial envelope mass fraction of about 15-30\% to explain the observations. For comparison, our formation simulations predict via Eq. \ref{eq:fitme0fromsynt} an initial envelope mass fraction of about 7\%, also much more than the 0.4\% that is ruled out from simulation M1.

Second, from the incompatibility of the simulations E1 and E2 we can deduce that evaporation in the Kepler-36 system was not as strong as in E2 (where also planet c would lose its H/He), but also not as weak as in E1 (where also planet b would keep it), but within one order of magnitude of the nominal evaporation model employed here. 

Third and finally, the simulation I1 (icy cores) also produces an incompatible $\mbare$. While it is not possible to constrain well the ice mass fraction of planet c because of the degeneracy with the H/He envelope mass fraction, the fact that the high density of planet b is consistent with a rocky Earth-like composition, with about 30\% of its mass in iron \citep{carteragol2012}, indicates that at least this planet indeed does not have an incompatible, ice-dominated composition.

Clearly, this simple analysis assumes that the other quantities and settings influencing the evaporative transition were at similar value as assumed in model. In reality degeneracies are possible, and one could construct models with modified evaporation rates and initial envelope masses that also lead to a $\mbare$ consistent with observations. But the analysis nevertheless shows than one can derive interesting (joint) constraints like the post-formation envelope mass or the strength of evaporation from such multi-planetary systems, making them very {\citep{owencampos2019}}. In the ideal case, studies investigate them should not only be limited to the interior structure aspect, but also take into account the global system architecture, i.e., the dynamical state in terms of eccentricities and mean motion resonances {\citep{poonnelson2020}}. These additional aspects may differ depending on the proposed mechanism proposed to explain a specific system.

\section{Summary}\label{sect:summary}
Precise observations of the radii of close-in extrasolar planets have revealed a depleted region at about 1.7 (1.5-2.0) $\rearth$ for solar-like stars separating smaller super-Earths from larger sub-Neptunes \citep{fultonpetigura2017,vaneylen2018,fultonpetigura2018}. This depletion can be explained as an evaporation valley between planets with and without H/He that is caused by atmospheric escape. This was predicted independently in a  similar fashion by several theoretical models \citep{owenwu2013,lopezfortney2013,jinmordasini2014,chenrogers2016}. The subsequently observationally found locus of the valley agrees with these theoretical predictions and is consistent with a mainly rocky, Earth-like composition without much ice as  shown by \citet{owenwu2017,jinmordasini2018}.

Building upon these works, in this paper we have conducted a systematic numerical and analytical study to further constrain the valley's location, and to understand how it depends on important planetary properties like the orbital distance, the planet's mass {and composition}, the H/He envelope mass, as well as the stellar XUV luminosity. An important goal of this evolutionary study is to derive constraints for planet formation models {\citep[e.g.,][]{carreraford2018,mordasini2018,brueggeralibert2018,poonnelson2020}}. Regarding formation models{, o}ne could {for example} a priori  think that the locus of the valley depends on the post-formation envelope-to-core mass ratio and its dependency on orbital distance. This would constrain the efficiency of gas accretion during the nebular phase \citep[e.g.,][]{ikomahori2012,leechiang2015,cimermankuiper2017} and the opacity in these protoplanetary envelopes \citep{ormel2014,mordasini2014} (which is not the case, as {discussed} below). 

In the first part of the paper (Sect. \ref{sect:model}) we have conducted systematic numerical simulations of the evolution of thousands of close-in low-mass planets starting with H/He that undergo atmospheric escape using the Bern planet evolution code \texttt{completo21} \citep[][]{mordasinialibert2012b,jinmordasini2014,lindermordasini2018}. To understand the behavior in the relevant parameter space, we have simulated the evolution of model planets on grids of core mass and orbital separation. We have calculated 1{4} grids, varying the post-formation H/He envelope mass, the strength of evaporation ({i.e.} different stellar $L_{\rm XUV}$, durations of the saturated phase $t_{\rm sat}$ or efficiency factors $\varepsilon$ of energy-limited {evaporation}), the opacity in the planetary atmosphere, {the envelope metallicity}, and the core composition. 

The results of the numerical study (Table \ref{tab:sims}) can be summarized as follows:
\begin{enumerate}
\item \textit{Position of the valley.} The position of the bottom of the valley in the $a-R$ diagram which corresponds to the largest core at a given orbital distance (or period) that has completely lost its H/He can in all grids be well approximated by a power law of the form $R_{\rm bare}(a)/\rearth= R_{b,0p1} (a/0.1{\rm AU})^{-e_{\rm r}}$ (Sect. \ref{sect:numresults}). In the reference simulation M3, $R_{b,0p1}$=1.65 $\rearth$ and $e_{\rm r}$=0.27. Except for one case (M1)  all other simulations exhibit similar exponents $e_{\rm r}$. If the dependency on orbital period $P$ instead of semimajor axis is considered, this corresponds with Kepler's law to $R_{\rm bare}\propto P^{0.18}$. 
\item \textit{Dependency on the post-formation envelope mass.} The locus of the valley is only a  weak function of the post-formation H/He envelope mass, except for envelope masses that are much lower or higher than what is predicted by formation models (Sects. \ref{sect:M1M4} and \ref{sect:N1N2}). We can write the post-formation envelope mass as $M_{\rm e,0}\propto M_{\rm e,1} M_{\rm c}^{p_{\rm e}}$ as suggested by the Kelvin-Helmholtz timescale of envelope cooling during the nebular phase \citep[Sect. \ref{sect:enveandcoremass},][]{mordasiniklahr2014}. We have then varied the normalization constant  $M_{\rm e,1}$ by one order of magnitude (Simulations N1, N2), and explored  exponents $p_{\rm e}$=1,2,3 (Simulations M2, M3, M4). While we find a slight decrease of $R_{b,0p1}$ with increasing envelope mass, $R_{b,0p1}$ remained in all these simulations  close to the nominal value, namely between 1.60 and 1.68 $\rearth$.  The reason is that at a given core mass, a higher envelope mass means that there is more {gas} to evaporate, but also that the planet has a larger radius. Since the (total) mass is for these planets essentially given by the (constant) core mass, this means that the planet has a lower mean density. A lower mean density implies a higher evaporation rate, meaning that the planet with more H/He also loses more. 
\item \textit{Dependency on the strength of evaporation.} In the Simulations E1 and E2 (Sect. \ref{sect:evaprateE1E2}) we have {varied} the evaporation rate relative to the nominal model by a factor 10. This change in the evaporation rate could be due to a variations of the stellar $L_{\rm XUV}$, the duration of the saturated phase, or uncertainties in the evaporation model, like for example variations of the efficiency factor 
 $\varepsilon$ in the energy limited approximation \citep[e.g.,][]{kubyshkinafossati2018,kubyshkinafossati2018b}. Increasing (decreasing) the escape rate leads as expected to a shift of the valley to larger (smaller) radii, but the shift is moderate, namely to $R_{b,0p1}$=1.97 and 1.39 $\rearth$, respectively.  This corresponds to an increase of the bare core radius  by only a factor 1.42 for a factor 100 increase of the evaporation rate. This is to be compared with the observed spread  in the $L_{\rm XUV}$ levels of young solar-like stars (e.g., \citealt{tujohnstone2015}) which exhibit a spread of  about a factor $\sim$30 in the first $\sim$100 Myr. Most stars should however  pile up around a mean value of  $L_{\rm XUV}\approx 10^{30}$ erg/s during the first $\sim$100 Myr  \citep{johnstoneguedel2015}.
\item \textit{Dependency on the opacity.} We usually assumed that the opacity in the planetary atmospheres corresponds to a condensate-free solar-composition gas \citep{freedmanlustig-yaeger2014}. However, planet formation models suggest an increasing enrichment with decreasing planet mass \citep{fortneymordasini2013}. In Simulation O1 we have therefore studied how the valley locus changes if the atmospheric opacities are uniformly increased by a factor 10. A higher opacity means that the planets contract less, increasing via the reduced density mass loss. Therefore, $R_{b,0p1}$ increases to 1.78 $\rearth$.
\item {\textit{Dependency on the envelope composition} In Simulation O1 two important aspects were neglected: a higher metallicity atmosphere does not only imply a higher opacity, but also more coolants, which reduces the evaporation rate \citep[e.g.,][]{salzczesla2016}. It also implies a higher mean molecular weight. This leads to a higher planetary density, which reduces evaporation as well.  To have a more self-consistent picture than in Simulation O1, we have also conducted three simulations Z1, Z2, Z3 where the opacity, equation of state of the envelope gas, and evaporation rate were self-consistently changed. A wide range of envelope metal mass fractions $Z_{\rm enve}$=0.1, 0.3, and 0.5 were simulated.  It is found that the evaporation valley shifts gradually downwards with increasing $Z_{\rm enve}$ to $R_{b,0p1}$=1.58, 1.43, and 1.32  $\rearth$, respectively. The effects of the EOS and the reduced evaporation rate thus overwhelm the effect of the opacity. Whether this effect leads to an observable dependency of the valley position as a function of host star [Fe/H] depends on the unknown relation between  atmospheric composition of  planets near the valley and  host star [Fe/H]. Under the (simplistic) assumption that the planetary atmospheric $Z_{\rm enve}$ is directly given by the stellar $Z$, a shift of the valley by just 0.02 $\rearth$ would occur, as $Z_{\rm enve}$ would in this case only cover a small range from about 0.3 to 3\%. Observationally, the valley position does not change by $\gtrsim$15 \% in radius (about 0.3 $\rearth$) over a wide range of host star metallicity \citep{owenmurray2018}. Our results show that for the photoevaporative hypothesis of the valley, the planetary $Z_{\rm enve}$ could thus change even much more strongly with stellar [Fe/H] and still remain consistent with the observed absence of a (clear) correlation of  gap position and host star [Fe/H].   } 
\item \textit{Dependency on the core composition.} Finally, we have again studied {\citep{owenwu2017,jinmordasini2018,guptaschlichting2018}} how the valley's location  changes if we assume a (hypothetical) completely icy core composition. In this case, the bottom of the valley shifts to $R_{b,0p1}$=2.68 $\rearth$, which is the largest difference relative to the nominal case of all simulations conducted here. This illustrates again that the valley position is a good diagnostic of the core composition \citep{lopezfortney2013}. As  discussed in Sect. \ref{sect:evaprateE1E2} and recapitulated below, also the fact that the valley is not completely empty, but to some extend filled (as observed, \citealt{fultonpetigura2018}) does not  mean that ice-rich compositions are {necessarily} needed to explain the observations; a non-empty gap is actually even expected also for a homogeneous (rocky) composition. 
\end{enumerate}

In the second part of the paper (Sect. \ref{sect:analyticalstudy}) we have developed a simple analytical theory (see also \citealt{owenwu2017}) based on the comparison of the relevant energies, in the same spirit as originally done by \citet{lecavelier2007} for giant planets. In this analytical model complete evaporation of the envelope occurs if the temporal integral over the stellar XUV irradiation absorbed by a planet (modulo the efficiency factor $\varepsilon$) is larger than the binding energy of the envelope in the gravitational potential of the core, given the special core mass - envelope mass - radius relation of close-in low-mass planets with H/He \citep{lopezfortney2014}.

This approach quickly leads to the main result of the analytical calculation which is the radius of the largest completely stripped core which corresponds to the \textit{bottom} of the evaporation valley. As a function of orbital period $P$, it is given as (Sect. \ref{sect:rbare})
\begin{align}\label{eq:rbarelateperiod}
R_{\rm bare,late}\approx 1.64 \left(1+\frac{1}{2}f_{\rm ice}\right) \left(\frac{P}{10 \, \mathrm{day} }\right)^{-0.18}    \left(\frac{\varepsilon}{0.1}\right)^{0.135}   \nonumber \\   \times \ \ 
 \left(\frac{L_{\rm sat}}{10^{30}\, {\rm erg/s^{-1}}}\right)^{0.135} 
\left(\frac{t_{\rm sat}}{{\rm 100 \, Myr}}\right)^{0.135}   \rearth.
\end{align}
for planets around a solar-like star and assuming a constant efficiency factor of evaporation. The \textit{middle} of the valley lies about 0.3 $\rearth$ above this value. 

The numerical results (1), (2), (3), and (5) listed above are recovered and understood very well with this expression: 

\begin{itemize}
\item \textit{Position of the valley.} The analytical model shows that the exponent of the period $P$ dependency, $P^{-0.18}$ is $P^{-2 p_{\rm c}/3}$ where $p_{c}$$\approx$0.27 is the power law exponent in the mass-radius relation of the solid cores, $R_{\rm c} \propto M_{\rm c}^{p_{c}}$ (Eq. \ref{eq:rcoremcoreearthlike}). It is certainly remarkable that two fundamental properties of solid planets are imprinted into the valley's locus {(see also \citealt{owenwu2017,guptaschlichting2018})}: first, their mass-radius scaling in terms of $d \log R / d \log M = p_{\rm c}$ that is predicted by interior structure model of solid planets \citep[e.g.,][]{valenciaoconnell2006}. It determines  the valley's distance dependency. Second, the {cores'} ice mass fraction $f_{\rm ice}$, or more generally speaking, the bulk composition, which determines the absolute position of the valley at a fixed period.

It is likely that in reality, the evaporation efficiency factor $\varepsilon$ is not constant in contrast to what was assumed here in the nominal model. A variable $\varepsilon$ leads to a shallower slope \citep{owenwu2017,wu2018}, closer to the observational result (Sect. \ref{sect:compobs}). In reality, the situation might in any case be more complex, because of the following: XUV-driven atmospheric {photoevaporation} might not be the only mechanism driving the loss of H/He envelopes. Other mechanisms could be impacts \citep{liuhori2015,SchlichtingSari2015,wyattkral2019} or  mass loss powered by the luminosity of a planet's cooling core \citep{ginzburgschlichting2018,guptaschlichting2018,guptaschlichting2019}. Additionally, some planets especially at larger distances may only {formed} after the dispersal of the gas disk, more similar to the way the Earth likely formed {\citep{owenmurray2018,swainestrela2019}}. \citet{lopezrice2018} found that a such a formation in a gas-free environment should lead to a transition radius from solid super-Earth to sub-Neptunes  that increases with orbital distance, i.e., with the opposite dependency  than evaporation. If these process are at work in parallel with a certain contribution of each, this could lead to a different (in particular shallower)  slope that could not be explained with one mechanism only.

\item \textit{Dependency on the envelope mass.} As shown by the equation, in the analytical model the locus of the valley is not just weakly dependent on the post-formation envelope mass (as found numerically), but even independent of it. As becomes clear when deriving an approximative core mass - envelope mass - radius relation for low-mass planets with H/He (Sect. \ref{sect:analyticmrrelation}), this relation is such that the two aforementioned effects occurring when increasing the envelope mass (more mass to evaporate, but also a higher escape rate) balance each other. Note that these results are only valid in the (important) domain where the envelope is massive enough that the total planet radius  is clearly larger than the core radius, but not so massive that the envelope's self-gravity becomes relevant (Sect. \ref{sect:limitationsapprox}).

\item \textit{Dependency on the strength of evaporation.} The analytical model shows that  quantities describing the strength of evaporation ($L_{\rm sat}$,  $t_{\rm sat}$ and $\varepsilon$) enter the expression for $R_{\rm bare,late}$ only with a  very {weak} exponent, 0.135=$p_{\rm c}/2$, explaining why such a weak dependency was found numerically.

\item \textit{Dependency on the core composition.} $R_{\rm bare}$ is found analytically to increase linearly, i.e., with an in comparison  strong dependency, with ice mass fraction.

\end{itemize}

\section{Conclusions}\label{sect:conclusions} 
From the points {summarized above}, we can draw three important conclusions: 
\begin{enumerate}
\item Because of its weak dependency on the post-formation envelope mass, the position of the valley cannot be used to constrain well the efficiency of gas accretion during the nebular phase (except of course for the fact that its sheer existence shows that these planets have accreted H/He.). But it actually appears likely that it is  because of this independency on the post-formation envelope mass that the valley is clearly (or even at all) {visible} observationally: it means that the a spread in the post-formation envelope masses resulting from the formation epoch does not matter. 
\item In a similar vain, the weak dependency of the valley's position on the strength of evaporation also has important implications:  First, this explains why  on the theoretical side various  models of evaporation \citep[like][]{owenwu2013,lopezfortney2013,jinmordasini2014,chenrogers2016} found similar locations of the valley, i.e., why the details ({one} could also say uncertainties) of these models  do not matter  much. Second, on the observational side, the  weak  dependency on the strength of evaporation probably {also} explains why we see a  {valley} at all: the  observed spread in the $L_{\rm XUV}$ and saturation times of young stars does not  alter the location of the valley to an extent that it completely blurs away.  Combined, these two points mean that the evaporation valley is a  robust feature.
\item  At the same time, given the significant spread of $L_{\rm XUV}$ of young stars, the dependency of the valley's location on it is still strong enough to explain why the valley is not completely empty: as discussed in the context of the Simulations M3, E1, E2  (Sect. \ref{sect:evaprateE1E2})  the location of the valley likely varies from star to star because of different $L_{\rm XUV}$, but also different efficiency factors by about 0.5 $\rearth$ around the mean value. The intrinsic valley width has a similar magnitude (\citealt{owenwu2017}). Therefore, we can infer that by overlaying these simulations with different strength of evaporation, there would still be a depleted region, but not a completely empty one. This  corresponds to the observational result \citep{fultonpetigura2018}. The individual numerical simulations presented here  give the  misleading impression that the valley should be empty, but this is simply an artifact of assuming an identical $L_{\rm XUV}$ for all stars in one simulation. Clearly, in the actual  observations, we in contrast see the overlay of all individual star/system-specific  $L_{\rm XUV}$ and thus valleys which are not exactly at the same position. As we mentioned above, the fact that the valley is not completely empty, but to some extend filled does thus not mean that a compositional diversity of the cores is {needed} to explain the observations. Rather, a partially filled valley is (also) a natural consequence of evaporation given the spread of stellar rotation rates and thus $L_{\rm XUV}$.

\end{enumerate}

Looking ahead, to further understand the valley and its origin, and in particular atmospheric escape as a possible explanation, this hypothesis can be put to a number of observational tests: first, simply by determining precisely the density of transiting planets on both sides of the valley for example with TESS and CHEOPS plus high-accuracy radial velocities \citep[e.g.,][]{dumusqueturner2019}. This would reveal whether the densities are indeed consistent with on one side super-Earth planets with an Earth-like composition without H/He, and sub-Neptunes with H/He on the other side. While this test is obvious and simple in  principle, it is not so trivial in practice, because the observational error bars on the masses and thus the density are often significant given the low masses of the planets. As shown in \citet{jinmordasini2018} where a simple analysis of the densities of planets on both sides of the valley was done, the currently available sample of planets with sufficiently well-constrained densities is small. But it  at least does not contain planets with densities inconsistent with evaporation, namely planets with a density only explicable with H/He below the valley, or high density planets without volatile above it. The second test would be to obtain UV observations of atmospheric escape rates of planets on both sides with the HST. As shown for example in \citet[][their Figs. 5 and 8]{jinmordasini2014}, at an age of 5 Gyr, close-in planets just above the valley should have evaporation rates of $10^{9}$ to $10^{10}$ g/s (and much more at $\sim$100 Myr), while below it, the escape rates should be orders of magnitude lower. For comparison, the escape rate of the warm Neptune-mass planet GJ 3470b was  measured to be about $10^{10}$ g/s \citep{bourrierlecavelier2018}. Third, studying the dependency of the valleys on the spectral type of the host star (see {\citealt{fultonpetigura2018,wu2018,guptaschlichting2019,mcdonaldkreidberg2019,cloutiermenou2019}}) should in particular allow to disentangle the different proposed loss mechanisms. Finally, with PLATO, it may even become possible to measure the valley's locus as  a function of time, and to compare with the {various} theoretical predictions.

\acknowledgements{I would like to thank Sheng Jin, Paul Molli\`ere,  Gabriel-Dominique Marleau, Apurva Oza, {Alexandre Emsenhuber, Jonas Haldemann, and Julia Venturini.  I thank the referee for helpful comments and}  acknowledge the support from the Swiss National Science Foundation under grant BSSGI0$\_$155816 ``PlanetsInTime''. Parts of this work have been carried out within the framework of the NCCR PlanetS supported by the Swiss National Science Foundation.}

\appendix
%\newpage
\section{Analytical fits for the luminosity of low-mass planets as a function of core mass, envelope mass, and time}\label{appendix:lumifit}
In this appendix we use results from a high number of numerical calculations of planetary evolution to derive a fit for the mean luminosity of low-mass planets as a function of their core mass, envelope mass, and time. 

In a similar spirit as the analytical expressions derived in \citet{lopezfortney2013b} for the radii of close-in low-mass planets, these fits are not meant to be an exact replacement of a full cooling calculation of an individual planet. But such fits are of interest for a first rough estimate of the luminosity, and can in particular be used as input for static (time-independent) interior structure models like \citet[][]{rogersbodenheimer2011,dornventurini2017,lozovskyhelled2018}. Such models need the luminosity as a function of planet properties and time as an input parameter.  

To derive the fit, we consider the luminosity of non-evaporating low-mass planets in a population that repeats the coupled formation and evolution calculations presented in \citet{mordasinialibert2012c}, but now includes also the cooling model of the solid core described in \citet{lindermordasini2018}. The luminosity thus  includes the contribution of the cooling and contraction of the core and envelope, plus the radiogenic luminosity. To derive the fit, planets with a core mass $1\leq\mcore/\mearth\leq20$, a envelope mass $\menve/\mearth\leq20$ and an ice mass fraction of $\fice\leq0.1$ are included.

\begin{figure}
\begin{center}
\includegraphics[width=\columnwidth]{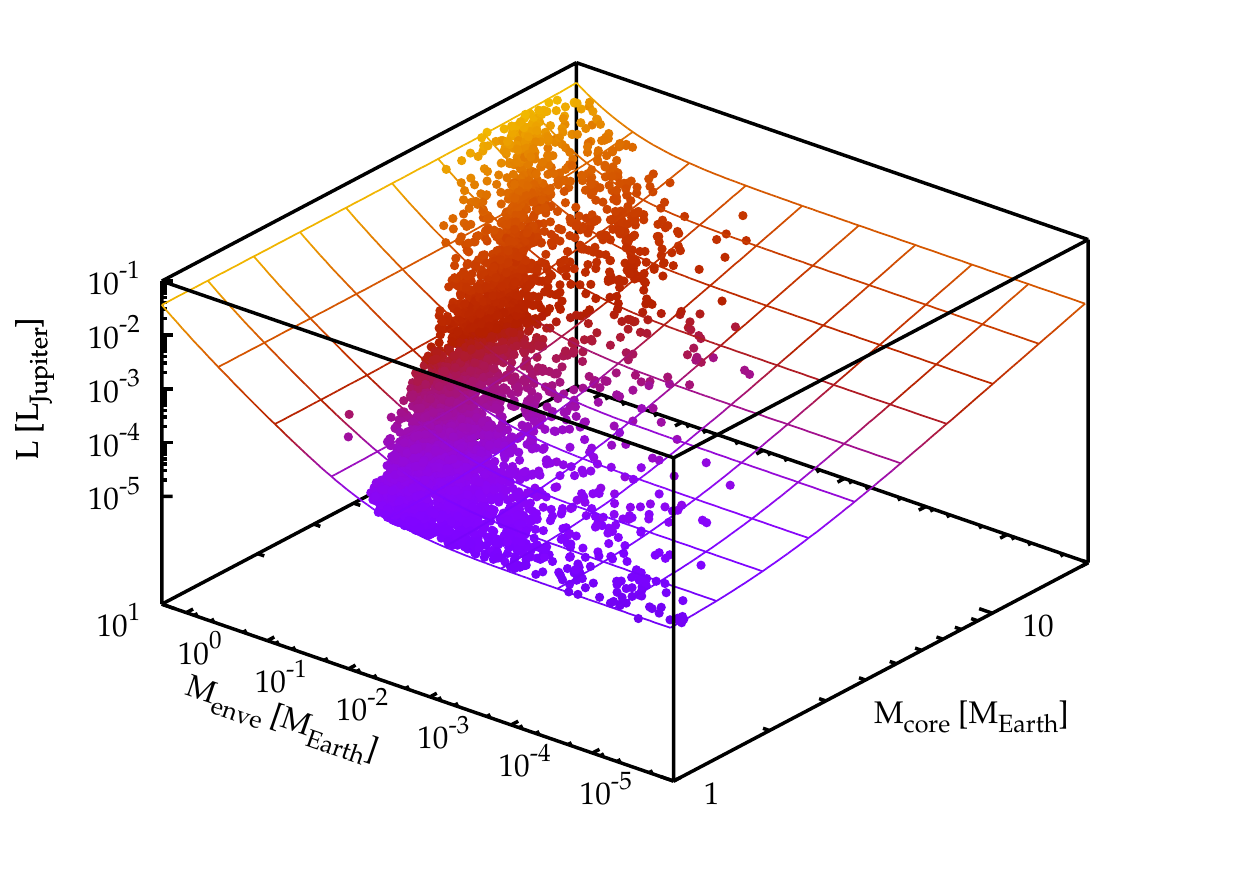}
\caption{Luminosity of low-mass planets with H/He as a function of core and envelope mass at 5 Gyrs. The points are the luminosities of synthetic planets in a population around a solar-like star. The grid shows the corresponding biquadratic fit.}\label{fig:lumimcme}
\end{center}
\end{figure}

Figure \ref{fig:lumimcme} shows the luminosity of the synthetic planets as a function of core and envelope mass at an age of 5 Gyrs. The plot shows that the luminosity can be approximated with a power law. We chose a biquadratic fit in $\mcore$ and $\menve$, giving  a fit of the form
\beq\label{eq:lumifitappendix}
\frac{L}{\lj}=a_{0}+b_{1} \left(\frac{\mcore}{\mearth}\right)+b_{2} \left(\frac{\mcore}{\mearth}\right)^{2}+c_{1} \left(\frac{\menve}{\mearth}\right)+c_{2} \left(\frac{\menve}{\mearth}\right)^{2}.
\eeq
The parameters $a_{0},b_{1},b_{2},c_{1}$ and $c_{2}$ are derived by the least square method separately at 14 different moments in time between 0.1 and 10 Gyrs. Note that despite the fact that the expression gives $L$  as a function of $\mcore$ and $\menve$ separately, these two quantities are statistically not independent of each other in the population that was used to derive it, as is visible in Fig. \ref{fig:lumimcme}. Since the cooling of the core and envelope are interdependent, this means that the fit is strictly speaking only applicable to planets that follow a similar $ \mcore-\menve$ relation. 

Note also that the expression is more complex than the scaling of the post-formation luminosity (at 10 Myr) used in the parameter study presented above (Eq. \ref{eq:l0par}). However, except for very low-mass (a few Earth masses) planets with  massive envelopes with a mass fraction of $\gtrsim 10$\% which are not usually forming in the formation simulations, the expressions yield roughly speaking comparable luminosities at 100 Myr when accounting for the fact that the luminosities decreases by approximately one order of magnitude from 10 to 100 Myr. But the fit of  Eq.\ref{eq:lumifitappendix} should of course be preferred when possible.

It is interesting to consider the luminosity predicted by the fit for specific planets. For an Earth-like planet ($\mcore/\mearth=1$, $\menve/\mearth=0$), the expressions yields a luminosity of 0.6 $L_{\oplus}$ at 5 Gyr, where $L_{\oplus}$ is the intrinsic luminosity of the Earth originating from radiogenic decay and delayed secular cooling, that is estimated to be about $4\times10^{20}$ erg/s \citep{kamland2011}, or about $10^{-4} \lj$. For a Neptune-like planet (assuming $\mcore/\mearth=15.4$, $\menve/\mearth=1.7$) of the same age, the expression yields 0.007 $\lj$, while the observed value is about 0.01 $ \lj$ \citep{guillotgautier2014}. Finally, for a Jovian-like planet with $\mcore/\mearth=17.8$ and $\menve/\mearth=300$ which is 15 times higher than the maximal envelope mass used to derive the fit (meaning that the fit is in principle not usable), it still predicts with 2.78 $\lj$ at 4.5 Gyr a value that is not orders of magnitudes off. 

\begin{table*}[htp]
\caption{Parameters for the analytical fit of the luminosity of low-mass planets.}\label{tab:lumifit}
\begin{center}
\begin{tabular}{lccccc}
Time [Gyr]   &	$a_{0}$ 	 & 			$ b_{1}$	&	 $b_{2}$  & 	$c_{1}$ &  $c_{2}$  \\ \hline
0.1 &	0.002529060&-0.002002380&0.001044080&0.05864850&0.000967878\\
0.3  &	0.001213240&-0.000533601&0.000360703&0.02141140&0.000368533 \\
0.5  &	0.000707416&-0.000394131&0.000212475&0.01381380&0.000189456\\
0.8  &	0.000423376&-0.000187283&0.000125872&0.00887292&0.000117141\\
1.0  &	0.000352187&-0.000141480&9.94382$\times10^{-5}$&0.00718831&9.20563$\times10^{-5}$ \\
2.0 &	0.000175775&-4.07832$\times10^{-5}$&4.58530$\times10^{-5}$&0.00357941&5.52851$\times10^{-5}$ \\
3.0 &	0.000114120&-2.09944$\times10^{-5}$&2.91169$\times10^{-5}$&0.00232693&4.00546$\times10^{-5}$ \\
4.0  &	8.81462$\times10^{-5}$&-2.07673$\times10^{-5}$&2.12932$\times10^{-5}$&0.00171412&2.90984$\times10^{-5}$ \\
5.0  &	6.91819$\times10^{-5}$&-1.90159$\times10^{-5}$&1.62128$\times10^{-5}$&0.00134355&2.30387$\times10^{-5}$ \\
6.0 &    5.49615$\times10^{-5}$&-1.68620$\times10^{-5}$&1.29045$\times10^{-5}$&0.00109019&1.96163$\times10^{-5}$ \\
7.0 &    4.50320$\times10^{-5}$&-1.51951$\times10^{-5}$&1.05948$\times10^{-5}$&0.00091005&1.70934$\times10^{-5}$ \\
8.0  &	3.80363$\times10^{-5}$&-1.40113$\times10^{-5}$&8.93639$\times10^{-6}$&0.00077687&1.50107$\times10^{-5}$\\ 
9.0  &	3.30102$\times10^{-5}$&-1.31146$\times10^{-5}$&7.69121$\times10^{-6}$&0.000675243&1.32482$\times10^{-5}$ \\ 
10.0 &	2.92937$\times10^{-5}$&-1.24023$\times10^{-5}$&6.73922$\times10^{-6}$&0.000595191&1.17809$\times10^{-5}$ \\ \hline
\end{tabular}
\end{center}
\label{default}
\end{table*}%

Finally, we compare the luminosity as a function of time as predicted by the analytic fit with the direct cooling calculations. One cannot expect that the two approaches yield identical results for an individual planet, because the fit reproduces only the statistical mean behavior, and a subpopulation of synthetic planets only was used to derive the $\mcore-\menve-t$ relation. This in particular means that individual planets may have different mass fractions of ice in the core, as well as different $\mcore$-$\menve$ relations than the ones used to derive the fit.

\begin{figure}
\begin{center}
\includegraphics[width=\columnwidth]{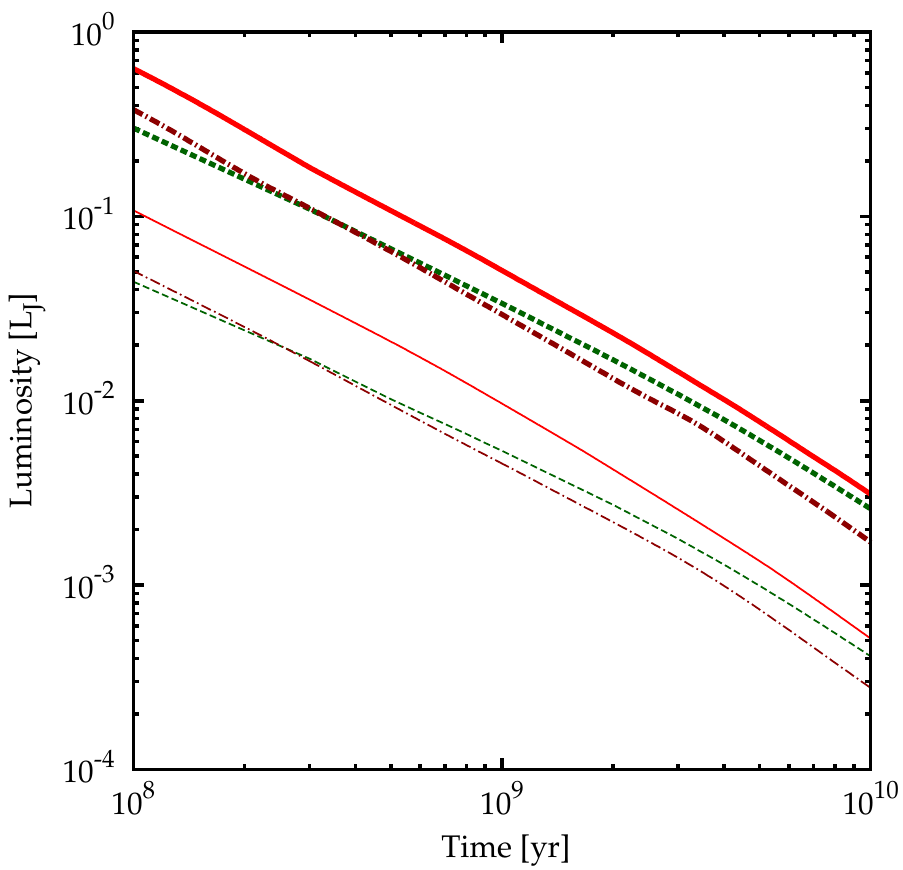}
\caption{Comparison of the luminosity as a function of time as predicted by direct cooling calculations and the fit. The upper three thicker lines are for a 15 $\mearth$ planet while the three thin lines below are for a 5 $\mearth$ planet. The green dashed line is the fit of Eq. \ref{eq:lumifitappendix}. The other lines are direct simulations. The brown dashed-dotted and the red solid lines assume a rocky core, and a core consisting of 50\% ice and 50\% rocky material, respectively.}\label{fig:compLt15Me5Me}
\end{center}
\end{figure}

Figure \ref{fig:compLt15Me5Me} shows the luminosity as a function of time for a 5 and 15 $\mearth$ planet (total mass) with an envelope of 0.5 and 2.7 $\mearth$, respectively. For both planets, three lines are shown: the  cooling curve obtained numerically for a core consisting of 50\% ice and 50\% rocky material, the cooling curve as found with identical settings except for a completely rocky core instead, and finally the luminosity predicted by Eq. \ref{eq:lumifitappendix}, interpolating linearly in $\log(L(\log(t)))$ using the parameters for the temporal dependency from Table \ref{tab:lumifit}.  We see that in both cases, the fit and numerical result yield roughly similar result, but that at early times, the luminosity predicted by the fit is approximately a factor two smaller than in the nominal model. The agreement with the numerical cooling curve  is clearly better for  planets with a rocky core, with relative differences that are on the order of 30\% or less. This better agreement is not surprising, since the fit was derived for planets with $\fice\leq0.1$, i.e., with (mainly) rocky cores.  We thus see that for otherwise identical parameters, a planet with a rocky core has a lower luminosity. This is due to the lower thermal expansion coefficient and the lower specific heat capacity of silicates relative to ice \citep{baraffechabrier2008,lindermordasini2018}.

A simple Fortran program implementing the interpolation routine can be found at 
\url{https://www.space.unibe.ch/research/research_groups/planets_in_time/numerical_data/index_eng.html}

\section{{ANEOS equation of state  for H$_{2}$O}}\label{appendix:aneos}
{ANEOS \citep{thompson1990} is an equation of state that yields the thermodynamic properties of various materials  derived from an analytical expression of the Helmholtz free energy. It has been used widely in the context of planetary interiors (for example \citealt{baraffechabrier2008,thorngrenfortney2016,lopez2017,lozovskyhelled2018}). While ANEOS has a number of limitations (see \citealt{baraffechabrier2014} for a discussion), and while there are  more modern water EOSs based on ab initio simulations  \citep[e.g.,][]{nettelmannholst2008,frenchmattson2009,soubiranmilitzer2015,mazevetlicari2019}, we are here interested mainly in an EOS that covers a wide range of validity as opposed to a very high degree of accuracy: in the current paper, H$_{2}$O is a placeholder for heavy elements in general. Additionally, ANEOS has the advantage of yielding a large number of  thermodynamical quantities of interest, including the density $\rho$, specific internal energy $u$, specific entropy $s$, specific thermal capacity at constant volume $c_{\rm v}$, sound speed $c_{\rm s}$,  and the radiative opacity $\kappa$. It also provides the derivative of the pressure with respect to temperature at constant density, and the derivative of the pressure with respect to density at constant temperature. This allows in particular to calculate via the Maxwell relations the adiabatic gradient $\nabla_{\rm ad}$ which is needed for planetary interior structure calculations as \citep{leconte2011}}
\beq
\nabla_{\rm ad}=\left.\frac{\partial \ln T}{\partial \ln P}\right|_{\rm s}=\frac{P}{T}\left( \left.\frac{\partial P}{\partial T}\right|_{\rho} + \left.\frac{\partial P}{\partial \rho}\right|_{T} \left.\frac{\partial \rho}{\partial T}\right|_{s} \right)^{-1}=\frac{P\left.\frac{\partial P}{\partial T}\right|_{\rho}}{c_{\rm v} \rho^{2}\left.\frac{\partial P}{\partial \rho}\right|_{T} + T \left.\frac{\partial P}{\partial T}\right|_{\rho}^{2}}.
\eeq

{The fundamental approach of ANEOS is to write an analytical expression of the Helmholtz free energy $F$ in the system. $F$ is minimal at equilibrium for a system with constant $T$ and $V$ (or $\rho$). Therefore, ANEOS takes $T$ and $\rho$ as input. For the current application for planetary interiors, where $T$ and $P$ are the independent variables, we have re-tabulated the ANEOS output.} {In ANEOS, $F(\rho,T)$ is written as sum of three contributions \citep{melosh1990}}
\beq
F(\rho,T)=F_{\rm cold}(\rho)+F_{\rm thermal}(\rho,T)+F_{\rm electronic}(\rho,T)
\eeq
{where the first term represents a ``cold'' component of atomic interactions independent of temperature, e.g., the interatomic potential (modeled as a Morse potential). The second is a ``thermal'' component for the thermal energy of the nuclei and their associated electrons. The third term is an ``electronic'' contribution that represents the energies associated with the ionization of atoms. For some of these three contributions, internally different models can be used. The ``thermal'' contribution is a combination of the free energy of a generalized Debye solid and the monoatomic gas limit at high temperatures. In a Debye solid, the energy is calculated as the phonon energy times the average number of phonons times the number of modes. Ionization can be treated with either via a Saha or Thomas Fermi model. At very high pressures, ANEOS approaches the Thomas-Fermi limit  of a free electron gas with embedded nuclei, where the pressure of the degenerate electron gas is obtained via the Thomas-Fermi-Dirac approximation. An advantage of ANEOS is thus a correct behavior in the asymptotic limit of very high pressures. }
 
{ Once $F$ is known, all other thermodynamic quantities can be calculated. For example, the entropy  and pressure are obtained as}
\beq
S=-\left.\frac{\partial F}{\partial T}\right|_{\rho} \ \ \ \    \\  \ \ \ \ \ P=\rho^{2}\left.\frac{\partial F}{\partial \rho}\right|_{T}.
 \eeq

{Because of its first principles approach, ANEOS includes a treatment of phase changes, which makes it, at least in principle,  superior to some empirical EOS or tabulated EOS which are often only available for a certain phase or $P-T$ domain. Several models for phase changes are included in ANEOS. Here we use the solid-liquid-gas with ionization mode (ANEOS Mode 4).}

{ANEOS requires an extended set of coefficients for any given material which can in part be derived from laboratory experiments, like  shock wave data. For  H$_{2}$O, we use the following coefficients (see also \citealt{turtlepierazzo2001}):}
{\tiny 
\begin{verbatim}
ANEOS1 -2  'H2O' THUG=-1   RHUG=-1  LONG
* nel  neos  rhoref   tref  pref  Bref   Grun.  Tdebye
ANEOS2  2  4  1.11  0.02008  0  -1.7e5  0.58  -0.045
*  Tg 3C24  Es  Tmelt      C53     C54   H0   C41
ANEOS3  .9   2  3.34e10  0.0224      0       0    0    0
*     rhomin  D1    D2    D3    D4    D5    Hf     -rholiq
ANEOS4   0.0   0.0   0.0   0.0   0.0   0.0  4.85e9  -1.
*       Up    Lo   alpha  beta  gamma  C60   C61   C62
ANEOS5   0.0  0.0   0.0    0.0   0.0   0.8   0.     0.26
* Ioniz. model  Reactive chem. Molec. Clusters   Pc
* Flag Eshift Sshift Rbond  Ebind  IntDOF  flag  exp
ANEOS6   1  0.  0.  0.E-8    3.2   0.  1.   1.8
*       Zi      fi
ANEOS7   1      2./3.
ANEOS8   8      1./3.
\end{verbatim}}

\begin{figure*}
\begin{center}
\includegraphics[width=\textwidth]{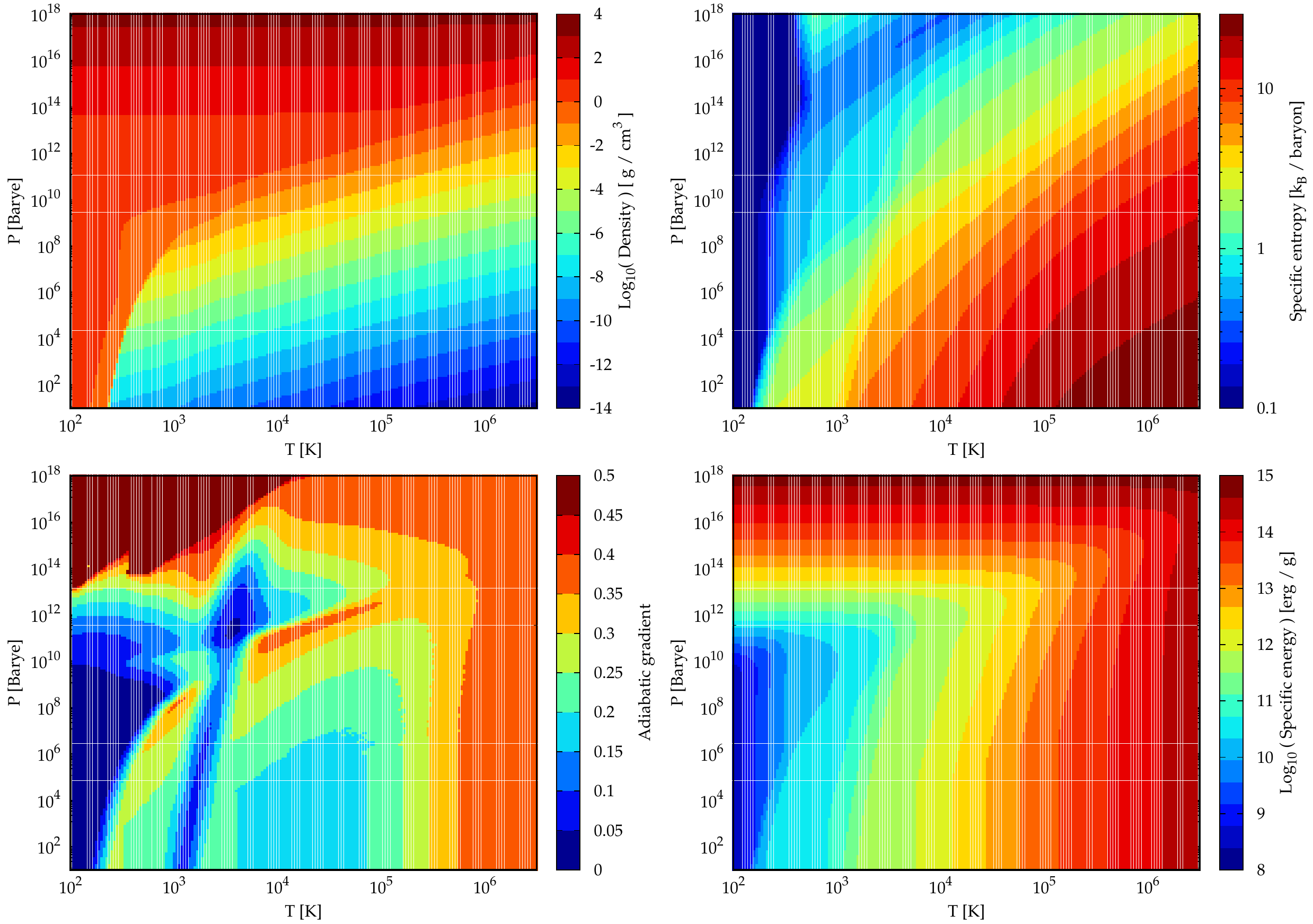}
\caption{{Thermodynamical properties of H$_{2}$O as predicted by ANEOS for a very large range in pressure $P$ and temperature $T$. The density (top left), specific entropy (top right), adiabatic gradient (bottom left), and specific internal energy (bottom right) are shown. One sees the effects of phase changes and of the asymptotic behavior under extreme conditions (ideal gas limit at high $T$ and low $P$; fully degenerate limit at very high $P$ and low $T$). }}\label{fig:aneos}
\end{center}
\end{figure*}

{Figure \ref{fig:aneos} shows the density $\rho$, specific entropy $s$, adiabatic gradient $\nabla_{\rm ad}$, and specific internal energy $u$ as a function of pressure and temperature. A very large range in $P-T$ is covered, ranging from the low values found in planetary atmospheres ($\sim$100 K, milibars) to extremely high values exceeding those expected in the center of a 5 Gyr old 20 Jupiter mass planet (about $3\times10^{5}$ K, $2\times10^{16}$ Barye, \citealt{mordasinialibert2012b}) . In the panel showing the density, one sees the various phases (solid, liquid, gaseous) as in a normal water phase diagram. Note that ANEOS predicts that the density slightly decreases when passing at a pressure of 1 bar ($10^{6}$ Barye) from the solid into the liquid state - clearly, it cannot capture the special behavior (anomaly) of H$_{2}$O where the density of liquid water is higher then the one of ice. At the phase transition, we also see a strong increase of the entropy when passing from the liquid into the gaseous phase, as expected. In the density panel, one can also see the diagonal structure of the iso-density contours in the lower right part of the $P-T$ domain. In this part, the water vapor (and plasma, after ionization) behaves like an ideal gas, meaning that $p\propto T$ on iso-density contours.  At very high pressures, we see that the density becomes in contrast independent of temperature, leading to horizontal iso-density lines. This is the expected behavior for completely degenerate  matter. We have found that ANEOS fails to provide physical values for the adiabatic gradient at very high pressures and low temperatures, such that we have limited $\nabla_{\rm ad}$ in this regime to 0.5, as visible in the top left corner of the $\nabla_{\rm ad}$ panel. Fortunately, this regime is not occurring in planetary interiors at the current age of the universe. }

%\url{http://www.space.unibe.ch/research/research_groups/planets_in_time/numerical_data/index_eng.html}
\bibliographystyle{aa} % style aa.bst 
\bibliography{/Users/chris/Dropbox/work/ownpublications/BibTeX/libcomb20162018.bib} 
\end{document}